%% file: main.tex
\documentclass[sigconf,screen,table]{acmart}

\setcopyright{acmlicensed}
\acmPrice{15.00}
\acmDOI{10.1145/3236024.3236025}
\acmYear{2018}
\copyrightyear{2018}
\acmISBN{978-1-4503-5573-5/18/11}
\acmConference[ESEC/FSE '18]{Proceedings of the 26th ACM Joint European Software Engineering Conference and Symposium on the Foundations of Software Engineering}{November 4--9, 2018}{Lake Buena Vista, FL, USA}
\acmBooktitle{Proceedings of the 26th ACM Joint European Software Engineering Conference and Symposium on the Foundations of Software Engineering (ESEC/FSE '18), November 4--9, 2018, Lake Buena Vista, FL, USA}

\usepackage{booktabs}   %% For formal tables:
\usepackage{subcaption} %% For complex figures with subfigures/subcaptions
\usepackage{amsmath}
\usepackage{amssymb}
\usepackage{amsthm}
\usepackage{stmaryrd}
\usepackage{adjustbox}
\usepackage{multirow}
\usepackage{mathtools}
\usepackage{todonotes}
\usepackage{enumitem}

\usepackage{listings}
\definecolor{sh_comment}{rgb}{0.12, 0.38, 0.18 } %adjusted, in Eclipse: {0.25, 0.42, 0.30 } = #3F6A4D
\definecolor{sh_keyword}{rgb}{0.37, 0.08, 0.25}  % #5F1441
\definecolor{sh_string}{rgb}{0.06, 0.10, 0.98} % #101AF9
\usepackage{fancyvrb}
\usepackage{pdfrender}
\usepackage{tikz}
\usetikzlibrary{calc,decorations.pathmorphing,shapes}
\usetikzlibrary{decorations.markings}
\usetikzlibrary{decorations.pathreplacing}
\tikzset{snake it/.style={decorate, decoration=snake}}

\input{macros}

\makeatletter
\newtheorem*{rep@theorem}{\rep@title}
\newcommand{\newreptheorem}[2]{%
\newenvironment{rep#1}[1]{%
 \def\rep@title{#2 \ref{##1}}%
 \begin{rep@theorem}}%
 {\end{rep@theorem}}}
\makeatother

\makeatletter
\def\PYG@reset{\let\PYG@it=\relax \let\PYG@bf=\relax%
    \let\PYG@ul=\relax \let\PYG@tc=\relax%
    \let\PYG@bc=\relax \let\PYG@ff=\relax}
\def\PYG@tok#1{\csname PYG@tok@#1\endcsname}
\def\PYG@toks#1+{\ifx\relax#1\empty\else%
    \PYG@tok{#1}\expandafter\PYG@toks\fi}
\def\PYG@do#1{\PYG@bc{\PYG@tc{\PYG@ul{%
    \PYG@it{\PYG@bf{\PYG@ff{#1}}}}}}}
\def\PYG#1#2{\PYG@reset\PYG@toks#1+\relax+\PYG@do{#2}}

\expandafter\def\csname PYG@tok@gd\endcsname{\def\PYG@tc##1{\textcolor[rgb]{0.63,0.00,0.00}{##1}}}
\expandafter\def\csname PYG@tok@gu\endcsname{\let\PYG@bf=\textbf\def\PYG@tc##1{\textcolor[rgb]{0.50,0.00,0.50}{##1}}}
\expandafter\def\csname PYG@tok@gt\endcsname{\def\PYG@tc##1{\textcolor[rgb]{0.00,0.27,0.87}{##1}}}
\expandafter\def\csname PYG@tok@gs\endcsname{\let\PYG@bf=\textbf}
\expandafter\def\csname PYG@tok@gr\endcsname{\def\PYG@tc##1{\textcolor[rgb]{1.00,0.00,0.00}{##1}}}
\expandafter\def\csname PYG@tok@cm\endcsname{\let\PYG@it=\textit\def\PYG@tc##1{\textcolor[rgb]{0.25,0.50,0.50}{##1}}}
\expandafter\def\csname PYG@tok@vg\endcsname{\def\PYG@tc##1{\textcolor[rgb]{0.10,0.09,0.49}{##1}}}
\expandafter\def\csname PYG@tok@vi\endcsname{\def\PYG@tc##1{\textcolor[rgb]{0.10,0.09,0.49}{##1}}}
\expandafter\def\csname PYG@tok@vm\endcsname{\def\PYG@tc##1{\textcolor[rgb]{0.10,0.09,0.49}{##1}}}
\expandafter\def\csname PYG@tok@mh\endcsname{\def\PYG@tc##1{\textcolor[rgb]{0.40,0.40,0.40}{##1}}}
\expandafter\def\csname PYG@tok@cs\endcsname{\let\PYG@it=\textit\def\PYG@tc##1{\textcolor[rgb]{0.25,0.50,0.50}{##1}}}
\expandafter\def\csname PYG@tok@ge\endcsname{\let\PYG@it=\textit}
\expandafter\def\csname PYG@tok@vc\endcsname{\def\PYG@tc##1{\textcolor[rgb]{0.10,0.09,0.49}{##1}}}
\expandafter\def\csname PYG@tok@il\endcsname{\def\PYG@tc##1{\textcolor[rgb]{0.40,0.40,0.40}{##1}}}
\expandafter\def\csname PYG@tok@go\endcsname{\def\PYG@tc##1{\textcolor[rgb]{0.53,0.53,0.53}{##1}}}
\expandafter\def\csname PYG@tok@cp\endcsname{\def\PYG@tc##1{\textcolor[rgb]{0.74,0.48,0.00}{##1}}}
\expandafter\def\csname PYG@tok@gi\endcsname{\def\PYG@tc##1{\textcolor[rgb]{0.00,0.63,0.00}{##1}}}
\expandafter\def\csname PYG@tok@gh\endcsname{\let\PYG@bf=\textbf\def\PYG@tc##1{\textcolor[rgb]{0.00,0.00,0.50}{##1}}}
\expandafter\def\csname PYG@tok@ni\endcsname{\let\PYG@bf=\textbf\def\PYG@tc##1{\textcolor[rgb]{0.60,0.60,0.60}{##1}}}
\expandafter\def\csname PYG@tok@nl\endcsname{\def\PYG@tc##1{\textcolor[rgb]{0.63,0.63,0.00}{##1}}}
\expandafter\def\csname PYG@tok@nn\endcsname{\let\PYG@bf=\textbf\def\PYG@tc##1{\textcolor[rgb]{0.00,0.00,1.00}{##1}}}
\expandafter\def\csname PYG@tok@no\endcsname{\def\PYG@tc##1{\textcolor[rgb]{0.53,0.00,0.00}{##1}}}
\expandafter\def\csname PYG@tok@na\endcsname{\def\PYG@tc##1{\textcolor[rgb]{0.49,0.56,0.16}{##1}}}
\expandafter\def\csname PYG@tok@nb\endcsname{\def\PYG@tc##1{\textcolor[rgb]{0.00,0.50,0.00}{##1}}}
\expandafter\def\csname PYG@tok@nc\endcsname{\let\PYG@bf=\textbf\def\PYG@tc##1{\textcolor[rgb]{0.00,0.00,1.00}{##1}}}
\expandafter\def\csname PYG@tok@nd\endcsname{\def\PYG@tc##1{\textcolor[rgb]{0.67,0.13,1.00}{##1}}}
\expandafter\def\csname PYG@tok@ne\endcsname{\let\PYG@bf=\textbf\def\PYG@tc##1{\textcolor[rgb]{0.82,0.25,0.23}{##1}}}
\expandafter\def\csname PYG@tok@nf\endcsname{\def\PYG@tc##1{\textcolor[rgb]{0.00,0.00,1.00}{##1}}}
\expandafter\def\csname PYG@tok@si\endcsname{\let\PYG@bf=\textbf\def\PYG@tc##1{\textcolor[rgb]{0.73,0.40,0.53}{##1}}}
\expandafter\def\csname PYG@tok@s2\endcsname{\def\PYG@tc##1{\textcolor[rgb]{0.73,0.13,0.13}{##1}}}
\expandafter\def\csname PYG@tok@nt\endcsname{\let\PYG@bf=\textbf\def\PYG@tc##1{\textcolor[rgb]{0.00,0.50,0.00}{##1}}}
\expandafter\def\csname PYG@tok@nv\endcsname{\def\PYG@tc##1{\textcolor[rgb]{0.10,0.09,0.49}{##1}}}
\expandafter\def\csname PYG@tok@s1\endcsname{\def\PYG@tc##1{\textcolor[rgb]{0.73,0.13,0.13}{##1}}}
\expandafter\def\csname PYG@tok@dl\endcsname{\def\PYG@tc##1{\textcolor[rgb]{0.73,0.13,0.13}{##1}}}
\expandafter\def\csname PYG@tok@ch\endcsname{\let\PYG@it=\textit\def\PYG@tc##1{\textcolor[rgb]{0.25,0.50,0.50}{##1}}}
\expandafter\def\csname PYG@tok@m\endcsname{\def\PYG@tc##1{\textcolor[rgb]{0.40,0.40,0.40}{##1}}}
\expandafter\def\csname PYG@tok@gp\endcsname{\let\PYG@bf=\textbf\def\PYG@tc##1{\textcolor[rgb]{0.00,0.00,0.50}{##1}}}
\expandafter\def\csname PYG@tok@sh\endcsname{\def\PYG@tc##1{\textcolor[rgb]{0.73,0.13,0.13}{##1}}}
\expandafter\def\csname PYG@tok@ow\endcsname{\let\PYG@bf=\textbf\def\PYG@tc##1{\textcolor[rgb]{0.67,0.13,1.00}{##1}}}
\expandafter\def\csname PYG@tok@sx\endcsname{\def\PYG@tc##1{\textcolor[rgb]{0.00,0.50,0.00}{##1}}}
\expandafter\def\csname PYG@tok@bp\endcsname{\def\PYG@tc##1{\textcolor[rgb]{0.00,0.50,0.00}{##1}}}
\expandafter\def\csname PYG@tok@c1\endcsname{\let\PYG@it=\textit\def\PYG@tc##1{\textcolor[rgb]{0.25,0.50,0.50}{##1}}}
\expandafter\def\csname PYG@tok@fm\endcsname{\def\PYG@tc##1{\textcolor[rgb]{0.00,0.00,1.00}{##1}}}
\expandafter\def\csname PYG@tok@o\endcsname{\def\PYG@tc##1{\textcolor[rgb]{0.40,0.40,0.40}{##1}}}
\expandafter\def\csname PYG@tok@kc\endcsname{\let\PYG@bf=\textbf\def\PYG@tc##1{\textcolor[rgb]{0.00,0.50,0.00}{##1}}}
\expandafter\def\csname PYG@tok@c\endcsname{\let\PYG@it=\textit\def\PYG@tc##1{\textcolor[rgb]{0.25,0.50,0.50}{##1}}}
\expandafter\def\csname PYG@tok@mf\endcsname{\def\PYG@tc##1{\textcolor[rgb]{0.40,0.40,0.40}{##1}}}
\expandafter\def\csname PYG@tok@err\endcsname{\def\PYG@bc##1{\setlength{\fboxsep}{0pt}\fcolorbox[rgb]{1.00,0.00,0.00}{1,1,1}{\strut ##1}}}
\expandafter\def\csname PYG@tok@mb\endcsname{\def\PYG@tc##1{\textcolor[rgb]{0.40,0.40,0.40}{##1}}}
\expandafter\def\csname PYG@tok@ss\endcsname{\def\PYG@tc##1{\textcolor[rgb]{0.10,0.09,0.49}{##1}}}
\expandafter\def\csname PYG@tok@sr\endcsname{\def\PYG@tc##1{\textcolor[rgb]{0.73,0.40,0.53}{##1}}}
\expandafter\def\csname PYG@tok@mo\endcsname{\def\PYG@tc##1{\textcolor[rgb]{0.40,0.40,0.40}{##1}}}
\expandafter\def\csname PYG@tok@kd\endcsname{\let\PYG@bf=\textbf\def\PYG@tc##1{\textcolor[rgb]{0.00,0.50,0.00}{##1}}}
\expandafter\def\csname PYG@tok@mi\endcsname{\def\PYG@tc##1{\textcolor[rgb]{0.40,0.40,0.40}{##1}}}
\expandafter\def\csname PYG@tok@kn\endcsname{\let\PYG@bf=\textbf\def\PYG@tc##1{\textcolor[rgb]{0.00,0.50,0.00}{##1}}}
\expandafter\def\csname PYG@tok@cpf\endcsname{\let\PYG@it=\textit\def\PYG@tc##1{\textcolor[rgb]{0.25,0.50,0.50}{##1}}}
\expandafter\def\csname PYG@tok@kr\endcsname{\let\PYG@bf=\textbf\def\PYG@tc##1{\textcolor[rgb]{0.00,0.50,0.00}{##1}}}
\expandafter\def\csname PYG@tok@s\endcsname{\def\PYG@tc##1{\textcolor[rgb]{0.73,0.13,0.13}{##1}}}
\expandafter\def\csname PYG@tok@kp\endcsname{\def\PYG@tc##1{\textcolor[rgb]{0.00,0.50,0.00}{##1}}}
\expandafter\def\csname PYG@tok@w\endcsname{\def\PYG@tc##1{\textcolor[rgb]{0.73,0.73,0.73}{##1}}}
\expandafter\def\csname PYG@tok@kt\endcsname{\def\PYG@tc##1{\textcolor[rgb]{0.69,0.00,0.25}{##1}}}
\expandafter\def\csname PYG@tok@sc\endcsname{\def\PYG@tc##1{\textcolor[rgb]{0.73,0.13,0.13}{##1}}}
\expandafter\def\csname PYG@tok@sb\endcsname{\def\PYG@tc##1{\textcolor[rgb]{0.73,0.13,0.13}{##1}}}
\expandafter\def\csname PYG@tok@sa\endcsname{\def\PYG@tc##1{\textcolor[rgb]{0.73,0.13,0.13}{##1}}}
\expandafter\def\csname PYG@tok@k\endcsname{\let\PYG@bf=\textbf\def\PYG@tc##1{\textcolor[rgb]{0.00,0.50,0.00}{##1}}}
\expandafter\def\csname PYG@tok@se\endcsname{\let\PYG@bf=\textbf\def\PYG@tc##1{\textcolor[rgb]{0.73,0.40,0.13}{##1}}}
\expandafter\def\csname PYG@tok@sd\endcsname{\let\PYG@it=\textit\def\PYG@tc##1{\textcolor[rgb]{0.73,0.13,0.13}{##1}}}

\def\PYGZob{\char`\{}
\def\PYGZcb{\char`\}}

\def\PYGZlt{\char`\<}

\def\PYGZdq{\char`\"}

\makeatother

\makeatletter
\def\PYGdefault@reset{\let\PYGdefault@it=\relax \let\PYGdefault@bf=\relax%
    \let\PYGdefault@ul=\relax \let\PYGdefault@tc=\relax%
    \let\PYGdefault@bc=\relax \let\PYGdefault@ff=\relax}
\def\PYGdefault@tok#1{\csname PYGdefault@tok@#1\endcsname}
\def\PYGdefault@toks#1+{\ifx\relax#1\empty\else%
    \PYGdefault@tok{#1}\expandafter\PYGdefault@toks\fi}
\def\PYGdefault@do#1{\PYGdefault@bc{\PYGdefault@tc{\PYGdefault@ul{%
    \PYGdefault@it{\PYGdefault@bf{\PYGdefault@ff{#1}}}}}}}
\def\PYGdefault#1#2{\PYGdefault@reset\PYGdefault@toks#1+\relax+\PYGdefault@do{#2}}

\expandafter\def\csname PYGdefault@tok@gd\endcsname{\def\PYGdefault@tc##1{\textcolor[rgb]{0.63,0.00,0.00}{##1}}}
\expandafter\def\csname PYGdefault@tok@gu\endcsname{\let\PYGdefault@bf=\textbf\def\PYGdefault@tc##1{\textcolor[rgb]{0.50,0.00,0.50}{##1}}}
\expandafter\def\csname PYGdefault@tok@gt\endcsname{\def\PYGdefault@tc##1{\textcolor[rgb]{0.00,0.27,0.87}{##1}}}
\expandafter\def\csname PYGdefault@tok@gs\endcsname{\let\PYGdefault@bf=\textbf}
\expandafter\def\csname PYGdefault@tok@gr\endcsname{\def\PYGdefault@tc##1{\textcolor[rgb]{1.00,0.00,0.00}{##1}}}
\expandafter\def\csname PYGdefault@tok@cm\endcsname{\let\PYGdefault@it=\textit\def\PYGdefault@tc##1{\textcolor[rgb]{0.25,0.50,0.50}{##1}}}
\expandafter\def\csname PYGdefault@tok@vg\endcsname{\def\PYGdefault@tc##1{\textcolor[rgb]{0.10,0.09,0.49}{##1}}}
\expandafter\def\csname PYGdefault@tok@vi\endcsname{\def\PYGdefault@tc##1{\textcolor[rgb]{0.10,0.09,0.49}{##1}}}
\expandafter\def\csname PYGdefault@tok@vm\endcsname{\def\PYGdefault@tc##1{\textcolor[rgb]{0.10,0.09,0.49}{##1}}}
\expandafter\def\csname PYGdefault@tok@mh\endcsname{\def\PYGdefault@tc##1{\textcolor[rgb]{0.40,0.40,0.40}{##1}}}
\expandafter\def\csname PYGdefault@tok@cs\endcsname{\let\PYGdefault@it=\textit\def\PYGdefault@tc##1{\textcolor[rgb]{0.25,0.50,0.50}{##1}}}
\expandafter\def\csname PYGdefault@tok@ge\endcsname{\let\PYGdefault@it=\textit}
\expandafter\def\csname PYGdefault@tok@vc\endcsname{\def\PYGdefault@tc##1{\textcolor[rgb]{0.10,0.09,0.49}{##1}}}
\expandafter\def\csname PYGdefault@tok@il\endcsname{\def\PYGdefault@tc##1{\textcolor[rgb]{0.40,0.40,0.40}{##1}}}
\expandafter\def\csname PYGdefault@tok@go\endcsname{\def\PYGdefault@tc##1{\textcolor[rgb]{0.53,0.53,0.53}{##1}}}
\expandafter\def\csname PYGdefault@tok@cp\endcsname{\def\PYGdefault@tc##1{\textcolor[rgb]{0.74,0.48,0.00}{##1}}}
\expandafter\def\csname PYGdefault@tok@gi\endcsname{\def\PYGdefault@tc##1{\textcolor[rgb]{0.00,0.63,0.00}{##1}}}
\expandafter\def\csname PYGdefault@tok@gh\endcsname{\let\PYGdefault@bf=\textbf\def\PYGdefault@tc##1{\textcolor[rgb]{0.00,0.00,0.50}{##1}}}
\expandafter\def\csname PYGdefault@tok@ni\endcsname{\let\PYGdefault@bf=\textbf\def\PYGdefault@tc##1{\textcolor[rgb]{0.60,0.60,0.60}{##1}}}
\expandafter\def\csname PYGdefault@tok@nl\endcsname{\def\PYGdefault@tc##1{\textcolor[rgb]{0.63,0.63,0.00}{##1}}}
\expandafter\def\csname PYGdefault@tok@nn\endcsname{\let\PYGdefault@bf=\textbf\def\PYGdefault@tc##1{\textcolor[rgb]{0.00,0.00,1.00}{##1}}}
\expandafter\def\csname PYGdefault@tok@no\endcsname{\def\PYGdefault@tc##1{\textcolor[rgb]{0.53,0.00,0.00}{##1}}}
\expandafter\def\csname PYGdefault@tok@na\endcsname{\def\PYGdefault@tc##1{\textcolor[rgb]{0.49,0.56,0.16}{##1}}}
\expandafter\def\csname PYGdefault@tok@nb\endcsname{\def\PYGdefault@tc##1{\textcolor[rgb]{0.00,0.50,0.00}{##1}}}
\expandafter\def\csname PYGdefault@tok@nc\endcsname{\let\PYGdefault@bf=\textbf\def\PYGdefault@tc##1{\textcolor[rgb]{0.00,0.00,1.00}{##1}}}
\expandafter\def\csname PYGdefault@tok@nd\endcsname{\def\PYGdefault@tc##1{\textcolor[rgb]{0.67,0.13,1.00}{##1}}}
\expandafter\def\csname PYGdefault@tok@ne\endcsname{\let\PYGdefault@bf=\textbf\def\PYGdefault@tc##1{\textcolor[rgb]{0.82,0.25,0.23}{##1}}}
\expandafter\def\csname PYGdefault@tok@nf\endcsname{\def\PYGdefault@tc##1{\textcolor[rgb]{0.00,0.00,1.00}{##1}}}
\expandafter\def\csname PYGdefault@tok@si\endcsname{\let\PYGdefault@bf=\textbf\def\PYGdefault@tc##1{\textcolor[rgb]{0.73,0.40,0.53}{##1}}}
\expandafter\def\csname PYGdefault@tok@s2\endcsname{\def\PYGdefault@tc##1{\textcolor[rgb]{0.73,0.13,0.13}{##1}}}
\expandafter\def\csname PYGdefault@tok@nt\endcsname{\let\PYGdefault@bf=\textbf\def\PYGdefault@tc##1{\textcolor[rgb]{0.00,0.50,0.00}{##1}}}
\expandafter\def\csname PYGdefault@tok@nv\endcsname{\def\PYGdefault@tc##1{\textcolor[rgb]{0.10,0.09,0.49}{##1}}}
\expandafter\def\csname PYGdefault@tok@s1\endcsname{\def\PYGdefault@tc##1{\textcolor[rgb]{0.73,0.13,0.13}{##1}}}
\expandafter\def\csname PYGdefault@tok@dl\endcsname{\def\PYGdefault@tc##1{\textcolor[rgb]{0.73,0.13,0.13}{##1}}}
\expandafter\def\csname PYGdefault@tok@ch\endcsname{\let\PYGdefault@it=\textit\def\PYGdefault@tc##1{\textcolor[rgb]{0.25,0.50,0.50}{##1}}}
\expandafter\def\csname PYGdefault@tok@m\endcsname{\def\PYGdefault@tc##1{\textcolor[rgb]{0.40,0.40,0.40}{##1}}}
\expandafter\def\csname PYGdefault@tok@gp\endcsname{\let\PYGdefault@bf=\textbf\def\PYGdefault@tc##1{\textcolor[rgb]{0.00,0.00,0.50}{##1}}}
\expandafter\def\csname PYGdefault@tok@sh\endcsname{\def\PYGdefault@tc##1{\textcolor[rgb]{0.73,0.13,0.13}{##1}}}
\expandafter\def\csname PYGdefault@tok@ow\endcsname{\let\PYGdefault@bf=\textbf\def\PYGdefault@tc##1{\textcolor[rgb]{0.67,0.13,1.00}{##1}}}
\expandafter\def\csname PYGdefault@tok@sx\endcsname{\def\PYGdefault@tc##1{\textcolor[rgb]{0.00,0.50,0.00}{##1}}}
\expandafter\def\csname PYGdefault@tok@bp\endcsname{\def\PYGdefault@tc##1{\textcolor[rgb]{0.00,0.50,0.00}{##1}}}
\expandafter\def\csname PYGdefault@tok@c1\endcsname{\let\PYGdefault@it=\textit\def\PYGdefault@tc##1{\textcolor[rgb]{0.25,0.50,0.50}{##1}}}
\expandafter\def\csname PYGdefault@tok@fm\endcsname{\def\PYGdefault@tc##1{\textcolor[rgb]{0.00,0.00,1.00}{##1}}}
\expandafter\def\csname PYGdefault@tok@o\endcsname{\def\PYGdefault@tc##1{\textcolor[rgb]{0.40,0.40,0.40}{##1}}}
\expandafter\def\csname PYGdefault@tok@kc\endcsname{\let\PYGdefault@bf=\textbf\def\PYGdefault@tc##1{\textcolor[rgb]{0.00,0.50,0.00}{##1}}}
\expandafter\def\csname PYGdefault@tok@c\endcsname{\let\PYGdefault@it=\textit\def\PYGdefault@tc##1{\textcolor[rgb]{0.25,0.50,0.50}{##1}}}
\expandafter\def\csname PYGdefault@tok@mf\endcsname{\def\PYGdefault@tc##1{\textcolor[rgb]{0.40,0.40,0.40}{##1}}}
\expandafter\def\csname PYGdefault@tok@err\endcsname{\def\PYGdefault@bc##1{\setlength{\fboxsep}{0pt}\fcolorbox[rgb]{1.00,0.00,0.00}{1,1,1}{\strut ##1}}}
\expandafter\def\csname PYGdefault@tok@mb\endcsname{\def\PYGdefault@tc##1{\textcolor[rgb]{0.40,0.40,0.40}{##1}}}
\expandafter\def\csname PYGdefault@tok@ss\endcsname{\def\PYGdefault@tc##1{\textcolor[rgb]{0.10,0.09,0.49}{##1}}}
\expandafter\def\csname PYGdefault@tok@sr\endcsname{\def\PYGdefault@tc##1{\textcolor[rgb]{0.73,0.40,0.53}{##1}}}
\expandafter\def\csname PYGdefault@tok@mo\endcsname{\def\PYGdefault@tc##1{\textcolor[rgb]{0.40,0.40,0.40}{##1}}}
\expandafter\def\csname PYGdefault@tok@kd\endcsname{\let\PYGdefault@bf=\textbf\def\PYGdefault@tc##1{\textcolor[rgb]{0.00,0.50,0.00}{##1}}}
\expandafter\def\csname PYGdefault@tok@mi\endcsname{\def\PYGdefault@tc##1{\textcolor[rgb]{0.40,0.40,0.40}{##1}}}
\expandafter\def\csname PYGdefault@tok@kn\endcsname{\let\PYGdefault@bf=\textbf\def\PYGdefault@tc##1{\textcolor[rgb]{0.00,0.50,0.00}{##1}}}
\expandafter\def\csname PYGdefault@tok@cpf\endcsname{\let\PYGdefault@it=\textit\def\PYGdefault@tc##1{\textcolor[rgb]{0.25,0.50,0.50}{##1}}}
\expandafter\def\csname PYGdefault@tok@kr\endcsname{\let\PYGdefault@bf=\textbf\def\PYGdefault@tc##1{\textcolor[rgb]{0.00,0.50,0.00}{##1}}}
\expandafter\def\csname PYGdefault@tok@s\endcsname{\def\PYGdefault@tc##1{\textcolor[rgb]{0.73,0.13,0.13}{##1}}}
\expandafter\def\csname PYGdefault@tok@kp\endcsname{\def\PYGdefault@tc##1{\textcolor[rgb]{0.00,0.50,0.00}{##1}}}
\expandafter\def\csname PYGdefault@tok@w\endcsname{\def\PYGdefault@tc##1{\textcolor[rgb]{0.73,0.73,0.73}{##1}}}
\expandafter\def\csname PYGdefault@tok@kt\endcsname{\def\PYGdefault@tc##1{\textcolor[rgb]{0.69,0.00,0.25}{##1}}}
\expandafter\def\csname PYGdefault@tok@sc\endcsname{\def\PYGdefault@tc##1{\textcolor[rgb]{0.73,0.13,0.13}{##1}}}
\expandafter\def\csname PYGdefault@tok@sb\endcsname{\def\PYGdefault@tc##1{\textcolor[rgb]{0.73,0.13,0.13}{##1}}}
\expandafter\def\csname PYGdefault@tok@sa\endcsname{\def\PYGdefault@tc##1{\textcolor[rgb]{0.73,0.13,0.13}{##1}}}
\expandafter\def\csname PYGdefault@tok@k\endcsname{\let\PYGdefault@bf=\textbf\def\PYGdefault@tc##1{\textcolor[rgb]{0.00,0.50,0.00}{##1}}}
\expandafter\def\csname PYGdefault@tok@se\endcsname{\let\PYGdefault@bf=\textbf\def\PYGdefault@tc##1{\textcolor[rgb]{0.73,0.40,0.13}{##1}}}
\expandafter\def\csname PYGdefault@tok@sd\endcsname{\let\PYGdefault@it=\textit\def\PYGdefault@tc##1{\textcolor[rgb]{0.73,0.13,0.13}{##1}}}

\makeatother

\newreptheorem{theorem}{Theorem}
\newreptheorem{lemma}{Lemma}

\begin{document}

%% Title information
\title{Data Race Detection on Compressed Traces}    
% \title{Exploiting Trace Compression for Faster Dynamic Race Detection}
% \title{Accelerating Dynamic Race Detection Using Trace Compression}  
% \title{Dynamic Data Race Detection on Compressed Traces}   
%% [Short Title] is optional;
                                        %% when present, will be used in
                                        %% header instead of Full Title.
% \titlenote{with title note}             %% \titlenote is optional;
                                        %% can be repeated if necessary;
                                        %% contents suppressed with 'anonymous'
% \subtitle{Enhancing the Predictive Power of Happens-Before Based Dynamic Analysis}                     %% \subtitle is optional
% \subtitlenote{with subtitle note}       %% \subtitlenote is optional;
                                        %% can be repeated if necessary;
                                        %% contents suppressed with 'anonymous'

%% Author information
%% Contents and number of authors suppressed with 'anonymous'.
%% Each author should be introduced by \author, followed by
%% \authornote (optional), \orcid (optional), \affiliation, and
%% \email.
%% An author may have multiple affiliations and/or emails; repeat the
%% appropriate command.
%% Many elements are not rendered, but should be provided for metadata
%% extraction tools.

%% Author with single affiliation.
\author{Dileep Kini}
\affiliation{
  \institution{Akuna Capital LLC}
  \country{USA}
}

\author{Umang Mathur}
% \authornote{with author1 note}          %% \authornote is optional;
                                        %% can be repeated if necessary
\orcid{0000-0002-7610-0660}             %% \orcid is optional
\affiliation{
  % \position{Position1}
  % \department{Department of Computer Science}              %% \department is recommended
  \institution{University of Illinois, Urbana Champaign}            %% \institution is required
  % \streetaddress{Street1 Address1}
  % \city{City1}
  % \state{State1}
  % \postcode{Post-Code1}
  \country{USA}                    %% \country is recommended
}
\email{umathur3@illinois.edu}          %% \email is recommended

%% Author with two affiliations and emails.
\author{Mahesh Viswanathan}
% \authornote{with author2 note}          %% \authornote is optional;
%                                         %% can be repeated if necessary
% \orcid{nnnn-nnnn-nnnn-nnnn}             %% \orcid is optional
\affiliation{
  % \position{Position1}
  % \department{Department of Computer Science}              %% \department is recommended
  \institution{University of Illinois, Urbana Champaign}            %% \institution is required
  % \streetaddress{Street1 Address1}
  % \city{City1}
  % \state{State1}
  % \postcode{Post-Code1}
  \country{USA}                    %% \country is recommended
}
\email{vmahesh@illinois.edu}          %% \email is recommended
% \affiliation{
%   \position{Position2b}
%   \department{Department2b}             %% \department is recommended
%   \institution{Institution2b}           %% \institution is required
%   \streetaddress{Street3b Address2b}
%   \city{City2b}
%   \state{State2b}
%   \postcode{Post-Code2b}
%   \country{Country2b}                   %% \country is recommended
% }
% \email{first2.last2@inst2b.org}         %% \email is recommended

%% Abstract
%% Note: \begin{abstract}...\end{abstract} environment must come
%% before \maketitle command
\begin{abstract}
\input{abstract}
\end{abstract}

% %% 2012 ACM Computing Classification System (CSS) concepts
% %% Generate at 'http://dl.acm.org/ccs/ccs.cfm'.
% <concept_id>10011007.10011006.10011008</concept_id>
% <concept_desc>Software and its engineering~General programming languages</concept_desc>
% <concept_id>10003456.10003457.10003521.10003525</concept_id>
% <concept_desc>Social and professional topics~History of programming languages</concept_desc>
% <concept_significance>300</concept_significance>

% \ccsdesc[500]{Software and its engineering~General programming languages}
% \ccsdesc[300]{Social and professional topics~History of programming languages}
% %% End of generated code

%% Keywords
%% comma separated list
% \keywords{dynamic program analysis, race detection, compression}  %% \keywords are mandatory in final camera-ready submission

%% \maketitle
%% Note: \maketitle command must come after title commands, author
%% commands, abstract environment, Computing Classification System
%% environment and commands, and keywords command.
\maketitle

\input{intro}

\input{prelim}

\input{hbalgorithm}

\input{lockset}

\input{experiments}

\input{conclusions}

%% Acknowledgments
\begin{acks}                            %% acks environment is optional
  %                                       %% contents suppressed with 'anonymous'
  % %% Commands \grantsponsor{<sponsorID>}{<name>}{<url>} and
  % %% \grantnum[<url>]{<sponsorID>}{<number>} should be used to
  % %% acknowledge financial support and will be used by metadata
  % %% extraction tools.
  % This material is based upon work supported by the
  % \grantsponsor{GS100000001}{National Science
  %   Foundation}{http://dx.doi.org/10.13039/100000001} under Grant
  % No.~\grantnum{GS100000001}{nnnnnnn} and Grant
  % No.~\grantnum{GS100000001}{mmmmmmm}.  Any opinions, findings, and
  % conclusions or recommendations expressed in this material are those
  % of the author and do not necessarily reflect the views of the
  % National Science Foundation.
We gratefully acknowledge National Science Foundation for supporting
 Umang Mathur (grant NSF CSR 1422798) and Mahesh Viswanathan (NSF CPS 1329991).
\end{acks}

% \clearpage
\bibliography{refs}

% %% Appendix
\clearpage 

\appendix

\input{app-complexity-prelim}
% \clearpage
\input{app-proofs}
% \clearpage
\input{app-hb-algo}
% \clearpage
\input{app-lockset-algo}

\end{document}

%% file: macros.tex
\newenvironment{myparagraph}[1]{\vspace*{0.1in}\noindent{\bf #1.}}{}

\newcommand{\figlabel}[1]{\label{fig:#1}}
\newcommand{\figref}[1]{Figure~\ref{fig:#1}}
\newcommand{\seclabel}[1]{\label{sec:#1}}
\newcommand{\secref}[1]{Section~\ref{sec:#1}}
\newcommand{\exlabel}[1]{\label{ex:#1}}
\newcommand{\exref}[1]{Example~\ref{ex:#1}}
\newcommand{\deflabel}[1]{\label{def:#1}}
\newcommand{\defref}[1]{Definition~\ref{def:#1}}
\newcommand{\thmlabel}[1]{\label{thm:#1}}
\newcommand{\thmref}[1]{Theorem~\ref{thm:#1}}
\newcommand{\proplabel}[1]{\label{prop:#1}}
\newcommand{\propref}[1]{Proposition~\ref{prop:#1}}
\newcommand{\lemlabel}[1]{\label{lem:#1}}
\newcommand{\lemref}[1]{Lemma~\ref{lem:#1}}
\newcommand{\itemlabel}[1]{\label{itm:#1}}
\newcommand{\itemref}[1]{(\ref{itm:#1})}
\newcommand{\equlabel}[1]{\label{eq:#1}}
\newcommand{\equref}[1]{Equation~(\ref{eq:#1})}
\newcommand{\applabel}[1]{\label{app:#1}}
\newcommand{\appref}[1]{ the Appendix}

\newcommand{\acq}[1]{\mathtt{acq}(#1)}
\newcommand{\rel}[1]{\mathtt{rel}(#1)}
\newcommand{\rd}[1]{\mathtt{r}(#1)}
\newcommand{\wt}[1]{\mathtt{w}(#1)}
\newcommand{\fork}[1]{\mathtt{fork}(#1)}
\newcommand{\join}[1]{\mathtt{join}(#1)}
\newcommand{\ev}[2]{\langle #1: #2\rangle}

\newcommand{\lk}{\ell}
\newcommand{\thr}{t}
\newcommand{\tr}{\sigma}
\newcommand{\dmlock}{\Lambda}
\newcommand{\dmlockthr}[1]{\Lambda_{#1}}

\newcommand{\execution}[2]{\begin{tikzpicture}%
                             \foreach \x in {1,...,#1}
                               \node[right] at (1.5*\x+0.5,0.25) {Thread \x};
                             \draw (1.9,0) -- (#1*1.5+1.9,0);%
                             \pgfmathsetmacro{\y}{1};%
                             #2%
                             \draw (1.9,0) -- (1.9,-0.4*\y);%
                             \draw (#1*1.5+1.9,0) -- (#1*1.5+1.9,-0.4*\y);%
                             \foreach \x in {2,...,#1}
                               \draw (2*\x -0.6,0) -- (2*\x - 0.6,-0.4*\y);%
                             \draw (1.9,-0.4*\y) -- (#1*1.5+1.9,-0.4*\y);%
                           \end{tikzpicture}}
\newcommand{\figev}[2]{\node [left] at (1.75,-0.4*\y) {\pgfmathprintnumber{\y}};%
                    \node [right] at (#1*1.5 + 0.6,-0.4*\y) {$ #2 $};%
                    \pgfmathsetmacro{\y}{\y+1};}

\newcommand{\grammar}[1]{\begin{tikzpicture}%
                           \pgfmathsetmacro{\y}{1};%
                           #1%
                         \end{tikzpicture}}
\newcommand{\gramrule}[2]{\node [right] at (0,-0.6*\y){$#1 \rightarrow #2$};%
                          \pgfmathsetmacro{\y}{\y+1};}

\newcommand{\s}[1]{ \mathsf{#1}}

\newcommand{\so}[2]{<_{#1}^{#2}}
\newcommand{\nso}[2]{\leq_{#1}^{#2}}
\newcommand{\HBS}[1][]{\so{\s{HB}}{#1}}
\newcommand{\HB}[1][]{\nso{\s{HB}}{#1}}
\newcommand{\TOS}[1][]{\so{\s{TO}}{#1}}
\newcommand{\TO}[1][]{\nso{\s{TO}}{#1}}
\newcommand{\tros}[1][]{\so{\s{tr}}{#1}}

\newcommand{\conflict}{\asymp}
\newcommand{\hbconc}{||_{\s{HB}}}

\newcommand{\hbrace}{$\s{HB}$-race}
\newcommand{\threads}{\s{Threads}}
\newcommand{\locks}{\s{Locks}}

\newcommand{\match}{\s{match}}
\newcommand{\after}[1]{\s{After}_{#1}}
\newcommand{\before}[1]{\s{Before}_{#1}}
\newcommand{\reads}[1]{\s{REvents}_{#1}}
\newcommand{\writes}[1]{\s{WEvents}_{#1}}

\newcommand{\tjevents}[1]{\s{ThEvents}^{\s{join}}_{#1}}
\newcommand{\tfevents}[1]{\s{ThEvents}^{\s{fork}}_{#1}}
\newcommand{\acquires}[1]{\s{AcqEvents}_{#1}}
\newcommand{\releases}[1]{\s{RelEvents}_{#1}}

\newcommand{\rvars}[1]{\s{Rd}({#1})}
\newcommand{\wvars}[1]{\s{Wr}({#1})}
\newcommand{\first}[1]{\s{First}_{#1}}
\newcommand{\last}[1]{\s{Last}_{#1}}
\newcommand{\af}[1]{\s{AF}_{#1}}
\newcommand{\bl}[1]{\s{BL}_{#1}}
\newcommand{\ar}[1]{\s{ALRd}_{#1}}
\newcommand{\aw}[1]{\s{ALWr}_{#1}}
\newcommand{\br}[1]{\s{BFRd}_{#1}}
\newcommand{\bw}[1]{\s{BFWr}_{#1}}
\newcommand{\race}{\s{Race?}}
\newcommand{\lockset}[1]{\s{LockSet}_{#1}}
\newcommand{\access}[1]{\s{Access}_{#1}}
\newcommand{\locksheld}[1]{\s{LocksHeld}_{#1}}
\newcommand{\oacq}[1]{\s{OpenAcq}_{#1}}
\newcommand{\orel}[1]{\s{OpenRel}_{#1}}

\newcommand{\glsr}[1]{\s{GLS}^{\s{R}}_{#1}}
\newcommand{\glsw}[1]{\s{GLS}^{\s{W}}_{#1}}

\newcommand{\sem}[1]{\llbracket #1 \rrbracket}

\newcommand{\proj}[2]{#1\!\upharpoonright_{#2}}
\newcommand{\set}[1]{\{#1\}}
\newcommand{\setpred}[2]{\set{#1\: |\: #2}}

\newcommand{\cD}{\mathcal{D}}

\newcommand{\tool}{\textsc{ZipTrack}}
\newcommand{\rapid}{\textsc{RAPID}}

\newcommand{\ucomment}[1]{}

\definecolor{darkorange}{rgb}{1.0, 0.55, 0.0}
\definecolor{darkmagenta}{rgb}{0.55, 0.0, 0.55}
\definecolor{darkyellow}{rgb}{0.97, 0.65, 0.0}

\newcommand{\red}[1]{\textcolor{black!10!red}{#1}}
\newcommand{\blue}[1]{\textcolor{black!10!blue}{#1}}
\newcommand{\green}[1]{\textcolor{black!30!green}{#1}}

\newcommand{\yellow}[1]{\textcolor{darkyellow}{#1}}

\newcommand\VRule[1][\arrayrulewidth]{\vrule width #1}

\definecolor{pblue}{rgb}{0.13,0.13,1}
\definecolor{pgreen}{rgb}{0,0.5,0}
\definecolor{pred}{rgb}{0.9,0,0}
\definecolor{pgrey}{rgb}{0.46,0.45,0.48}

\newcommand*{\boldsmile}{%
  \textpdfrender{
    TextRenderingMode=FillStroke,
    LineWidth=.7pt, % half of the line width is outside the normal glyph
  }{\smile}%
}

\newcommand*{\boldfrown}{%
  \textpdfrender{
    TextRenderingMode=FillStroke,
    LineWidth=.7pt, % half of the line width is outside the normal glyph
  }{\frown}%
}

%% file: abstract.tex
%!TEX root = main.tex

We consider the problem of detecting data races in program traces that
have been compressed using straight line programs (SLP), which are
special context-free grammars that generate exactly one string, namely
the trace that they represent. We consider two classical approaches to
race detection --- using the happens-before relation and the lockset
discipline. We present algorithms for both these methods that run in
time that is linear in the size of the compressed, SLP representation.
Typical program executions almost always exhibit patterns that
lead to significant compression.  Thus, our algorithms are expected to
result in large speedups when compared with analyzing the uncompressed
trace.  Our experimental evaluation of these new algorithms on
standard benchmarks confirms this observation.

%% file: intro.tex
%!TEX root = main.tex

\section{Introduction}
\seclabel{intro}

Dynamic analysis of multi-threaded programs is the problem of
discovering anomalies in a program by examining a single or multiple
traces of a program. While dynamic analysis is sometimes performed
online as the program is running, it is often performed offline, on a
stored trace of the program. 
The reasons for performing offline dynamic analysis are many. 
The overhead of analyzing the trace as the
program is running maybe large, causing undesirable slowdowns.  
This is especially true for expensive dynamic analysis techniques
which employ heavy-weight machinery such as the use of SMT solvers~\cite{swys11,hmr14},
graph based analysis~\cite{ffy08,bhsb14} or even vector clocks~\cite{ps03,ff09}.
% For example, in order to protect against
% \emph{metadata races}, dynamic data race detectors must introduce
% additional synchronization, which can give rise to additional
% performance overhead.  Recent experiments~\cite{fib2017,radish2012} to
% evaluate the overhead due to such synchronization in implementations of the
% \textsc{FastTrack}~\cite{ff09} algorithm suggest that the resulting
% slow-down is significant. For concurrent programs, these latencies
% may also sometimes introduce subtle synchronization errors that may
% otherwise not be present in the original program when run alone
% without dynamic analysis. 
Often, it is desirable to perform multiple,
different analyses on a single trace, and the kinds of analyses to be
performed may even be unknown at the time the program is being
observed. Finally, storing the trace and later replaying it in a
controlled environment, can help in debugging programs, in{}
understanding performance overheads and in performance
tuning. Trace-driven simulations are used widely in computer architecture
for quantitative evaluations of new ideas and
designs~\cite{ksw03,mm07}.

However program traces are often huge, recording millions and billions
of events. When debugging a large software application, long traces are
often necessary to ensure adequate code coverage. This is especially
acute\ucomment{??} for multi-threaded programs where subtle concurrency bugs are
often revealed only under specific thread schedules. Therefore, useful
traces are those that exercise the same program fragment multiple
times, under different scenarios \ucomment{What?}; this is substantiated by the
observation that some concurrency bugs only manifest themselves in
traces with millions of events~\cite{ep14}. In such circumstances, the
only way to alleviate the warehousing needs of storing such traces is
to compress them~\cite{ksw03,mm07}.

In this paper, we study the problem of detecting data races in
programs by examining compressed traces. Data races are the most
common symptom of a programming error in concurrent programming. The
na\"{i}ve approach to solving this problem would be to uncompress the
trace and then process it using any one of the many algorithms that
have been developed for dynamic data race
detection~\cite{fid88,mat88,sbnsa97,ps03,ff09,hmr14,kmv17}. But is
this necessary? Is this na\"{i}ve algorithm, asymptotically,
 the best one can hope for? Studying the complexity of problems where the input is
represented succinctly has a long history. Starting from the seminal
paper by Galperin and Wigderson~\cite{gw83}, where they studied the
complexity of graph problems when the input graph is represented by a
circuit, it has been observed that typically there is an exponential
blowup in the complexity of problems when they are solved on
compressed
inputs~\cite{gw83,py86,lb89,bal96,vei96,fkvv98,dst14}. Thus, often the
na\"{i}ve algorithm is the best algorithm asymptotically.

Our results in this paper, fortunately, are the\ucomment{an} exception to the above
rule. We consider two classical race detection approaches --- a
sound
\footnote{We say a race detector is sound if it never issues any warning on race-free programs or executions. This is often referred to as \emph{precise}~\cite{ff09} in the race detection literature.} 
 method based on computing Lamport's happens-before
relation~\cite{lam78}, and the lightweight lockset-based algorithm of
Eraser~\cite{sbnsa97} --- and extend them to work directly on the
compressed trace without first uncompressing it. Our algorithms run in
time that is linear in the size of the compressed trace. Thus, we show
that compression can in fact be used as an algorithmic principle to
speedup the analysis in this context.

\input{program}
To understand why compression actually speeds up the analysis,
consider the simple program shown in \figref{simple_java}. 
A single execution of this program generates
about 680 million events, taking about 1.3 GB disk space. 
However, when this trace is compressed
using the Sequitur algorithm~\cite{nw97,nev,sequitur}, 
the compressed representation only occupies about 34 MB of disk space.  The reason why the
trace could be compressed so effectively is because the program has a
simple loop structure that is executed multiple times. Thus the
program trace has a ``regular'' structure that the compression
algorithm exploits. An algorithm processing the uncompressed trace is
agnostic to this regularity, and is forced to repeat the same analysis
each time the sub-trace corresponding to the loop body is
encountered. Compression makes this regular structure ``visible'', and
an algorithm working on the compressed representation can exploit it
by only performing an analysis once for each distinct sub-trace.

We consider compression schemes that compress traces as straight
line programs (SLPs). SLPs are a special class of context-free grammars where the
language of the grammar consists of a single string, namely, the trace
being compressed. Several lossless compression schemes, like run-length
encoding and the family of Lempel-Ziv encodings~\cite{zl77}, can be
converted efficiently to SLPs of similar size. 
% Our algorithms on SLPs
% proceed inductively on the structure of the grammar that compute
% book-keeping information for each non-terminal in the grammar,
% compositionally. 
Our algorithms on SLPs
proceed inductively on the structure of the grammar, and compute,
in a compositional fashion,
book-keeping information for each non-terminal in the grammar. 
Thus, a sub-trace generated by a non-terminal that
may appear in many positions in the uncompressed trace, is analyzed
only once. For happens-before-based race detection, our algorithm is
inspired by the Goldilocks method~\cite{eqt07}, where the book-keeping
information is captured by a set of threads and locks.
\ucomment{And for lockset?}

We have implemented our algorithms in a tool called {\tool}. The traces
are compressed using a popular SLP-based compresssion algorithm called
Sequitur~\cite{sequitur}. Our experiments on standard benchmark
examples reveal that the algorithms on compressed traces perform well,
and on large traces, often have an order of magnitude improvement over
algorithms running in the uncompressed setting.

The rest of the paper is organized as follows. After discussing
closely related work, we introduce basic notation and classical race
detection algorithms in \secref{prelim}. In \secref{hb}, we present
our happens-before based data race detection algorithm on compressed traces. Our
algorithm for checking violations of the lockset discipline on compressed traces is presented in
\secref{lockset}. We present our experimental results in
\secref{experiments}.

% \vspace{-0.1in}
\myparagraph{Related Work}{
Type systems to prevent data races have been
developed~\cite{ff00,blr02,aff06}. Since the race detection problem is
undecidable, the several static analysis
techniques~\cite{yah01,ea03,naw06,vjl07,mqbbnn08,pfh11,rd13,zh16}
suffer from two problems --- they don't scale and they raise many
false alarms since they are typically conservative.  Dynamic race
detection techniques can be classified into three categories. There
are the unsound lockset-based techniques, which may raise false
alarms~\cite{sbnsa97}. Techniques like random testing~\cite{sen08} and
static escape analysis~\cite{pg01} can reduce the false alarms in such
algorithms, but not eliminate them. The second category of dynamic
analysis techniques are predictive runtime analysis
techniques~\cite{wkgg09,swys11,hmr14,hr16,ltz16}, where the race
detector explores all possible reorderings of the given trace to
search for a possible witness of a data race. Since the number of
interleavings of a given trace is very large, these do not scale to
long traces. The last category of dynamic race detection algorithms
are those based on identifying a partial order on the events of a
trace, and then searching for a pair of conflicting data accesses that
are unordered by the partial order. These techniques are sound and
scale to long traces since they typically run in linear time. The
simplest, and most commonly used partial order is
happens-before~\cite{lam78}. Early vector-clock based algorithms to
compute happens-before on traces~\cite{fid88,mat88} have been
subsequently optimized~\cite{ps03,ff09}. A lockset-based method for
computing the happens-before partial order was proposed
in~\cite{eqt07}. 
% Light-weight heuristic methods such as sampling~\cite{mmn09,bcm10}
% have been proposed to approximate the happens before relation.
% There are light-weight, heuristic methods for
% computing an approximation of the happens-before relation using
% hashing~\cite{mmn09,bcm10} and crowd sourcing~\cite{kzc13}. 
Structured parallelism has been exploited to optimize the memory overhead in
detecting happens before~\cite{fl97,cflrs98,rzsvy12,ss16,yng16}. More
recently, partial order that are weaker than happens before have been
proposed for detecting data races, including causal precedence~\cite{sesyf12} and weak causal
precedence~\cite{kmv17}.
% 
%  These partial orders therefore
% enable detecting more races and are nonetheless sound. Prominant among
% these is causal precedence~\cite{sesyf12} and weak causal
% precedence~\cite{kmv17}; the former is not know to have linear time
% algorithms, while the latter does. We will exploit weak causal
% precedence for detecting data races and deadlocks.
$\s{Sofya}$~\cite{kdr07} and \textsc{RoadRunner}~\cite{ff10} are tools
that provide a framework for implementing dynamic analysis tools.
}

%% file: program.tex
%!TEX root = main.tex

\begin{figure}[t]
\begin{Verbatim}[commandchars=\\\{\}, fontsize=\footnotesize, baselinestretch=0.5]
\PYG{k+kd}{public} \PYG{k+kd}{class} \PYG{n+nc}{Test} \PYG{k+kd}{extends} \PYG{n}{Thread}\PYG{o}{\PYGZob{}}

    \PYG{k+kd}{static} \PYG{k+kd}{final} \PYG{k+kt}{long} \PYG{n}{ITERS} \PYG{o}{=} \PYG{l+m+mi}{1000000000L}\PYG{o}{;}
    \PYG{k+kd}{static} \PYG{k+kt}{int} \PYG{n}{y}\PYG{o}{;}

    \PYG{k+kd}{public} \PYG{k+kt}{void} \PYG{n+nf}{inc}\PYG{o}{()} \PYG{o}{\PYGZob{}}
        \PYG{n}{y}\PYG{o}{++;}
    \PYG{o}{\PYGZcb{}}

    \PYG{n+nd}{@Override}
    \PYG{k+kd}{public} \PYG{k+kt}{void} \PYG{n+nf}{run}\PYG{o}{()} \PYG{o}{\PYGZob{}}
        \PYG{k}{for} \PYG{o}{(}\PYG{k+kt}{long} \PYG{n}{i} \PYG{o}{=} \PYG{l+m+mi}{0}\PYG{o}{;} \PYG{n}{i} \PYG{o}{\PYGZlt{}} \PYG{n}{ITERS}\PYG{o}{;} \PYG{n}{i}\PYG{o}{++)} \PYG{o}{\PYGZob{}}
            \PYG{n}{inc}\PYG{o}{();}
        \PYG{o}{\PYGZcb{}}
    \PYG{o}{\PYGZcb{}}

    \PYG{k+kd}{public} \PYG{k+kd}{static} \PYG{k+kt}{void} \PYG{n+nf}{main}\PYG{o}{(}\PYG{n}{String} \PYG{n}{args}\PYG{o}{[])} \PYG{k+kd}{throws} \PYG{n}{Exception} \PYG{o}{\PYGZob{}}
        \PYG{k+kd}{final} \PYG{n}{Test} \PYG{n}{t1} \PYG{o}{=} \PYG{k}{new} \PYG{n}{Test}\PYG{o}{();}
        \PYG{k+kd}{final} \PYG{n}{Test} \PYG{n}{t2} \PYG{o}{=} \PYG{k}{new} \PYG{n}{Test}\PYG{o}{();}
        \PYG{n}{t1}\PYG{o}{.}\PYG{n+na}{start}\PYG{o}{();}
        \PYG{n}{t2}\PYG{o}{.}\PYG{n+na}{start}\PYG{o}{();}
        \PYG{n}{t1}\PYG{o}{.}\PYG{n+na}{join}\PYG{o}{();}
        \PYG{n}{t2}\PYG{o}{.}\PYG{n+na}{join}\PYG{o}{();}
        \PYG{n}{System}\PYG{o}{.}\PYG{n+na}{out}\PYG{o}{.}\PYG{n+na}{println}\PYG{o}{(}\PYG{l+s}{\PYGZdq{}y (actual) = \PYGZdq{}} \PYG{o}{+} \PYG{n}{y}\PYG{o}{);}
        \PYG{n}{System}\PYG{o}{.}\PYG{n+na}{out}\PYG{o}{.}\PYG{n+na}{println}\PYG{o}{(}\PYG{l+s}{\PYGZdq{}y (expected) = \PYGZdq{}} \PYG{o}{+} \PYG{n}{ITERS}\PYG{o}{*}\PYG{l+m+mi}{2}\PYG{o}{);}
    \PYG{o}{\PYGZcb{}}
\PYG{o}{\PYGZcb{}}
\end{Verbatim}
\vspace{0.5cm}
\caption{\small A simple concurrent program in Java }
\figlabel{simple_java}
\end{figure}

% \begin{figure}
% \begin{lstlisting}
% public class Test extends Thread{

%     static final long ITERS = 100000000L;

%     static int y;

%     public void inc() {
%         y++;
%     }

%     @Override
%     public void run() {
%         for (long i = 0; i < ITERS; i++) {
%             inc();
%         }
%     }

%     public static void main(String args[]) throws Exception {
%         final Test t1 = new Test();
%         final Test t2 = new Test();
%         t1.start();
%         t2.start();
%         t1.join();
%         t2.join();
%         System.out.println("y (actual) = " + y);
%         System.out.println("y (expected) = " + ITERS*2);
%     }
% }
% \end{lstlisting}
% \caption{A simple concurrent program in Java }
% \figlabel{simple_java}
% \end{figure}

% \begin{figure}
% \begin{verbatim}
% public class Test extends Thread{

%     static final long ITERS = 1000000000L;

%     static int y;

%     public void inc() {
%     y++;
%     }

%     @Override
%     public void run() {
%         for (long i = 0; i < ITERS; i++) {
%             inc();
%         }
%     }

%     public static void main(String args[]) throws Exception {
%         final Test t1 = new Test();
%         final Test t2 = new Test();
%         t1.start();
%         t2.start();
%         t1.join();
%         t2.join();
%         System.out.println("y (actual) = " + y);
%         System.out.println("y (expected) = " + ITERS*2);
%     }
% }
% \end{verbatim}
% \caption{A simple concurrent program in Java }
% \figlabel{simple_java}
% \end{figure}

%% file: prelim.tex
%!TEX root = main.tex

\section{Preliminaries}
\seclabel{prelim}

In the section we introduce basic notation, our assumptions about
concurrent programs, the happens before ordering on events, and some
classical algorithms for race detection.

\begin{myparagraph}{Traces}
We will analyze traces of concurrent programs synchronizing through
locks while accessing shared memory locations (also referred to as
global variables, or simply, variables). Traces are (finite) sequences
of events of the form $\ev{\thr}{o}$, where $\thr$ is the thread
performing the operation $o$~\footnote{Formally, each event in a trace
  is assumed to have a unique event id. Thus, two occurences of a
  thread performing the same operation will be considered
  \emph{different} events. Even though we will implicitly assume the
  uniqueness of each event in a trace, to reduce notational overhead,
  we do not formally introduce event ids.}.
  % ; the empty trace will be denoted by $\emptr$. 
Operations can be one of the following: forking
of a new child thread ($\fork{\thr}$); joining of a child thread
($\join{\thr}$); acquiring and releasing a lock ($\acq{\lk}$ and
$\rel{\lk}$); and, reading and writing to a variable ($\rd{x}$ and
$\wt{x}$).
We will assume that a child thread is forked and joined by the same parent thread.
% \ucomment{Threads are joined and forked by the same thread. This is needed for the def of HB.}
%
Locks are assumed to be \emph{reentrant}. That is, a thread
$\thr$ may acquire a lock $\lk$ multiple times, as long as $\thr$
holds $\lk$. However, $\thr$ must release $\lk$, as many times as it
was acquired, before $\lk$ becomes available for being acquired by
some other thread. 
Therefore, with every release event $e = \ev{\thr}{\rel{\lk}}$, we can
associate a unique acquire event $e' = \ev{\thr}{\acq{\lk}}$, which is
the last $\acq{\lk}$-event in thread $\thr$ before $e$ that is not
matched with any $\rel{\lk}$ event in thread $\thr$ before $e$. This
$\acq{\lk}$ event $e'$ is said to be the matching acquire of $e$, and
is denoted by $\match(e)$. Similarly, for an acquire event $e'$ such
that $e' = \match(e)$, we will say that $e$ is the matching release of
$e'$, and we will also denote this by $\match(e')$.
For a trace $\tr$, $\proj{\tr}{\thr}$ will denote the subsequence of
events performed by thread $\thr$.
% , and $\proj{\tr}{\lk}$ will denote the
% subsequence of events pertaining to lock $\lk$.
% \todo{No def. of a valid trace?}
% A \emph{contiguous} subsequence $\s{cs} = \ev{\thr}{\acq{\lk}}, e_1,
% \ldots e_k, \ev{\thr}{\rel{\lk}}$ of $\proj{\tr}{\thr}$, is said to be
% a \emph{critical section} on lock $\lk$ in $\tr$, provided
% $\proj{\s{cs}}{\lk}$ belongs to the language generated by the
% context-free grammar $S \rightarrow \emptr\: |\: SS\: |\:
% \ev{\thr}{\acq{\lk}} S \ev{\thr}{\rel{\lk}}$. 
% \todo{This represents a seq. of critical sections.}
%\todo{In fact, if this stays the same, I will have to change some sentences in Section 4 (after def. of openacq).}
% Note, that since we
% consider locks to be reentrant, it is possible for one critical
% section on lock $\lk$ to be strictly contained in another critical
% section on the same lock.
%
% In a critical section $\s{cs} = e_1 = \ev{\thr}{\acq{\lk}}, e_2,
% \ldots e_{k-1}, e_k = \ev{\thr}{\rel{\lk}}$, $e_1$ and $e_k$ are said
% to be \emph{matching} events. 

% \todo{This is not acc. to the current def of CS}
% We denote this by saying $\match(e_1) =
% e_k$ and $\match(e_k) = e_1$. For any acquire (or release) event $e$,
% there is a unique event (if one exists) $e'$ such that $\match(e) =
% e'$.
\end{myparagraph}

\begin{myparagraph}{Notation}
Let us fix a trace $\tr$. For an event $e$, we will say $e \in \tr$ to
denote the fact that $e$ appears in the sequence $\tr$. The set of
locks acquired or released in $\tr$ will be denoted by $\locks(\tr)$.
$\threads(\tr)$ will denote the set of threads performing some event
in $\tr$; in the presence of forks and joins, this is a bit subtle and
we define it as
\begin{align*}
\threads(\tr) =  \setpred{\thr}{\exists e \in \tr.\ & e =
  \ev{\thr}{o} \mbox{ for some $o$, or }  e = \ev{\thr'}{\fork{\thr}} \\
  \mbox{ or } & e =
  \ev{\thr'}{\join{\thr}} \mbox{ for some thread $t'$} }.
\end{align*}
For a variable $x$, the set of $\wt{x}$-events will be denoted by
$\writes{\tr}(x)$ and the set of $\rd{x}$-events performed by thread
$\thr \in\threads(\tr)$ will be denoted by $\reads{\tr}(\thr,x)$.  We
will use $\rvars{\tr}$ to denote the set of pairs $(\thr,x)$ for which
% $\reads{\tr}(\thr,x) \neq \emptyset$.
$\reads{\tr}(\thr,x)$ $\neq \emptyset$.
 % $\set{(\thr,x) \in \threads(\tr)\times\vars(\tr) \,|\,
 %  \reads{\tr}(\thr,x) \neq \emptyset}$.  
  Similarly, we will use
$\wvars{\tr}$ to denote the set of variables $x$
for which $\writes{\tr}(x)$ is non-empty.
 % $\set{x \in \vars(\tr) \,|\,
 %  \writes{\tr}(x) \neq \emptyset}$.  
When $\tr$ is clear from the
context, we may drop it.

For a non-empty subset of events $S$, we will denote by
$\last{\tr}(S)$ the (unique) event $e \in S$, that is latest in $\tr$
among the events in $S$. Similarly, $\first{\tr}(S)$ is the event $e
\in S$ that is earliest in $\tr$ amongst the events in $S$. When $S$
is empty, we say both $\first{\tr}(S)$ and $\last{\tr}(S)$ are
undefined.
\end{myparagraph}

\input{trace1}

\begin{example}
\exlabel{basic-defn} We illustrate the above definitions on the
example trace $\tr_1$ shown in \figref{example-trace-grammar}. We will follow
the convention of representing events of a trace from top-to-bottom,
where temporally earlier events appear above the later ones. We use
$e_i$ to denote the $i$th event in $\tr_1$. Let $S_1 =
\reads{\tr_1}(2,x) = \set{e_3,e_{11}}$ and $S_2 = \writes{\tr_1}(y) =
\set{e_5,e_{10},e_{13},e_{16}}$. The set $\rvars{\tr_1} =
\set{(1,x),(2,x)}$ while $\wvars{\tr_1} = \set{x,y}$. Finally,
$\last{\tr_1}(S_1) = e_{11}$, and $\first{\tr_1}(S_2) = e_5$.
\end{example}

\begin{myparagraph}{Orders on Traces} 
Let us fix a trace $\tr$. 
% $=^\tr$ denotes the identity relation on the
% events of $\tr$. 
If an event $e_1$ appears earlier in the sequence
$\tr$ than $e_2$, then we say $e_1$ is \emph{trace ordered before}
$e_2$ and denote it as $e_1 \tros[\tr] e_2$. 
We say $e_1$ is
\emph{thread ordered before} $e_2$, denoted by $e_1 \TOS[\tr] e_2$, if
$e_1$ and $e_2$ are events performed by the same thread and $e_1
\tros[\tr] e_2$. Our race detection algorithm will rely on computing
the \emph{happens before} strict order, which we define next.
\begin{definition}[Happens Before]
\deflabel{happens-before} Event $e$ in trace $\tr$ said to
\emph{happen before} event $e' \in \tr$, denoted $e \HBS[\tr] e'$, if
and only if there is a sequence of events $e = e_1, e_2, e_3, \ldots
e_n = e'$ such that for every pair $(e_i,e_{i+1})$ ($i < n$), $e_i
\tros[\tr] e_{i+1}$ and {\bf one} of the following conditions hold.
% \ucomment{Just confirm if we are indeed computing strict order}
\begin{enumerate}
\item\itemlabel{thread-order} $e_i \TOS[\tr] e_{i+1}$,
\item\itemlabel{acq-rel} $e_i = \ev{\thr}{\rel{\lk}}$ and $e_{i+1} =
  \ev{\thr'}{\acq{\lk}}$ for some $\thr,\thr',\lk$,
\item\itemlabel{fork} $e_i = \ev{\thr}{\fork{\thr'}}$ and $e_{i+1} =
  \ev{\thr'}{o}$ for some $\thr,\thr',o$, or
\item\itemlabel{join} $e_i = \ev{\thr'}{o}$ and $e_{i+1} =
  \ev{\thr}{\join{\thr'}}$ for some $\thr,\thr',o$.
\end{enumerate}
\end{definition}
For any $P \in \set{\s{tr},\s{TO},\s{HB}}$, $\nso{P}{\tr}$ refers to
the partial relation $\so{P}{\tr} \cup =^{\tr}$,
where $=^\tr$ denotes the identity relation on the
events of $\tr$. 
When $\tr$ is clear
from the context we will drop the superscript from these relations;
for example, we will use $\HB$ instead of $\HB[\tr]$.

Finally, we say a pair of events $e_1, e_2$ are
\emph{concurrent} (w.r.t. happens before) if neither $e_1 \HB e_2$,
nor $e_2 \HB e_1$; we denote this by $e_1 \hbconc e_2$.

We now define races identified by the happens-before relation. A pair
of events $e_1 = \ev{t_1}{\mathtt{a}_1(x)}$ and $e_2 =
\ev{t_2}{\mathtt{a}_2(x)}$ (for some variable $x$) is said to be
\emph{conflicting}, denoted $e_1 \conflict e_2$, if $t_1 \neq t_2$ and
at least one out of $\mathtt{a}_1$ and $\mathtt{a}_2$ is
$\mathtt{w}$. A trace $\tr$ is said to have a \emph{happens before
  race} (\emph{{\hbrace}}, for short) if there is a pair of events
$e_1,e_2 \in \tr$ such that $e_1 \conflict e_2$ and $e_1 \hbconc e_2$.
\begin{example}
\exlabel{hb-defn} We illustrate the happens before relation through
the trace $\tr_1$ in \figref{example-trace-grammar}. $e_1 \HB e_3$ because
$e_2$ happens before every event in thread $2$ since it forks thread
$2$. Similarly, we can conclude that $e_{13} \HB e_{16}$ because the
join event $e_{15}$ is after every event in thread $2$. Another
interesting pair is $e_5 \HB e_{10}$. This is because
$e_4,e_5,e_6$ and $e_8,e_9$ are critical sections over the same lock
$\lk$, and thus, $e_6$ happens before $e_8$. Therefore, $e_5 \TO e_6 \HB e_8
\TO e_{10}$. It is useful to pay attention to a couple of concurrent
pairs of events. Events $e_3$ and $e_7$ are concurrent, but do not
constitute an {\hbrace} because $e_3$ and $e_7$ being read events are
not conflicting. However, there is an {\hbrace} between events
$e_{10}$ and $e_{13}$; they are concurrent and a conflicting pair of
events.
\end{example}
The standard \textsc{FastTrack} style vector clock
algorithm~\cite{lam78,fid88,mat88,ps03,ff09} detects if
a given trace has a race and runs in time $O(nT\log n)$
and uses space $O((V+L+T)T\log n)$ for a trace with $n$ events,
$T$ threads, $L$ locks and $V$ variables.
% \begin{theorem}
% Consider a trace $\tr$ of length $n$ with $T$ threads, $L$ locks, and
% $V$ variables. The vector clock algorithm for detecting {\hbrace}{s}
% runs in time $O(nT\log n)$ and uses space $O((V+L+T)T\log n)$.
% \thmlabel{vc_complexity}
% \end{theorem}
\end{myparagraph}

\begin{myparagraph}{Goldilocks Algorithm}
Goldilocks algorithm~\cite{eqt07} is another algorithm that detects
the presence of HB-races. 
% Roughly, for every prefix $\tr'$ of a given trace $\tr$,
% and for every variable $x$, the set of threads and locks that have an event
% that \emph{happens-after} an access event of $x$.
In order to formally describe the algorithm,
let us first fix some notations.
Consider the function $\after{\tr}$ defined as follows:
\begin{equation*}
\begin{array}{rll}
\after{\tr}(e) = & \setpred{\thr \in \threads(\tr)}{\exists e' =
  \ev{\thr}{o}.\ e \HB[\tr] e'}  \\ 
\quad\quad\cup & \setpred{\thr \in \threads(\tr)}{\exists e' =
  \ev{\thr'}{\fork{\thr}}.\ e \HB[\tr] e'}  \\
\cup & \setpred{\lk \in \locks(\tr)}{\exists e' = \ev{\thr}{\rel{\lk}}.\ e
  \HB[\tr] e'}
\end{array}
\equlabel{after}
\end{equation*}
Thus, informally, $\after{\tr}(e)$ is the set of all threads and locks
that have an event $\s{HB}$-after $e$.
% Also, let us define $\lr{\tr}(\thr,x)$ (resp. $\lw{\tr}(x)$)
% to be the event $e \in \reads{\tr}(\thr,x)$ (resp. $e \in \writes{\tr}(x)$ ) 
% such that for every event $e' \in \reads{\tr}(\thr,x)$
% (resp. $e' \in \writes{\tr}(x)$), we have that $e' \tro[\tr] e$.
% If $\reads{\tr}(\thr,x) = \emptyset$ (resp. $\writes{\tr}(x) = \emptyset$),
% we say that $\lr{\tr}(\thr,x)$ (resp. $\lw{\tr}(x)$) is undefined.
% Next, let us define $\lr{\tr}(\thr,x)$ to be the event 
% $e \in \reads{\tr}(\thr,x)$ such that for every event 
% $e' \in \reads{\tr}(\thr,x)$, we have that $e' \tro[\tr] e$.
% If $\reads{\tr}(\thr,x) = \emptyset$, we say that $\lr{\tr}(\thr,x)$ is undefined.
% Similarly, if $\writes{\tr}(x) \neq \emptyset$, we denote by 
% $\lw{\tr}(x)$ the event $e \in \writes{\tr}(x)$ 
% such that $\forall e' \in \writes{\tr}(x), e' \tro[\tr] e$;
% otherwise $\lw{\tr}(x)$ is undefined.

Then, for every prefix $\tr'$ of the trace, and for every thread $\thr$ and variable $x$ in $\tr'$,
the Goldilocks algorithm maintains the set $\glsr{\tr'}(t,x)$ defined by
\[
\glsr{\tr'}(t,x) = \after{\tr'}(\last{\tr'}(\reads{\tr'}(t,x)))
\]
and for every variable $x$ in $\tr'$, the set 
\[
\glsw{\tr'}(x) = \after{\tr'}(\last{\tr'}(\writes{\tr'}(x)))
\]
where, $\after{\tr'}(\s{undefined})$ is assumed to be the empty set.

% \ucomment{Should I add description of how these sets get updated?}
Finally, a race is declared after observing an event $e$
such that one of the following hold:
\begin{enumerate}
  \item $e = \ev{t}{\wt{x}}$ and either $t \not\in \glsw{\tr'}(x)$
   or $t \not\in \glsr{\tr'}(t',x)$ for some thread $t'\in \threads(\tr')$
  \item $e = \ev{t}{\rd{x}}$ and $t \not\in \glsw{\tr'}(x)$.
\end{enumerate}
where $\tr'$ is the prefix until the event $e$.  
This algorithm runs in time $O(n(L+TV))$ and  uses space
$O(TV(T+L))$ for a trace with $n$ events, $T$ threads,
$L$ locks and $V$ variables.
% space complexity of this algorithm is given below.
% \begin{theorem}
% \thmlabel{gls-complexity} Consider a trace $\tr$ length $n$ with $T$
% threads, $L$ locks, and $V$ variables. The Goldilocks algorithm 
%  runs in time $O(n(L+TV))$ and uses space
% $O(TV(T+L))$.
% \end{theorem}
% \ucomment{Example illustrating goldilocks?}
\end{myparagraph}

\begin{myparagraph}{Eraser's Lockset Algorithm}
The lockset algorithm~\cite{sbnsa97} is a low overhead technique to detect potential races.
The basic idea here, is to maintain, for
every variable $x$, the set of locks that protect each access to $x$,
and check if this set becomes empty as the execution proceeds.  We
recall the details of this technique here. We will assume that none of
the elements in the set $\cD = \set{\dmlock} \cup
\setpred{\dmlockthr{\thr}}{\thr \mbox{ is a thread}}$ are locks used
by the program. The elements of the set $\cD$ are ``dummy'' or fake
locks introduced by the algorithm to ensure that alarms are not raised
when a (global) variable is only read (and never written to), and when
a variable is accessed by only one thread~\cite{ps03}. For a read/write event
event $e = \ev{\thr}{\mathtt{a}(x)}$ (where $\mathtt{a}$ is either
$\mathtt{r}$ or $\mathtt{w}$) in trace $\tr$, $\s{LocksHeld}_{\tr}(e)$
is the set of locks held by $\thr$ when $e$ is performed. 
Using this, for an event $e=\ev{\thr}{\mathtt{a}(x)}$  
we define $\lockset{\tr}(e)$ to be
\[
\lockset{\tr}(e) = \left\{ \begin{array}{ll}
                  \set{\dmlock,\dmlockthr{\thr}} \cup 
                          \locksheld{\tr}(e) & 
                                     \mbox{if } \mathtt{a} = \mathtt{r}\\
                  \set{\dmlockthr{\thr}} \cup \locksheld{\tr}(e) & 
                                     \mbox{if } \mathtt{a} = \mathtt{w}
                                             \end{array} \right.
\]
For a variables $x$ and thread $\thr$, let $\access{\tr}(\thr,x)$
be the set of all events in $\proj{\tr}{\thr}$ whose corresponding operations are either
$\rd{x}$ or $\wt{x}$. Then,
\[
\lockset{\tr}(\thr,x) = \smashoperator[r]{\bigcap_{e \in \access{\tr}(\thr,x)}}
                               \lockset{\tr}(e).
\]
%\ucomment{Here also you have defined $\access{\tr}$, locally though.}
As per convention, when $\access{\tr}(\thr,x) = \emptyset$ (i.e.,
thread $\thr$ never accesses the variable $x$), the right hand side of
the above equation is assumed to be $\locks(\tr) \cup \cD$. A few
observations about these definitions are in order. First
$\lockset{\tr}(\thr,x)$ is always non-empty because $\dmlockthr{\thr}
\in \lockset{\tr}(\thr,x)$. Second, if all events in
$\access{\tr}(\thr,x)$ are read events, then $\dmlock \in
\lockset{\tr}(\thr,x)$. 
The lockset discipline is said to be \emph{violated} in trace $\tr$,
if for some variable $x$, 
\[\smashoperator[r]{\bigcap_{\thr \in \threads(\tr)}}
\lockset{\tr}(\thr,x) = \emptyset.\] 
Note that the Eraser algorithm crucially depends upon the 
accurate computation of $\locksheld{\tr}(e)$. 
To compute this for traces having reentrant locks, 
we need to record, for each thread $\thr$ and lock $\lk$,
the number of times $\lk$ has been acquired, without being released,
which can be maintained using an integer variable.

% We conclude this description by highlighting the importance of the
% locks in $\cD$ that were introduced. Take $\s{LS} = \cap_{\thr \in
%   \threads(\tr)} \lockset{\tr}(\thr,x)$. 
We briefly highlight the importance of the
locks in $\cD$ that were introduced. 
 Let  
$\s{LS}(x) = \cap_{\thr \in
  \threads(\tr)} \lockset{\tr}(\thr,x)$.
% Take $\s{LS} = \cap_{\thr \in
%   \threads(\tr)} \lockset{\tr}(\thr,x)$. 
%\footnote{The idea of using dummy locks to enhance the precision of 
%Eraser was also described in~\cite{multirace2003}}. 
If the variable $x$ is only accessed by a single thread $\thr_1$, then
$\s{LS}(x)$ is non-empty because it contains $\dmlockthr{\thr_1}$. And if
a variable $x$ is only read and never written to, then $\s{LS}(x)$ is
again non-empty because it contains $\dmlock$.
The Eraser algorithm~\cite{sbnsa97} checks for violation of the lockset
principle by maintaining the lockset for each thread-variable pair.
It runs in time $O(n(L + \log r))$ and uses space $O(TL\log r + V(T+L))$
where $n$, $T$, $L$ and $V$ are the number of events, threads, locks,
and variables respectively, and $r$ is the maximum number of
times a thread acquires a lock without releasing it.
% \begin{theorem}
% \thmlabel{eraser-complexity} 
% Let $\tr$ be a trace $\tr$ of length $n$ with $T$
% threads, $L$ locks, and $V$ variables. Let $r$ be the maximum number of
% times a thread acquires a lock without releasing it.
% The Eraser algorithm detects violations of the lockset 
% discipline on $\tr$ in time $O(n(L + \log r))$ and uses space 
% % $O(TL(V + \log r))$.
% $O(TL\log r + V(T+L))$.
% \end{theorem}
\end{myparagraph}

\input{trace2}

\begin{example}
\exlabel{eraser} We illustrate the lockset algorithm on a couple of
examples. Consider the trace $\tr_2$ in \figref{eraser-trace-grammar}. The
relevant locksets are as follows.
\[
\begin{array}{ll}
\lockset{\tr_2}(1,x) = \set{\dmlock,\dmlockthr{1}} & 
\lockset{\tr_2}(2,x) = \set{\dmlock,\dmlockthr{2}} \\
\lockset{\tr_2}(1,y) = \set{\dmlockthr{1},\lk} &
\lockset{\tr_2}(2,y) = \set{\dmlockthr{2},\lk} \\
\lockset{\tr_2}(1,z) = \set{\dmlockthr{1}} &
\lockset{\tr_2}(2,z) = \set{\dmlock,\dmlockthr{1},\dmlockthr{2},\lk}\\
\end{array}
\]
% \[
% \begin{array}{ll}
% \lockset{\tr_2}(1,x) = \set{\dmlock,\dmlockthr{1}} & 
% \lockset{\tr_2}(1,y) = \set{\dmlockthr{1},\lk} \\
% \lockset{\tr_2}(1,z) = \set{\dmlockthr{1}} &
% \lockset{\tr_2}(2,x) = \set{\dmlock,\dmlockthr{2}} \\
% \lockset{\tr_2}(2,y) = \set{\dmlockthr{2},\lk} &
% \lockset{\tr_2}(2,z) = \set{\dmlock,\dmlockthr{1},\dmlockthr{2},\lk}\\
% \end{array}
% \]
Observe that $\lockset{\tr_2}(2,z)$ is the set of all locks because
thread $2$ does not access $z$. The trace $\tr_2$ does not violate the
lockset discipline. Informally, the reason for this is because
variable $x$ is only read by both threads, accesses to variable $y$ is
always protected by lock $\lk$, and variable $z$ is local to thread
$1$. Trace $\tr_2$ also contains no {\hbrace}.

For trace $\tr_1$ from \figref{example-trace-grammar}, 
\[
\begin{array}{ll}
\lockset{\tr_1}(1,x) = \set{\dmlockthr{1}} &
\lockset{\tr_1}(2,x) = \set{\dmlock,\dmlockthr{2}} \\
\lockset{\tr_1}(1,y) = \set{\dmlockthr{1}} &
\lockset{\tr_1}(2,y) = \set{\dmlockthr{2},\lk}.
\end{array}
\] 
The lockset discipline is violated on both variables $x$ and $y$. On
the other hand, there is an {\hbrace} only on variable $y$ (events 10
and 13; see \exref{hb-defn}). Thus, the lockset discipline may falsely
conclude the presence of races; it is only a lightweight approximate
approach.
\end{example}
%\ucomment{Maybe say a line or two saying that Eraser is unsound but complete,
%HB is sound but incomplete, and what does soundness/incomplete mean?
%Implication: whenever HB detects race, eraser does too, 
%and if eraser does not report, then so does HB.
%Future work: Similar to~\cite{choi:2003:HDD:781498.781528}, use a hybrid algorithm to
%improve precision}

\begin{myparagraph}{Straight Line Programs (SLP)} 
We consider traces that are compressed using special context-free
grammars called straight line programs (SLP). 
Recall that a context-free grammar (in Chomsky Normal Form) is $G =
(T,N,S,R)$, where $T$ is the set of terminals, $N$ the set of
non-terminals, $T\cup N$ is the set of symbols, $S \in N$ is the start symbol, and $R$ is the set of
rules in which each rule in $R$ is either $A \rightarrow a$ or $A
\rightarrow BC$, for $A,B,C \in N$ and $a \in T$. 
A \emph{straight line program} is a context free grammar such that (a)
for every non-terminal $A$, there is exactly one rule where $A$
appears on the left, and (b) the non-terminals are ranked in such that
way that in every rule, the non-terminals on the right are of larger
rank than the non-terminal on the left of the rule, i.e., for rules $A
\rightarrow BC$, $A \prec B$ and $A \prec C$.
It is easy to observe that the language of the grammar contains a
single string, 
namely, the one that is being succinctly represented by the SLP.
% \ucomment{and we say that the SLP succinctly represents this unique string.} 
Without loss of generality, we will assume that every
non-terminal in the SLP is \emph{useful}, i.e., every non-terminal in
the grammar appears in some sentential form in the unique derivation
in the grammar. Thus, the language associated with any non-terminal
$A$ has a single string. 
We will call this (unique) string generated by non-terminal $A$ a
\emph{chunk}, and denote it by $\sem{A}$. We will often abuse notation
and refer to both $\sem{A}$ and $A$ as ``$A$''. For example,
$\locks(A)$ will mean $\locks(\sem{A})$.

The \emph{size} of an SLP $G = (T,N,S,R)$ will be taken to be
$|T|+|N|$; note that this measure of size is linearly related to other
measures of size one might consider like $|R|$ or sum of the sizes of
all the rules in $R$.
We make a couple of observations about the size of an SLP versus the
size of the trace it represents.
First, every trace $\tr = e_1, e_2, \ldots e_n$ can be represented by
a ``trivial'' SLP of size $O(n)$ as follows. The non-terminals are
$\setpred{A_{[i,i]}}{1 \leq i \leq n} \cup \setpred{A_{[1,i]}}{1 \leq
  i \leq n}$ with start symbol $A_{[1,n]}$. Intuitively, $A_{[i,i]}$
represents the string $e_i$, while $A_{[1,i]}$ represents the prefix of
length $i$. This is accomplished by the rules --- $A_{[i,i]}
\rightarrow e_i$ and $A_{[1,i]} \rightarrow A_{[1,i-1]}A_{[i,i]}$ for
each $1 \leq i \leq n$.
Second, the SLP representation of a string $\tr$ maybe exponentially
smaller than $\tr$ itself. For example, take $\tr = a^{2^n}$. An
$O(n)$ SLP representation for $\tr$ is as follows: $N =
\setpred{A_i}{0 \leq i \leq n}$ with rules $A_0 \rightarrow a$, and
$A_{i+1} \rightarrow A_iA_i$. One can inductively observe that
$\sem{A_i} = a^{2^i}$, and so $\sem{A_n} = \tr$.
\begin{example}
\exlabel{SLP} \figref{example-trace-grammar} describes an 
SLP representation of trace $\tr_1$. The
rules for $E,F,D$, and $G$ are not strictly in the format of an SLP,
but it can easily be converted into one; the representation in
\figref{example-trace-grammar} is sufficient for our illustrative
purposes. We will again use $e_i$ to denote the $i$th event of
$\tr_1$. Chunk $E$ represents $e_1,e_2$, $F$ represents
$e_3,e_4,e_5,e_6$ and $e_{11},\ldots e_{14}$, $D$ represents
$e_7,e_8,e_9,e_{10}$, and $G$ represents $e_{15},e_{16}$. The
sub-traces represented by the other non-terminals can be similarly
discovered. As mentioned before, we will confuse the notation
distinguishing between a non-terminal and the string it
represents. Thus, for example, $\threads(E) = \set{1,2}$.

Similarly, the SLP for $\tr_2$ is shown in \figref{eraser-trace-grammar}. 
The sub-traces
represented by non-terminals need not conform to thread and critical
section boundaries. For example, the chunk $\sem{X}$ has partial
critical sections of different threads.
\end{example}
% \ucomment{Add paragraph on online algorithms that construct SLP compressions
% of traces. In paryicular discuss Sequitr -- What do you mean by online?}.
% We will consider traces succinctly represented by SLPs.  
Several well known algorithms for SLP based compression are known in the
literature.  The most basic and popular one is
Sequitur~\cite{nev,nw97}.  Sequitur takes a string as an input, and
generates an SLP representing the trace.  It runs in time and space
linear in the size of the input string.  The Sequitur algorithm works
in an online incremental fashion; it reads the input string one
character at a time, and updates the SLP generated so-far.  It
maintains a list of digrams (symbol pairs) that occur somewhere in the
SLP so-far.  On seeing a new character, the algorithm appends it at
the end of the rule corresponding to the start symbol.  The new digram
formed (by appending the new character to the last symbol of the rule)
is added to the list of digrams, if it is not already present.
Otherwise, a new rule, with a fresh nonterminal generating the digram,
is added to the SLP, and every occurence of the digram is replaced by
the freshly introduced non-terminal.  At every step, non-terminals,
that are not useful, are also removed.  Other popular grammar based
compression schemes include Sequential~\cite{sequential2000},
LZ77~\cite{zl77}, LZW~\cite{lzw}, Bisection~\cite{bisection}, longest
match~\cite{longestmatch2000} and Re-Pair~\cite{repair2000}.
\end{myparagraph}

% \begin{myparagraph}{ZDDs}
% Postponed adding text.
% \end{myparagraph}

%% file: trace1.tex
%!TEX root = main.tex

% \begin{figure*}[h]
% \begin{minipage}[b]{0.7\textwidth}
% % \centering
% \begin{minipage}[b]{0.25\textwidth}
%   \centering
% \scalebox{0.8}{
%   \execution{2}{
%     \figev{1}{\wt{x}}
%     \figev{1}{\fork{2}}
%     \figev{2}{\rd{x}}
%     \figev{2}{\acq{\lk}}
%     \figev{2}{\wt{y}}
%     \figev{2}{\rel{\lk}}
%     \figev{1}{\rd{x}}
%     \figev{1}{\acq{\lk}}
%     \figev{1}{\rel{\lk}}
%     \figev{1}{\wt{y}}
%     \figev{2}{\rd{x}}
%     \figev{2}{\acq{\lk}}
%     \figev{2}{\wt{y}}
%     \figev{2}{\rel{\lk}}
%     \figev{1}{\join{2}}
%     \figev{1}{\wt{y}}
%   }
%   }
%   \caption{Example trace $\tr_1$}
%   \figlabel{example-trace}
% \end{minipage}
% \begin{minipage}[b]{0.27\textwidth}
%   \centering
% \scalebox{0.8}{
%   \grammar{
%     \gramrule{S}{AB}
%     \gramrule{A}{CD}
%     \gramrule{C}{EF}
%     \gramrule{E}{\ev{1}{\wt{x}} \ev{1}{\fork{2}}}
%     \gramrule{F}{\ev{2}{\rd{x}} \ev{2}{\acq{\lk}} \ev{2}{\wt{y}} \ev{2}{\rel{\lk}}}
%     \gramrule{D}{\ev{1}{\rd{x}} \ev{1}{\acq{\lk}} \ev{1}{\rel{\lk}} \ev{1}{\wt{y}}}
%     \gramrule{B}{FG}
%     \gramrule{G}{\ev{1}{\join{2}} \ev{1}{\wt{y}}}
%   }
%   }
%   \caption{Example SLP representation of trace in
%     \figref{example-trace}.}
%   \figlabel{example-grammar}
% \end{minipage}
% \end{minipage}
% \end{figure*}

\begin{figure}[t]
\scalebox{0.75}{
  \execution{2}{
    \figev{1}{\wt{x}}
    \figev{1}{\fork{2}}
    \figev{2}{\rd{x}}
    \figev{2}{\acq{\lk}}
    \figev{2}{\wt{y}}
    \figev{2}{\rel{\lk}}
    \figev{1}{\rd{x}}
    \figev{1}{\acq{\lk}}
    \figev{1}{\rel{\lk}}
    \figev{1}{\wt{y}}
    \figev{2}{\rd{x}}
    \figev{2}{\acq{\lk}}
    \figev{2}{\wt{y}}
    \figev{2}{\rel{\lk}}
    \figev{1}{\join{2}}
    \figev{1}{\wt{y}}
  }
  \hspace{0.2in}
  \grammar{
    \gramrule{S}{AB}
    \gramrule{A}{CD}
    \gramrule{C}{EF}
    \gramrule{E}{\ev{1}{\wt{x}} \ev{1}{\fork{2}}}
    \gramrule{F}{\ev{2}{\rd{x}} \ev{2}{\acq{\lk}} \ev{2}{\wt{y}} \ev{2}{\rel{\lk}}}
    \gramrule{D}{\ev{1}{\rd{x}} \ev{1}{\acq{\lk}} \ev{1}{\rel{\lk}} \ev{1}{\wt{y}}}
    \gramrule{B}{FG}
    \gramrule{G}{\ev{1}{\join{2}} \ev{1}{\wt{y}}}
  }
  }
  \vspace{0.7cm}
  \caption{\small Example trace $\tr_1$ and its SLP representation}
  \figlabel{example-trace-grammar}
\end{figure}

%% file: trace2.tex
%!TEX root = main.tex

% \begin{figure*}[h]
% \centering
% \begin{minipage}[b]{.45\textwidth}
%   \centering
%   \scalebox{0.8}{
%   \execution{2}{
%     \figev{1}{\rd{x}}
%     \figev{1}{\acq{\lk}}
%     \figev{1}{\wt{y}}
%     \figev{1}{\rel{\lk}}
%     \figev{2}{\acq{\lk}}
%     \figev{2}{\rd{x}}
%     \figev{2}{\wt{y}}
%     \figev{2}{\rel{\lk}}
%     \figev{2}{\rd{x}}
%     \figev{1}{\wt{z}}
%   }
%   }
%   \caption{Example trace $\tr_2$}
%   \figlabel{eraser-trace}
% \end{minipage}
% % \begin{minipage}[b]{.45\textwidth}
% %   \centering
% %   \grammar{
% %     \gramrule{S}{WX}
% %     \gramrule{W}{YZ}
% %     \gramrule{Y}{\ev{1}{\rd{x}} \ev{1}{\acq{\lk}}} 
% %     \gramrule{Z}{\ev{1}{\wt{y}} \ev{1}{\rel{\lk}} \ev{2}{\acq{\lk}} \ev{2}{\wt{y}}}
% %     \gramrule{X}{\ev{2}{\rel{\lk}} \ev{2}{\rd{x}} \ev{1}{\wt{z}}}
% %   }
%   \begin{minipage}[b]{.45\textwidth}
%   \centering
%   \scalebox{0.8}{
%   \grammar{
%     \gramrule{S}{UV}
%     \gramrule{U}{WX}
%     \gramrule{W}{\ev{1}{\rd{x}} \ev{1}{\acq{\lk}}}
%     \gramrule{X}{\ev{1}{\wt{y}} \ev{1}{\rel{\lk}} \ev{2}{\acq{\lk}}}
%     \gramrule{V}{YZ}
%     \gramrule{Y}{\ev{2}{\rd{x}} \ev{2}{\wt{y}}} 
%     \gramrule{Z}{\ev{2}{\rel{\lk}} \ev{2}{\rd{x}} \ev{1}{\wt{z}}}
%   }
%   }
%   \caption{SLP representation of trace $\tr_2$ in
%     \figref{eraser-trace}.}
%   \figlabel{eraser-grammar}
% \end{minipage}
% \end{figure*}

\begin{figure}[t]
\scalebox{0.75}{
  \execution{2}{
    \figev{1}{\rd{x}}
    \figev{1}{\acq{\lk}}
    \figev{1}{\wt{y}}
    \figev{1}{\rel{\lk}}
    \figev{2}{\acq{\lk}}
    \figev{2}{\rd{x}}
    \figev{2}{\wt{y}}
    \figev{2}{\rel{\lk}}
    \figev{2}{\rd{x}}
    \figev{1}{\wt{z}}
  }
  \hspace{0.2in}
    \grammar{
    \gramrule{S}{UV}
    \gramrule{U}{WX}
    \gramrule{W}{\ev{1}{\rd{x}} \ev{1}{\acq{\lk}}}
    \gramrule{X}{\ev{1}{\wt{y}} \ev{1}{\rel{\lk}} \ev{2}{\acq{\lk}}}
    \gramrule{V}{YZ}
    \gramrule{Y}{\ev{2}{\rd{x}} \ev{2}{\wt{y}}} 
    \gramrule{Z}{\ev{2}{\rel{\lk}} \ev{2}{\rd{x}} \ev{1}{\wt{z}}}
  }
}
\vspace{0.7cm}
\caption{\small Example trace $\tr_2$ and its SLP representation}
\figlabel{eraser-trace-grammar}
\end{figure}

%% file: hbalgorithm.tex
%!TEX root = main.tex

\section{{\hbrace}s in Compressed traces}
\seclabel{hb}

In this section, we will present our algorithm for detecting
{\hbrace}s in compressed traces represented by SLPs. The algorithm's
running time will be linear in the size of the SLP 
(as opposed to algorithms analyzing uncompressed traces 
with running times at least linear in the size 
of the uncompressed trace). 
While it is very different from the classical
vector clock algorithm, it is similar in flavor to the Goldilocks
Algorithm.

\subsection{Detecting Cross-Races}
\seclabel{hb-cross-race}

Our algorithm will proceed inductively. Starting
from the non-terminals of largest rank, we will proceed to determine
for each non-terminal $A$, whether there is an {\hbrace} amongst the
events in the chunk that $A$ generates. In other words, for each
non-terminal $A$, we will determine the predicate $\race(A)$ which is
true if and only if there is an {\hbrace} between events in
$\sem{A}$. For a non-terminal $A$, whose (only) rule is of the form $A
\rightarrow a$, where $a$ is an event, $\race(A)$ is clearly false,
because $\sem{A}$, in this case, has only one event. 

Let us now consider the case when the rule corresponding to
$A$ has the form $A \rightarrow BC$,
where $B$ and $C$ are non-terminals of higher rank. If there is a race
in chunk $\sem{A}$ between events (say) $e$ and $e'$, then it is one
of two kinds. The first case is when $e$ and $e'$ both belong to chunk
$\sem{B}$ or both belong to chunk $\sem{C}$. The existence of such
races can be determined by computing (inductively) the predicates
$\race(B)$ and $\race(C)$. The other possibility is that $e \in
\sem{B}$ while $e' \in \sem{C}$. How we discover the presence of such
\emph{cross-races}, is the main challenge we need to overcome. 

\input{scenarios_hb}

Consider two events $e,e'$ such that $e \in \sem{B}$ and $e' \in
\sem{C}$. Suppose $e \HB[BC] e'$. Then, %by \defref{happens-before},
there is a sequence $e = e_1, e_2, \ldots e_n = e'$ that satisfies the
conditions in \defref{happens-before}. 
Thus, for $1\leq i \leq n-1$, we have the trace order $e_i\tros[BC] e_{i+1}$.
Also, $e = e_1 \in \sem{B}$, and $e_n = e' \in \sem{C}$.
This means that there exists $j$ such that for all $i \leq j$, $e_i \in \sem{B}$, and
for all $i \geq j+1$, $e_i \in \sem{C}$. 
In other words,
$(e_j,e_{j+1})$ is how the sequence $e_1, \ldots e_n$ ``crosses'' the
$B$-$C$ boundary (see \figref{hb-ind-step}). Observe that we have
$e=e_1 \HB[B] e_j$ and $e_{j+1} \HB[C] e_n = e'$. It is important to
note that the relationship between $e$ and $e_j$ (and $e_{j+1}$ and
$e'$) only depends on the events in chunk $\sem{B}$
($\sem{C}$). Depending on which of the conditions
\itemref{thread-order}, \itemref{acq-rel}, \itemref{fork}, and
\itemref{join} of \defref{happens-before} hold for the pair
$(e_j,e_{j+1})$, we have one of the following: either $e_j$ and
$e_{j+1}$ are events of the same thread, or $e_j$ is a release event
and $e_{j+1}$ is an acquire event on the same lock, or $e_j$ is a fork
event and $e_{j+1}$ is an event of the child thread, or $e_j$ is a
join event and $e_{j+1}$ is an event of the parent thread. These
scenarios are illustrated in \figref{hb-ind-step}. Thus, if an event
$e \in B$ \emph{happens-before} an event $e' \in C$ then there is a
common thread or a common lock through which the ordering is ``communicated''
across the $B$-$C$ boundary. The converse of this observation is also
true. We now make this intuition precise.

For a trace $\tr$, and event $e \in \tr$, recall the
 function $\after{\tr}$:
\begin{equation*}
\begin{array}{rll}
\after{\tr}(e) = & \hspace{-0.1in} \setpred{\thr \in \threads(\tr)}{\exists e' =
  \ev{\thr}{o}.\ e \HB[\tr] e'}  \\ 
\quad\quad\cup & \hspace{-0.1in}\setpred{\thr \in \threads(\tr)}{\exists e' =
  \ev{\thr'}{\fork{\thr}}.\ e \HB[\tr] e'}  \\
\cup & \hspace{-0.1in} \setpred{\lk \in \locks(\tr)}{\exists e' = \ev{\thr}{\rel{\lk}}.\ e
  \HB[\tr] e'}
\end{array}
% \equlabel{after}
\end{equation*}
% Thus, informally, $\after{D}(e)$ is the set of all threads and locks
% that have an event $\s{HB}$-after $e$. 
We can, dually, define the set
of locks/threads that have an event $\s{HB}$-before $e$ in $\tr$.
\begin{equation*}
\begin{array}{ll}
\before{\tr}(e) = & \hspace{-0.15in} \setpred{\thr \in \threads(\tr)}{\exists e' =
  \ev{\thr}{o}.\ e' \HB[\tr] e} \\
\quad\quad\quad\quad\;\;\cup & \hspace{-0.15in} \setpred{\thr \in \threads(\tr)}{\exists e' =
  \ev{\thr'}{\join{\thr}}.\ e' \HB[\tr] e} \\
\quad\quad\quad\quad\;\;\cup & \hspace{-0.15in} \setpred{\lk \in \locks(\tr)}{\exists e' = \ev{\thr}{\acq{\lk}}.\ e'
  \HB[\tr] e}
\end{array}
\equlabel{before}
\end{equation*}
The main observation that underlies the algorithm is that $\after{}$
and $\before{}$ sets can be used to discover $\s{HB}$ ordering between
events across chunks.
\begin{lemma}
\lemlabel{hb-before-after}
Consider events $e \in \sem{B}$ and $e' \in \sem{C}$. $e \HB[BC] e'$
iff $\after{B}(e) \cap \before{C}(e') \neq \emptyset$.
\end{lemma}

\lemref{hb-before-after} suggests that cross races in chunk $BC$ can be
discovered by maintaining the after and before sets of data access
events. However, we don't need to maintain these sets for all access
events; instead, we can do it only for the first and last events. This
is the content of the next lemma.
% \begin{lemma}
% \lemlabel{first_and_last_events}
% If both $\race(B)$ and $\race(C)$ are false,
% then there is an HB race between events $e \in \sem{B}$ and $e' \in \sem{C}$
% iff there is an HB race between $\s{last}_B^e$ and $\s{first}_C^{e'}$.
% \end{lemma}
% Here, for an access event $e \in \sem{D}$, the events 
% $\s{first}_D^e$ and $\s{last}_D^e$ are the first and last events
% in $D$ of the same \emph{type} as $e$. Formally,
% \begin{equation*}
% \s{first}_{D}^e = 
%   \left\{\begin{array}{ll}
%     \first{D}(\reads{D}(\thr, x)) & \text{if } e = \ev{\thr}{\rd{x}} \\
%     \first{D}(\writes{D}(x)) & \text{if } e = \ev{\thr}{\wt{x}}
%   \end{array} \right.
% \end{equation*}
% \begin{equation*}
% \s{last}_{D}^e = 
%   \left\{\begin{array}{ll}
%     \last{D}(\reads{D}(\thr, x)) & \text{if } e = \ev{\thr}{\rd{x}} \\
%     \last{D}(\writes{D}(x)) & \text{if } e = \ev{\thr}{\wt{x}}
%   \end{array} \right.
% \end{equation*}
\begin{lemma}
\lemlabel{first_and_last_events} If there is no {\hbrace} in $\sem{B}$
or in $\sem{C}$, and if there is an {\hbrace} between events $e \in
\sem{B}$ and $e' \in \sem{C}$ then, there is an {\hbrace} between
$\s{last}_B^e$ and $\s{first}_C^{e'}$, where
\begin{equation*}
\s{last}_{B}^e = 
  \left\{\begin{array}{ll}
    \last{B}(\reads{B}(\thr, x)) & \text{if } e = \ev{\thr}{\rd{x}} \\
    \last{B}(\writes{B}(x)) & \text{if } e = \ev{\thr}{\wt{x}}
  \end{array} \right.
\end{equation*}
and
\begin{equation*}
\s{first}_{C}^{e'} = 
  \left\{\begin{array}{ll}
    \first{C}(\reads{C}(\thr, x)) & \text{if } e' = \ev{\thr'}{\rd{x}} \\
    \first{C}(\writes{C}(x)) & \text{if } e' = \ev{\thr'}{\wt{x}}
  \end{array} \right.
\end{equation*}
\end{lemma}

% \ucomment{Although, there is some subtlety involved here. Naively
% comparing all pairs of events in $B\times C$ will take time proportional
% to the product of the number of events in $B$ and $C$, which is undesirable.
% Our algorithm is cleverly crafted to avoid this problem.}

\lemref{first_and_last_events} suggests that in order to check for
\emph{cross} races, it is enough to inductively maintain the after
sets of the last read/write events and the before sets of the first
read/write events of each variable and thread.  We will denote these
sets by $\ar{}$, $\aw{}$, $\br{}$ and $\bw{}$.  Formally,
% \begin{enumerate}[leftmargin=*]
%   \item $\ar{D}(\thr,x) = \after{D}(\last{D}(\reads{D}(\thr,x)))$,
%   \item $\aw{D}(x) = \after{D}(\last{D}(\writes{D}(x)))$,
%   \item $\br{D}(\thr,x) = \before{D}(\first{D}(\reads{D}(\thr,x)))$,
%   \item $\bw{D}(x) = \after{D}(\first{D}(\writes{D}(x)))$
% \end{enumerate}
\begin{equation}
\equlabel{ar-aw-br-bw}
\begin{array}{ll}
  \ar{D}(\thr,x) &= \after{D}(\last{D}(\reads{D}(\thr,x))) \\
  \aw{D}(x) &= \after{D}(\last{D}(\writes{D}(x))) \\
  \br{D}(\thr,x) &= \before{D}(\first{D}(\reads{D}(\thr,x))) \\
  \bw{D}(x) &= \before{D}(\first{D}(\writes{D}(x)))
\end{array}
\end{equation}
where we set both $\after{D}(\s{undefined})$ and $\before{D}(\s{undefined})$
to be $\emptyset$.

Based on all of these observations we can conclude
that for a non-terminal $A$ with rule $A \rightarrow BC$, we have,
\begin{equation}
\begin{array}{ll}
\race(A) = & \race(B) \vee \race(C) \vee\\
 \bigvee_{x \in \wvars{B} \cap \wvars{C}} 
   &  \aw{B}(x) \cap \bw{C}(x) = \emptyset \\ %\s{RaceWW}_{B,C}(x)\\
\bigvee_{x \in \wvars{B}, (\thr, x) \in \rvars{C}}
  & \aw{B}(x) \cap \br{C}(\thr, x) = \emptyset \\
\bigvee_{(\thr,x) \in \rvars{B}, x \in \wvars{C}}
  & \ar{B}(\thr, x) \cap \bw{C}(x) = \emptyset\\
\end{array}
\equlabel{race}
\end{equation}
% where
% \begin{itemize}[leftmargin=*]
%   \item $\s{RaceWW}_{B,C}(x) \equiv \aw{B}(x) \neq \emptyset \land \bw{C}(x) \neq \emptyset \land \aw{B}(x) \cap \bw{C}(x) = \emptyset$,
%    \item $\s{RaceWR}_{B,C}(\thr,x) \equiv \aw{B}(x) \neq \emptyset \land \br{C}(\thr,x) \neq \emptyset \land \aw{B}(x) \cap \br{C}(\thr, x) = \emptyset$, and
%     \item $\s{RaceRW}_{B,C}(\thr, x) \equiv \ar{B}(x) \neq \emptyset \land \bw{C}(x) \neq \emptyset \land \ar{B}(\thr, x) \cap \bw{C}(x) = \emptyset$
% \end{itemize}

Thus, our race detection algorithm will be complete if we can
effectively compute the sets $\ar{B}(\thr, x)$, $\aw{B}(x)$, $\br{C}(\thr, x)$, and
$\bw{C}(x)$.
We embark on this challenge in the next section. 

Our definition of the predicate
$\race$ is correct and we state this next.

\begin{theorem}
\thmlabel{race-correctness} For any non-terminal $A$, $\race(A) =
\s{true}$ if and only if there are events $e_1,e_2 \in \sem{A}$ such
that $e_1 \conflict e_2$ and $e_1 \hbconc e_2$.
\end{theorem}

\begin{example}
\exlabel{hb-cross-race} 
Let us illustrate the ideas presented in this section through some
examples. We will consider traces $\tr_1$ and its SLP in \figref{example-trace-grammar}, 
and $\tr_2$ with its SLP in \figref{eraser-trace-grammar}. 
% Again, we
% will use $e_i$ to denote the $i$th event in either $\tr_1$ or $\tr_2$;
% which one among the two will be clear based on the name of the chunk.

We begin by giving examples of $\before{}$ and $\after{}$ sets. 
\[
\begin{array}{ll}
\after{E}(e_1) = \set{1,2} & \after{C}(e_1) = \set{1,2,\lk} \\
\before{G}(e_{16}) = \set{1,2} & \before{B}(e_{16}) = \set{1,2,\lk}\\
\after{W}(e_1) = \set{1} & \after{U}(e_1) = \set{1,2, \lk} \\
\before{X}(e_3) = \set{1} & \before{U}(e_3) = \set{1,\lk}
\end{array}
\]
Let us highlight the significant aspects of these examples. $2 \in
\after{E}(e_1)$ because of $e_2 = \ev{1}{\fork{2}}$ and $\lk \in
\after{C}(e_1)$ because of event $e_6 = \ev{2}{\rel{\lk}}$. On the
other hand, $\lk \not\in \after{W}(e_1)$ because there is no
$\rel{\lk}$ event in chunk $\sem{W}$ (of $\tr_2$). But when
considering the chunk $\sem{U}$ (of $\tr_2$), we have $\lk \in
\after{W}(e_1)$ because of the event $e_4 = \ev{1}{\rel{\lk}}$. Next,
$2 \in \before{G}(e_{16})$ because of the join event $e_{15}$, and $\lk
\in \before{B}(e_{16})$ because of acquire event $e_{12}$.  In trace
$\tr_2$, $\lk \in \before{U}(e_3)$ because of acquire event $e_2$.

Now let us consider the computation of cross-races for the chunks in
\figref{example-trace-grammar}. For $M \in \set{D,E,F,G}$, it is easy to see
that $\race(M) = \s{false}$, because each of these chunks only contain
events of one thread. Let us look at the interesting pairs of events
we considered in \exref{hb-defn}. The absence of race between $e_1$
and $e_3$ can be seen because $\aw{E}(x) = \set{1,2}$ and
$\br{F}(2, x) = \set{2}$, both of which have the thread $2$ in common,
and thus the intersection $\aw{E}(x) \cap \br{F}(x)$ is
non-empty. In fact, what this reasoning demonstrates is that 
there is no race between any $\wt{x}$-event in $E$ and 
any $\rd{x}$-event in
$F$. Similarly, the absence of a race between $e_{13}$ and $e_{16}$
can be seen because $\aw{F}(y) \cap \br{G}(1, y)$ $= \set{2,\lk} \cap \set{1,2} = \set{2} \neq \emptyset$. 
To reason about the events $e_5$ and $e_{10}$,
observe that $\aw{F}(y) = \set{2,\lk}$ and $\bw{D}(y) =
\set{1,\lk}$, both of which have the $\lk$ in common.
Thus, we can conclude there is no race between any 
pair of $\wt{y}$-events crossing the chunk $FD$.

Our reasoning also reveals the existence of $\s{HB}$-concurrent
events. For example, $\ar{F}(2, x) = \set{2,\lk}$, and $\br{D}(1, x) =
\set{1}$. Since these sets are disjoint, it reveals that there are a
pair of $\rd{x}$-events (namely, $e_3$ and $e_7$) that are
$\s{HB}$-concurrent; it is not a {\hbrace} because these events are
not conflicting (none of $e_3$ and $e_7$ is a write event).  The race
between $e_{10}$ and $e_{13}$ can be seen as follows. $\aw{A}(y) =
\set{1}$, and $\bw{B}(y) = \set{2,\lk}$.  We can see that there is a
cross race in chunk $AB$, because these two sets are disjoint.
\end{example}

\subsection{Computing $\before{}$ and $\after{}$ sets}
\seclabel{hb-before-after}

Our discussion in \secref{hb-cross-race} suggests that if we manage to
inductively compute the sets $\ar{}$, $\aw{}$, $\br{}$, and $\bw{}$
(\equref{ar-aw-br-bw}) for each chunk in the grammar, 
then we can use \equref{race} to
determine if a chunk has a race.
In this section
we present such an inductive computation for these sets. We will only
describe the computation of sets $\ar{}$ and $\br{}$. The computation
of the sets $\aw{}$ and $\bw{}$ is similar and is presented
in\appref{hb-algo}.

The base case for non-terminals with rule $A \rightarrow a$, where $a$
is an event, is straightforward. To conserve space, this definition is
skipped here, but presented in\appref{hb-algo}. So we focus on the
inductive step when we have a non-terminal with rule $A \rightarrow
BC$.

First consider the case of $\ar{A}(\thr, x)$, 
which is equal to the set $\after{A}(e)$, 
where $e$ is the last event amongst the read events $\reads{A}(\thr, x)$. 
% If there are no read events in $A$ by thread $\thr$ on variable $x$
If variable $x$ is never read by thread $t$ in the chunk $A$
(i.e., $\reads{A}(\thr, x) = \emptyset$), 
we will have $\ar{A}(\thr, x) = \emptyset$.
Otherwise, depending upon where the last read event $e$ occurs 
in the chunk $\sem{A}$, we have two cases to consider. 
In the first case, this last read event $e$ 
belongs to the chunk $\sem{C}$.
In this, clearly, $e = \last{A}(\reads{A}(t, x))$.
% $\ar{A}(\thr, x) = \after{A}(e)$.
% 
Observe that since $\setpred{e'}{e \HB[C] e'} =
\setpred{e'}{e \HB[A] e'}$, we have $\after{A}(e) = \after{C}(e)$.
Thus, in this case,   $\ar{A}(\thr, x) = \ar{C}(\thr, x)$.
% In this case, the set $\ar{C}(\thr, x)$ will clearly not be $\s{undefined}$.
The interesting case is when $\reads{C}(\thr, x)$ is empty
and $\reads{B}(\thr, x) \neq \emptyset$, i.e.,
the last read event $e$ belongs to the chunk $B$.
Since $\setpred{e' \in \sem{B}}{e \HB[B] e'} \subseteq \setpred{e' \in \sem{A}}{e \HB[A]
  e'}$, we have $\after{B}(e) \subseteq \after{A}(e)$. 
Consider $e' \in \sem{C}$ such that $e \HB[A] e'$. As in the discussion on
cross-races in \secref{hb-cross-race}, this means there is a pair of
events $e_1 \in \sem{B}$ and $e_2 \in \sem{C}$ such that $e \HB[B]
e_1$, $e_2 \HB[C] e'$, and either (1) $e_1,e_2$ are events of the same
thread, or (2) $e_1$ is a fork event and $e_2$ is an event of the child
thread, or (3) $e_1$ is an event of a child thread 
and $e_2$ is a join event, or (4) $e_1$ is a release event 
and $e_2$ is an acquire event on the same lock. 
In each of these cases, $e_1$ witnesses the membership
of some thread/lock $u$ in $\after{B}(e)$, 
and $e'$ is $\s{HB}$-after the ``first'' 
event (namely $e_2$) of $u$ in chunk $C$. 
% The definition
% of what it means for an event to be the ``first'' event of a
% thread/lock $u$ is subtle, and its definition is key in accurately
% capturing the intuitions just outlined. 
The definition of what it means for an event to be ``after'' the 
``first'' event of a thread/lock $u$ is subtle, 
and is key in accurately
capturing the intuitions just outlined. 

For a non-terminal $D$
and  thread $\thr$, define 
\begin{equation}
\af{D}(t) = \after{D}(\first{D}(\tjevents{D}(\thr)))
\equlabel{after-first-thread}
\end{equation}
where
$\tjevents{D}(\thr) = \setpred{e \in D}{e = \ev{\thr}{o} \text{ or } e = \ev{\thr'}{\join{\thr}}}$.

Similarly, for a lock $\lk$, define 
\begin{equation}
\af{D}(\lk) = \after{D}(\first{D}(\acquires{D}(\lk)))
\equlabel{after-first-lock}
\end{equation}
where
$\acquires{D}(\lk)$ is the set $\setpred{e \in D}{e = \ev{\thr}{\acq{\lk}} }$.
As before, we set $\after{D}(\s{undefined}) = \emptyset$.

We now formalize our intuitions in the following lemma. 
\begin{lemma}
\lemlabel{after-cross-chunk}
Let $A$ be a non-terminal with rule $A \rightarrow BC$ and let $e \in
\sem{B}$. Then
\[
\after{A}(e) = \after{B}(e) \cup \smashoperator[r]{\bigcup\limits_{u \in \after{B}(e)}} \af{C}(u)
\]
\end{lemma}
The proof of \lemref{after-cross-chunk} is postponed to\appref{hb-correct}. 
Its statement gives us the following inductive
definition for $\ar{A}(\thr, x)$.
% \begin{align*}
% \ar{A}(x) & = \ar{C}(x) \\ & \cup 
%     \setpred{v \cup \bigcup\limits_{u \in v} \af{C}(u)}{v \in \ar{B}(x)}
% \end{align*}
\begin{equation}
\ar{A}(\thr, x)  = \left\{ 
      \begin{array}{ll}
        \ar{C}(\thr, x) &  \mbox{if } \ar{C}(\thr, x) %\\
                        % &  
                        \neq \emptyset \\
        \ar{B}(\thr, x)  \cup &  \mbox{otherwise} \\
        \bigcup\limits_{u \in \ar{B}(\thr, x)}  \af{C}(u)
      \end{array} \right.
\equlabel{ar-ind}
\end{equation}
Notice that the second expression is $\emptyset$ if $\ar{B}(\thr, x) = \emptyset$.

To complete the formal definition of $\ar{A}(x)$, we need to give an
inductive definition for the sets $\af{}$. Again defining $\af{A}$ for
$A \rightarrow a$ is straightforward from \equref{after-first-thread}
and \equref{after-first-lock} and is
deferred to\appref{hb-algo}. Consider the inductive step, of a
non-terminal $A$ with rule $A \rightarrow BC$ and let $t$ be some
thread. If $\first{A}(\tjevents{A}(\thr)) \in \sem{B}$ then
\lemref{after-cross-chunk} forms the basis of our definition. However,
it is possible that the set $\tjevents{B}(\thr)$ is empty, while $\first{A}(\tjevents{A}(\thr)) \in
\sem{C}$. In this case, $\af{A}(t) = \af{C}(t)$. 
A similar reasoning applies for a lock $\lk$ as well.
Putting all these
observations together, we get
% \[
\begin{equation}
\af{A}(u) = \af{B}(u)  \cup 
    \smashoperator[r]{\bigcup\limits_{u' \in \set{u} \cup \af{B}(u)}} \af{C}(u')
\end{equation}
% \]
% \ucomment{$\s{undefined}$ can be a problem.}

Let us now discuss the inductive definition of the set 
$\br{A}(\thr,x)$ for thread $\thr$ and variable $x$. 
As before, the first event $e \in \sem{A}$ of the kind $\ev{\thr}{\rd{x}}$
can either belong to $\sem{B}$ or to $\sem{C}$.
In the former case, we have $\before{A}(e) =\before{B}(e)$.
On the other hand, if $e \in \sem{C}$, in a manner
similar to the case for $\after{B}(e)$, we need to ``compose''
$\before{C}(e)$ with the ``before'' sets associated with the \emph{last}
events of threads/locks in chunk $B$. 
% To carry this out, we need to define the set of events ``before'' the
% last event of a lock/thread in a chunk. 

For a non-terminal $D$, thread $\thr$ and lock $\lk$,
\begin{equation}
\bl{D}(\thr) = \before{D}(\last{D}(\tfevents{D}(\thr)))
\equlabel{before-last-thread}
\end{equation}
\begin{equation}
\bl{D}(\lk) = \before{D}(\last{D}(\releases{D}(\lk)))
\equlabel{before-last-lock}
\end{equation}
where
$\tfevents{D}(\thr) = \setpred{e \in D}{e = \ev{\thr}{o} \text{ or } e = \ev{\thr'}{\fork{\thr}}}$ and
$\releases{D}(\lk) = \setpred{e \in D}{e = \ev{\thr}{\rel{\lk}} }$.
As before, we assume $\before{D}(\s{undefined}) = \emptyset$.

The dual of \lemref{after-cross-chunk} is the following lemma.
\begin{lemma}
\lemlabel{before-cross-chunk}
Let $A$ be a non-terminal with rule $A \rightarrow BC$ and let $e \in
\sem{C}$. Then
\[
\before{A}(e) = \before{C}(e) \cup 
      \smashoperator[r]{\bigcup\limits_{u \in \before{C}(e)}} \bl{B}(u)
\]
\end{lemma}
Using \lemref{before-cross-chunk}, the inductive definition of
$\br{A}(\thr, x)$ is
\begin{equation}
\br{A}(\thr, x)  = \left\{ 
      \begin{array}{ll}
        \br{B}(\thr, x) & \mbox{if } \br{B}(\thr, x) \neq \emptyset \\
        \br{C}(\thr, x)  \cup &  \mbox{otherwise} \\
        \bigcup\limits_{u \in \br{C}(\thr, x)}  \bl{B}(u)
      \end{array} \right.
\equlabel{br-ind}
\end{equation}
% Again, the second expression is $\emptyset$
% if $\br{C}(\thr, x) = \emptyset$.

To complete the algorithm, we need to give the inductive definition of
$\bl{A}(u)$ for thread/lock $u$. Again the interesting case is the
inductive case of a non-terminal $A$ with rule $A \rightarrow BC$. 
A similar reasoning as in
the case of $\af{}$ sets gives the following definition.
% As
% in the case of $\af{}$-sets, \lemref{before-cross-chunk} needs to be
% adapted to account for the two possibilities: 
% when the last event of a thread/lock $u$ is in $\sem{B}$ or
% $\sem{C}$ to give us the following definition.
\[
\bl{A}(u) = \bl{C}(u) \cup 
    \smashoperator[r]{\bigcup\limits_{u' \in \set{u} \cup \bl{C}(u)}} \bl{B}(u')
\]
This completes the description of the $\s{HB}$-algorithm on compressed
traces. Its correctness is proved in\appref{hb-correct}. 
For a trace $\tr$ compressed as an SLP of size $g$,
this algorithm runs in time $ O(g(T+L)^2(L+TV))$ and uses space
$O(g(T+L)(L+TV))$, where $T$, $L$ and $V$ denote the number of
threads, locks and variables in $\tr$.
% We conclude
% this section by stating its complexity and giving an example
% illustrating the definitions introduced for computing before and after
% sets.
% %
% \begin{theorem}
% \thmlabel{hbalgo-complexity}
% % \ucomment{This bound needs to change. }
% Let $G$ be an SLP of size $g$ representing a trace $\tr$, having $T$
% threads, $L$ locks, and $V$ variables. 
% The inductive algorithm described in this section to 
% detect {\hbrace}s runs in time
% $ O(g(T+L)^2(L+TV))$ and uses space
% $O(g(T+L)(L+TV))$.
% \end{theorem}
%

\begin{example}
\exlabel{hb-before-after} We conclude this section by showing that the
before and after sets given in \exref{hb-cross-race} are computed
correctly using our inductive characterization. 
We will focus on trace
$\tr_1$ and its SLP grammar in \figref{example-trace-grammar}. 
Let us consider the computation of
$\aw{C}(x)$. Observe that the last $\wt{x}$-event in $C$ is
$e_1$. Further,
\[
\begin{array}{lll}
\aw{E}(x) = \set{1,2} & \af{F}(1) = \emptyset & \af{F}(2) = \set{2,\lk}
\end{array}
\]
Here $\af{F}(1) = \emptyset$ because there is no event of thread 1 in
$F$. Using the inductive definition similar to \equref{ar-ind}, we get
$\aw{C}(x) = \set{1,2,\lk}$ which is correct. 

Next, consider the computation $\bw{B}(y)$. Notice that the first
$\wt{y}$-event in $B$ is $e_{13}$,
which is in the chunk $F$.
This immediately gives $\bw{B}(y) = \bw{F}(y) = \set{2, \lk}$
using  a characterization similar to \equref{br-ind}.
\end{example}

%% file: scenarios_hb.tex
%!TEX root = main.tex

\begin{figure}[t]
\begin{tikzpicture}[decoration=snake]

% \begin{scope}[scale=1.2, every node/.append style={transform shape}]

%%% Thread order
\begin{scope}
\draw[densely dotted] (0,2.25) -- (0,1.25);
\draw (0,1.25) -- (0,0.25);
\draw (0,0.25) -- (1,0.25);
\draw (1,0.25) -- (1,1.25);
\draw[densely dotted] (1,1.25) -- (1,2.25);

\node (e) at (0.2, 2) {\tiny $\bullet$};
\node (ename) at (0.4, 2) {\tiny $e$};
\draw [-{stealth[scale=3.0]},decorate,decoration={snake,amplitude=.4mm,segment length=2mm,post length=1mm}, blue] (0.2,2) -- (0.5,0.55) ;
% \path [->, draw=blue,snake it] (0.1,2) -- (0.5,0.75) ;
\node (ej) at (0.5, 0.5) {\tiny $\bullet$};
\node (ejname) at (0.7, 0.5) {\tiny $e_j$};

\node (t) at (0.6, 1.5) {\tiny $t$};
\draw[dashed] (0.5, 1.5) -- (0.5, -1.5);
\begin{scope}[ thick, decoration={
    markings,
    mark=at position 0.5 with {\arrow{stealth}}}
    ] 
    \draw[postaction={decorate}] (0.5, 0.5) -- (0.5, -0.5);
\end{scope}

\node (ejplusone) at (0.5, -0.5) {\tiny $\bullet$};
\node (ejplusonename) at (0.25, -0.6) {\tiny $e_{j+1}$};
% \path [draw=blue,snake it] (0.5,-0.75) -- (0.9,-2) ;
\draw [-{stealth[scale=3.0]},decorate,decoration={snake,amplitude=.4mm,segment length=2mm,post length=1mm}, blue] (0.5,-0.5) -- (0.8,-1.95) ;
\node (eprime) at (0.8, -2) {\tiny $\bullet$};
\node (eprimename) at (0.6, -2) {\tiny $e'$};

\draw[densely dotted] (0,-2.25) -- (0,-1.25);
\draw (0,-1.25) -- (0,-0.25);
\draw (0,-0.25) -- (1,-0.25);
\draw (1,-0.25) -- (1,-1.25);
\draw[densely dotted] (1,-1.25) -- (1,-2.25);

\node (one) at (0.5, -2.6) {(1)};
\end{scope}

%%%% Release -> acquire
\begin{scope}[shift={(1.75,0)}]
\draw[densely dotted] (0,2.25) -- (0,1.25);
\draw (0,1.25) -- (0,0.25);
\draw (0,0.25) -- (1.5,0.25);
\draw (1.5,0.25) -- (1.5,1.25);
\draw[densely dotted] (1.5,1.25) -- (1.5,2.25);

\node (e) at (0.2, 2) {\tiny $\bullet$};
\node (ename) at (0.4, 2) {\tiny $e$};
\draw [-{stealth[scale=3.0]},decorate,decoration={snake,amplitude=.4mm,segment length=2mm,post length=1mm}, blue] (0.2,2) -- (0.5,0.55) ;
% \path [->, draw=blue,snake it] (0.1,2) -- (0.5,0.75) ;
\node (ej) at (0.5, 0.5) { $\boldsmile$};
\node (lkname) at (0.7, 0.5) {\tiny $\lk$};
\node (ejname) at (0.27, 0.45) {\tiny $e_j$};

\draw[dashed] (0.5, 1.5) -- (0.5, -1.5);
\node (t) at (0.6, 1.5) {\tiny $t$};
\begin{scope}[ thick, decoration={
    markings,
    mark=at position 0.5 with {\arrow{stealth}}}
    ] 
    \draw[postaction={decorate}] (0.5, 0.5) -- (1, -0.475);
\end{scope}

\draw[dashed] (1, 1.5) -- (1, -1.5);
\node (tprime) at (1.1, 1.5) {\tiny $t'$};

\node (ejplusone) at (1, -0.55) { $\boldfrown$};
\node (ejplusonename) at (1.3, -0.66) {\tiny $e_{j+1}$};
\node (ejplusonename) at (0.8, -0.5) {\tiny $\lk$};
% \path [draw=blue,snake it] (0.5,-0.75) -- (0.9,-2) ;
\draw [-{stealth[scale=3.0]},decorate,decoration={snake,amplitude=.4mm,segment length=2mm,post length=1mm}, blue] (1,-0.55) -- (1.3,-1.95) ;
\node (eprime) at (1.3, -2) {\tiny $\bullet$};
\node (eprimename) at (1.1, -2) {\tiny $e'$};

\draw[densely dotted] (0,-2.25) -- (0,-1.25);
\draw (0,-1.25) -- (0,-0.25);
\draw (0,-0.25) -- (1.5,-0.25);
\draw (1.5,-0.25) -- (1.5,-1.25);
\draw[densely dotted] (1.5,-1.25) -- (1.5,-2.25);

% \node (cap21) at (0.75, -2.75) {\tiny $e_j = \ev{t}{\rel{\lk}}$};
% \node (cap22) at (0.65, -3.0) {\tiny $e_{j+1} = \ev{t'}{\acq{\lk}}$};
\node (two) at (0.75, -2.6) {(2)};
\end{scope}

%%%% Fork - start
\begin{scope}[shift={(4,0)}]
\draw[densely dotted] (0,2.25) -- (0,1.25);
\draw (0,1.25) -- (0,0.25);
\draw (0,0.25) -- (1.5,0.25);
\draw (1.5,0.25) -- (1.5,1.25);
\draw[densely dotted] (1.5,1.25) -- (1.5,2.25);

\node (e) at (0.2, 2) {\tiny $\bullet$};
\node (ename) at (0.4, 2) {\tiny $e$};
\draw [-{stealth[scale=3.0]},decorate,decoration={snake,amplitude=.4mm,segment length=2mm,post length=1mm}, blue] (0.2,2) -- (0.5,0.55) ;
% \path [->, draw=blue,snake it] (0.1,2) -- (0.5,0.75) ;
\node (ej) at (0.5, 0.5) {\tiny $\bullet$};
\node (ejname) at (0.35, 0.5) { \tiny $e_j$};
\node (forkname) at (1.0, 0.5) { \tiny $\fork{t'}$};

\draw[dashed] (0.5, 1.5) -- (0.5, -1.5);
\node (t) at (0.4, -1.5) {\tiny $t$};
\begin{scope}[ thick, decoration={
    markings,
    mark=at position 0.5 with {\arrow{stealth}}}
    ] 
    \draw[postaction={decorate}] (0.5, 0.5) -- (1, -0.5);
\end{scope}

\draw[dashed] (1, -0.25) -- (1, -1.5);
\node (tprime) at (0.9, -1.5) {\tiny $t'$};

\node (ejplusone) at (1, -0.5) {\tiny $\bullet$};
\node (ejplusonename) at (0.75, -0.6) {\tiny $e_{j+1}$};
% \path [draw=blue,snake it] (0.5,-0.75) -- (0.9,-2) ;
\draw [-{stealth[scale=3.0]},decorate,decoration={snake,amplitude=.4mm,segment length=2mm,post length=1mm}, blue] (1,-0.5) -- (1.3,-1.95) ;
\node (eprime) at (1.3, -2) {\tiny $\bullet$};
\node (eprimename) at (1.1, -2) {\tiny $e'$};

\draw[densely dotted] (0,-2.25) -- (0,-1.25);
\draw (0,-1.25) -- (0,-0.25);
\draw (0,-0.25) -- (1.5,-0.25);
\draw (1.5,-0.25) -- (1.5,-1.25);
\draw[densely dotted] (1.5,-1.25) -- (1.5,-2.25);

% \node (cap31) at (0.75, -2.875) {\tiny $e_j = \ev{t}{\fork{t'}}$};
\node (three) at (0.75, -2.6) {(3)};
\end{scope}

%%%%% End -> Join
\begin{scope}[shift={(6.25,0)}]
\draw[densely dotted] (0,2.25) -- (0,1.25);
\draw (0,1.25) -- (0,0.25);
\draw (0,0.25) -- (1.5,0.25);
\draw (1.5,0.25) -- (1.5,1.25);
\draw[densely dotted] (1.5,1.25) -- (1.5,2.25);

\node (e) at (1.3, 2) {\tiny $\bullet$};
\node (ename) at (1.1, 2) {\tiny $e$};
\draw [-{stealth[scale=3.0]},decorate,decoration={snake,amplitude=.4mm,segment length=2mm,post length=1mm}, blue] (1.3,2) -- (1,0.55) ;
% \path [->, draw=blue,snake it] (0.1,2) -- (0.5,0.75) ;
\node (ej) at (1, 0.5) {\tiny $\bullet$};
% \node (ejname) at (0.5, 0.5) {\tiny \begin{tabular}{c} $e_j{=}$ \\$\langle{{u{:}o}}\rangle$ \end{tabular} };
% \node (ejname) at (0.5, 0.5) {\tiny $e_j{=} \langle{{u{:}o}}\rangle$ };
\node (ejname) at (1.2, 0.5) {\tiny $e_j$ };

\draw[dashed] (1, 1.5) -- (1, 0.25);
\node (tprime) at (0.9, 1.5) {\tiny $t'$};
\begin{scope}[ thick, decoration={
    markings,
    mark=at position 0.5 with {\arrow{stealth}}}
    ] 
    \draw[postaction={decorate}] (1, 0.5) -- (0.5, -0.5);
\end{scope}

\draw[dashed] (0.5, 1.5) -- (0.5, -1.5);
\node (t) at (0.4, 1.5) {\tiny $t$};

\node (ejplusone) at (0.5, -0.5) {\tiny $\bullet$};
% \node (ejplusonename) at (1.2, -0.6) {\tiny \begin{tabular}{cc} & $e_{j+1} = $ \quad \quad\\ & $\langle{{t{:}\join{t'}}}\rangle$ \end{tabular} };
\node (ejplusonename) at (0.25, -0.55) {\tiny $e_{j+1}$  };
\node (joinname) at (1.0, -0.6) {\tiny $\join{t'}$  };
% \path [draw=blue,snake it] (0.5,-0.75) -- (0.9,-2) ;
\draw [-{stealth[scale=3.0]},decorate,decoration={snake,amplitude=.4mm,segment length=2mm,post length=1mm}, blue] (0.5,-0.5) -- (0.2,-1.95) ;
\node (eprime) at (0.2, -2) {\tiny $\bullet$};
\node (eprimename) at (0.4, -2) {\tiny $e'$};

\draw[densely dotted] (0,-2.25) -- (0,-1.25);
\draw (0,-1.25) -- (0,-0.25);
\draw (0,-0.25) -- (1.5,-0.25);
\draw (1.5,-0.25) -- (1.5,-1.25);
\draw[densely dotted] (1.5,-1.25) -- (1.5,-2.25);

% \node (cap41) at (0.75, -2.875) {\tiny $e_{j+1} = \ev{t}{\join{t'}}$};
\node (four) at (0.75, -2.6) {(4)};
\end{scope}

% \end{scope}

% \draw [decorate,decoration={brace,amplitude=5pt},xshift=4pt,yshift=0pt]
% (7.75,2.25) -- (7.75,0.25) node [black,midway,xshift=+0.3cm] {$B$};

% \draw [decorate,decoration={brace,amplitude=5pt},xshift=4pt,yshift=0pt]
% (7.75,-0.25) -- (7.75,-2.25) node [black,midway,xshift=+0.3cm] {$C$};

% \draw [decorate,decoration={brace,amplitude=10pt},xshift=+5pt,yshift=0pt]
% (8,2.5) -- (8,-2.5) node [black,midway,xshift=+0.6cm] 
% {$A$};

\node at (-0.25, 2.2) {-};
\draw [-{stealth[scale=3.0]}] (-0.25, 1.25) -- (-0.25, 2.25);
\draw [-{stealth[scale=3.0]}] (-0.25, 1.25) -- (-0.25, 0.25);
\node[fill=white, inner sep=0.05cm] at (-0.25, 1.25) {\tiny $B$};
\node at (-0.25, 0.2) {-};

\node at (-0.25, -0.3) {-};
\draw [-{stealth[scale=3.0]}] (-0.25, -1.25) -- (-0.25, -2.25);
\draw [-{stealth[scale=3.0]}] (-0.25, -1.25) -- (-0.25, -0.25);
\node[fill=white, inner sep=0.05cm] at (-0.25, -1.25) {\tiny $C$};
\node at (-0.25, -2.3) {-};

\node at (8.0, 2.2) {-};
\draw [-{stealth[scale=3.0]}] (8.0, 0) -- (8.0, -2.25);
\draw [-{stealth[scale=3.0]}] (8.0, 0) -- (8.0, 2.25);
\node[fill=white, inner sep=0.05cm] at (8.0, 0) {\tiny $A$};
\node at (8.0, -2.3) {-};

\end{tikzpicture}
\bigskip
\caption{\small Illustrating the various scenarios that establish $e_j\HBS[BC]
  e_{j+1}$. In (2), `$\boldfrown \lk$' represents $\acq{\lk}$, `$\boldsmile \lk$' is $\rel{\lk}$.}  
\figlabel{hb-ind-step}
\end{figure}

%% file: lockset.tex
%!TEX root = main.tex

\section{LockSet Algorithm for Compressed Traces}
\seclabel{lockset}

Similar to our algorithm for detecting {\hbrace}s on compressed
traces, we will formulate an algorithm for detecting violations of
the lockset discipline on SLPs in an inductive fashion.  The challenge
here again is similar --- violations occurring inside a chunk $\sem{B}$ 
are also violations of any other chunk that contains $\sem{B}$, 
and detecting ``cross'' violations is, therefore, the key challenge.  
In this section, we will outline these ideas in detail.

\subsection{Cross violations}

Recall that, for a thread $\thr$ and variable $x$,
$\lockset{\tr}(\thr, x)$ is the set of all the locks (including the
dummy locks in $\cD$) that \emph{protect} every access event of $x$
performed by $\thr$, in $\tr$.

In this section, we show how to compute $\lockset{A}(\thr,x)$ for
every non-terminal $A$ and for every pair $(\thr,x)$ of thread and
variable, by inducting on the non-terminals in decreasing order of
their rank.  Checking if 
%\[\smashoperator[r]{\bigcap_{\thr \in
%    \threads(A)}} \lockset{A}(\thr,x) = \emptyset\] 
$\cap_{\thr \in \threads{A}} \lockset{A}(\thr,x) = \emptyset$
then follows
easily.

\input{scenarios_lockset}

The base case for non-terminals with rule $A \rightarrow a$ 
is straightforward, and is presented in\appref{app-lockset-algo}.
Now consider the inductive step for non-terminals having rules of the
form $A\rightarrow BC$. To understand what $\lockset{A}(\thr,x)$ will
be, it is useful to examine what $\locksheld{A}(e)$ for an event $e$
looks like. Consider a data access event $e \in \sem{B}$ performed by
thread $\thr$. Clearly, $\locksheld{B}(e) \subseteq
\locksheld{A}(e)$. But are they equal? The answer turns out to be no.
Suppose a lock $\lk$ which is released in $\sem{C}$ by thread $\thr$
but does not have a \emph{matching} acquire in $\sem{A}$ (and hence,
neither in $\sem{B}$). Such a lock $\lk$ will protect all the events
performed before it in $\sem{A}$.  Thus trivially, it will enclose all
the events performed by $\thr$ in chunk $\sem{B}$.  As a consequence,
$\lk$ must be included in the set $\locksheld{A}(e)$ for every event
$e \in \proj{B}{\thr}$. Lock $\lk_2$ in \figref{lockset-openlocks}(1)
illustrates this. Similarly, for an event $e \in \sem{C}$ performed by
thread $\thr$, the set $\locksheld{A}(e)$ must additionally include
locks which have been acquired by thread $\thr$ in $\sem{B}$ but have
not been matched in $\sem{A}$ (see lock $\lk_2$ in
\figref{lockset-openlocks}(2)). However, one must be careful. A lock
$\lk$ which was released by $\thr$ in $C$ (at event $e_{\rel{\lk}}$)
and whose matching acquire is in $B$ (event $e_{\acq{\lk}}$), does not
affect the locks held by any event in $B$ --- for those events $e\in
\proj{B}{t}$ which were after $e_{\acq{\lk}}$, $\lk$ was already in
$\locksheld{B}(e)$, while for the events $e$ before $e_{\acq{\lk}}$,
$\lk$ does not anyway protect $e$, and thus $\lk \not\in
\locksheld{B}(e)$. This is illustrated through lock $\lk_1$ in
\figref{lockset-openlocks} [(1) and (2)].

In the presence of re-entrant locks, we need to account for another
fact. Since locks can be acquired and released multiple times, a lock
that is released more times in $C$ (by thread $\thr$) than it is
acquired in $B$ (by thread $\thr$) will protect all events of $\thr$
in $B$, because the outermost release is still unmatched in $A$. The
same holds for locks that have been acquired more times than they are
released in $C$. Both these scenarios are shown in
\figref{lockset-openlocks} [(3) and (4)].

To formalize the above notions, we will now introduce some notation.
For a non-terminal $D$, let us first define the
number of \emph{unmatched} acquire events of lock $\lk$ in thread $\thr$ as
\begin{equation}
% \begin{array}[ll]
\oacq{D}(\thr,\lk) =  | \setpred{e = \ev{\thr}{\acq{\lk}} \in D}{\match(e) \not\in \sem{D}} |
\equlabel{open-acq}
\end{equation}
and the number of release events as
\begin{equation}
\orel{D}(\thr,\lk) =  | \setpred{e = \ev{\thr}{\rel{\lk}} \in D}{\match(e) \not\in \sem{D}} |
\equlabel{open-rel}
\end{equation}
% Informally, $\oacq{D}(\thr,\lk)$ (resp. $\orel{D}(\thr,\lk)$) denotes the number
% of release (acquire) events of lock $\lk$ by thread $\thr$
% that need to be appended (prepended) to $D$
% in order to complete the last (first) critical section of
% $\lk$ in $\proj{D}{t}$ \todo{What is `completing' a C.S.?}
% \footnote{Note that for a non-terminal $D$,
% only the first and the last critical section in $D$ 
% can be potentially incomplete; the rest are completely matched.}.

Our intuitions, as discussed above, can then be captured for the more
complex case of re-entrant locks as follows.
\begin{lemma}
\lemlabel{locksheld}
Let $A$ be a non-terminal with rule $A \rightarrow BC$.
Let $e \in \proj{B}{\thr}$ and $e' \in \proj{C}{\thr'}$
be read/write events performed by threads $\thr,\thr'$.
Then,
\vspace{-0.05in}
\begin{equation*}
\begin{array}{ll}
\locksheld{A}(e)  = & \locksheld{B}(e) \\
					& \cup \setpred{\lk}{\orel{C}(\thr,\lk) > \oacq{B}(\thr,\lk)} 
\end{array}
\end{equation*}
% \vspace{-0.05in}
% and,
\vspace{-0.05in}
\begin{equation*}
\begin{array}{ll}
\locksheld{A}(e') = & \locksheld{C}(e') \\
					& \cup \setpred{\lk}{\oacq{B}(\thr',\lk) > \orel{C}(\thr',\lk)} 
\end{array}
\end{equation*}
\end{lemma}
% The proof of Lemma~\ref{lem:locksheld} can be found in the appendix.

Building on Lemma~\ref{lem:locksheld}, we can now
state the inductive definition of $\lockset{}$
in terms of $\oacq{}$ and $\orel{}$.
\begin{equation}
\begin{array}{ll}
	\lockset{A}(\thr, x) \\ 
	= \big( \lockset{B}(\thr, x) \cup \setpred{\lk}{\orel{C}(\thr,\lk) > \oacq{B}(\thr,\lk)}\big) \\
	\cap \big( \lockset{C}(\thr, x) \cup \setpred{\lk}{\oacq{B}(\thr',\lk) > \orel{C}(\thr',\lk)}\big)
\end{array} 
\equlabel{lockset-ind}
\end{equation}
The base case for computing $\lockset{A}(\thr, x)$ is 
trivial and is presented in\appref{app-lockset-algo} for completeness.

\begin{example}
\exlabel{lockset-cross} 
Consider the SLP for $\tr_2$ from \figref{eraser-trace-grammar}.
%
%
% For the first example,
% $\oacq{U}(\thr, \lk) = \orel{U}(\thr, \lk) = 0$,
% for all non-terminals $U \in \{S,A,B,C,D,$ $E,F,G\}$,
% and for both threads $\thr \in \set{1,2}$.
% This is because the strings generated by each of them
% have complete critical sections over $\lk$.
$\oacq{}$ and $\orel{}$ for various non-terminals is given
below.
\[
\begin{array}{ll}
\oacq{W}(1, \lk) = 1 & \orel{X}(1, \lk) = 1\\
\oacq{X}(2, \lk) = 1 & \orel{Z}(2, \lk) = 1\\
\oacq{U}(2, \lk) = 1 & \orel{V}(2, \lk) = 1.
\end{array}
\]
The values for all other combinations are $0$.  Note how the unmatched
acquire in $W$ and the unmatched release in $X$ on thread $1$ get
matched in the concatenated chunk $U$, giving $\oacq{U}(1, \lk) =
\orel{U}(1, \lk) = 0$.  This is essentially the insight we will
explore in \secref{oacq_orel_ind}
to inductively define $\oacq{}$ and $\orel{}$.
% next, .

Let us now see how $\lockset{}$ computation takes place.  First,
$\lockset{W}(1, x) = \set{\dmlock,\dmlockthr{1}}$ since the only event
of $x$ in $W$ is a read by thread $1$.  Also, $\lockset{X}(1, y) =
\set{\lk,\dmlockthr{1}}$ as $e_3$ is protected by the (unmatched)
release $e_4$ in $X$.  In chunk $Y$, $\lockset{Y}(2, x) =
\set{\dmlock,\dmlockthr{2}}$ and $\lockset{Y}(2, y) =
\set{\dmlockthr{2}}$; interestingly, the locksets for $Y$ does not
reveal that both $e_6$ and $e_7$ are enclosed within the critical
section of lock $\lk$.  On the other hand, using the inductive
formulation discussed above, we can infer that $\lockset{V}(2, y) =
(\lockset{Y}(2, y) \cup \set{\lk}) \cap \top$ which evaluates to
$\set{\dmlockthr{2},\lk}$ as expected.  The universal set $\top$ is
described in \defref{lockset_def} (See \appref{app-lockset-algo}).
Again, the lock $\lk$ does not appear in $\lockset{U}(1,x)$ eventhough it
is unmatched in $X$, because it gets matched with $e_2$ in $W$.  This
also follows from the inductive definition of $\lockset{U}(1, x)$.
% because $\orel{X}(1, \lk) \not> \oacq{W}(1, \lk)$.
\end{example}
\vspace{-0.1in}
Given \equref{lockset-ind}, our
inductive formulation will be complete once we
can inductively compute the functions $\oacq{}$ and $\orel{}$.
We describe this next.

\subsection{Computing \texorpdfstring{$\oacq{}$}{\textsf{OpenAcq}} and \texorpdfstring{$\orel{}$}{\textsf{OpenRel}}}
\seclabel{oacq_orel_ind}

% The aim of this section is to give inductive definitions of the
% functions $\oacq{}$ and $\orel{}$.  
The base case for non-terminals
having rules of the form $A \rightarrow a$ is straightforward and can
be found in\appref{app-lockset-algo} for completeness.  
In the inductive case we have a non-terminal $A$ with production rule
of the form $A\rightarrow BC$.
For this case, let us first attempt to characterize
the acquire events in $\sem{A}$ that have not been matched.
Notice that if a lock  is acquired (without a matching release) 
in the chunk $\sem{C}$, it would remain unmatched in 
the bigger chunk $\sem{A}$.
In addition, the unmatched acquire events
acquired in $\sem{B}$ whose matching release
is not present in $\sem{C}$ will also contribute to the
unmatched acquire events in $\sem{A}$.
This reasoning is formalized below.
% \begin{lemma}
% \lemlabel{openacq-ind}
% Let $A$ be a non-terminal with rule $A \rightarrow BC$. Then,
\begin{equation}
\begin{array}{rcl}
\oacq{A}(\thr, \lk) \!\!\!\!\!\!&=& \!\!\!\!  \oacq{C}(\thr, \lk) \\ 
& + & \!\!\!\! \max\{0, \oacq{B}(\thr, \lk) - \orel{C}(\thr, \lk)\}
\end{array}
\equlabel{openacq-ind}
\end{equation}
% \end{lemma}
Notice the use of the $\max{}$ operator in \equref{openacq-ind}.
If the quantity $\oacq{B}(\thr, \lk) - \orel{C}(\thr, \lk)$ is negative,
then there are more unmatched $\rel{\lk}$-events in $\proj{C}{\thr}$,
which should be accounted for in $\orel{A}(t, \lk)$, 
instead of affecting the contribution of $\sem{C}$
towards the unmatched acquire events of $\proj{A}{\thr}$ .

Similar reasoning gives the inductive formulation for $\orel{}$
% \vspace{-0.05cm}
% \begin{lemma}
% \lemlabel{openrel-ind}
% Let $A$ be a non-terminal with rule $A \rightarrow BC$. Then,
\begin{equation}
\begin{array}{rcl}
\orel{A}(\thr, \lk) \!\!\!\!\!\!&=& \!\!\!\!  \orel{B}(\thr, \lk) \\ 
& + & \!\!\!\! \max\{0, \orel{C}(\thr, \lk) - \oacq{B}(\thr,\lk)\}
\end{array}
\equlabel{openrel-ind}
\end{equation}
% \end{lemma}
This completes the description of our algorithm for computing locksets
and checking violations of lockset discipline
for compressed traces.
For a trace $\tr$ compressed as an SLP of size $g$,
this algorithm runs in time $O(gTL(\log r + V))$ and uses space $O(gTL(\log r + V))$,
where $T$, $L$ and $V$ are the number of threads, 
locks and variables respectively in $\tr$,
and $r$ denotes the maximum number of
times a thread acquires a lock without releasing it in $\tr$.

% Let us now analyze the time and space complexity
% of the algorithm just described.

% \begin{theorem}
% \thmlabel{lockset-complexity}
% Let $G$ be an SLP of size $g$ representing a string $\tr$, having $T$
% threads, $L$ locks, and $V$ variables. 
% Let $r$ be the maximum number of
% times a thread acquires a lock without releasing it, in the trace generated by $G$.
% The inductive algorithm described in this section to detect violations of lockset discipline
% runs in time $O(gTL(\log r + V))$ and uses space $O(gTL(\log r + V))$.
% \end{theorem}

%% file: scenarios_lockset.tex
%!TEX root = main.tex

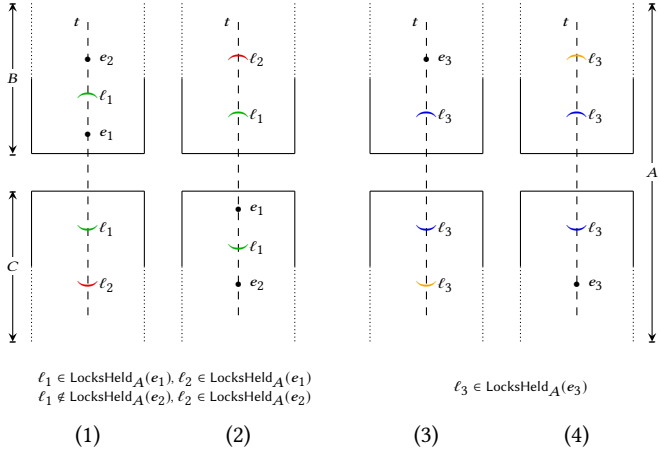
\begin{figure}[t]
\begin{tikzpicture}[decoration=snake]

% \begin{scope}[scale=1.2, every node/.append style={transform shape}]

%%%% Release in C
\begin{scope}[shift={(-0.25, 0)}]
\draw[densely dotted] (0,2.25) -- (0,1.25);
\draw (0,1.25) -- (0,0.25);
\draw (0,0.25) -- (1.5,0.25);
\draw (1.5,0.25) -- (1.5,1.25);
\draw[densely dotted] (1.5,1.25) -- (1.5,2.25);

\draw[dashed] (0.75, 2) -- (0.75, -2);
\node (t) at (0.6, 2) {\tiny $t$};

\node (e1) at (0.75, 0.5) {\tiny $\bullet$};
\node (e1name) at (1.0, 0.5) {\tiny $e_1$};

\node (acq1) at (0.75, 1.0) { \green{$\boldfrown$}};
\node (acq1name) at (1.0, 1.0) {\tiny $\lk_1$};

\node (e2) at (0.75, 1.5) {\tiny $\bullet$};
\node (e2name) at (1.0, 1.5) {\tiny $e_2$};

\node (rel1) at (0.75, -0.75) {  \green{$\boldsmile$}};
\node (rel1name) at (1, -0.75) {\tiny $\lk_1$};

\node (rel2) at (0.75, -1.5) { \red{$\boldsmile$}};
\node (rel2name) at (1, -1.5) {\tiny $\lk_2$};

\draw[densely dotted] (0,-2.25) -- (0,-1.25);
\draw (0,-1.25) -- (0,-0.25);
\draw (0,-0.25) -- (1.5,-0.25);
\draw (1.5,-0.25) -- (1.5,-1.25);
\draw[densely dotted] (1.5,-1.25) -- (1.5,-2.25);

\node (one) at (0.75, -3.5) {(1)};
\end{scope}

\begin{scope}[shift={(1.75,0)}]
\draw[densely dotted] (0,2.25) -- (0,1.25);
\draw (0,1.25) -- (0,0.25);
\draw (0,0.25) -- (1.5,0.25);
\draw (1.5,0.25) -- (1.5,1.25);
\draw[densely dotted] (1.5,1.25) -- (1.5,2.25);

\draw[dashed] (0.75, 2) -- (0.75, -2);
\node (t) at (0.6, 2) {\tiny $t$};

\node (acq1) at (0.75, 0.75) {  \green{$\boldfrown$}};
\node (acq1name) at (1, 0.75) {\tiny $\lk_1$};

\node (acq2) at (0.75, 1.5) { \red{$\boldfrown$}};
\node (acq2name) at (1, 1.5) {\tiny $\lk_2$};

\node (e1) at (0.75, -0.5) {\tiny $\bullet$};
\node (e1name) at (1.0, -0.5) {\tiny $e_1$};

\node (rel1) at (0.75, -1.0) { \green{$\boldsmile$}};
\node (rel1name) at (1.0, -1.0) {\tiny $\lk_1$};

\node (e2) at (0.75, -1.5) {\tiny $\bullet$};
\node (e2name) at (1.0, -1.5) {\tiny $e_2$};

\draw[densely dotted] (0,-2.25) -- (0,-1.25);
\draw (0,-1.25) -- (0,-0.25);
\draw (0,-0.25) -- (1.5,-0.25);
\draw (1.5,-0.25) -- (1.5,-1.25);
\draw[densely dotted] (1.5,-1.25) -- (1.5,-2.25);

\node (two) at (0.75, -3.5) {(2)};
\end{scope}

\node (one_and_two) at (1.65, -2.75) {\tiny $\lk_1 \in \locksheld{A}(e_1)$, $\lk_2 \in \locksheld{A}(e_1)$};
\node (one_and_two) at (1.65, -3.0) {\tiny $\lk_1 \not\in \locksheld{A}(e_2)$, $\lk_2 \in \locksheld{A}(e_2)$};

\begin{scope}[shift={(4.25,0)}]
\draw[densely dotted] (0,2.25) -- (0,1.25);
\draw (0,1.25) -- (0,0.25);
\draw (0,0.25) -- (1.5,0.25);
\draw (1.5,0.25) -- (1.5,1.25);
\draw[densely dotted] (1.5,1.25) -- (1.5,2.25);

\draw[dashed] (0.75, 2) -- (0.75, -2);
\node (t) at (0.6, 2) {\tiny $t$};

\node (e3) at (0.75, 1.5) {\tiny  $\bullet$};
\node (e3name) at (1, 1.5) {\tiny $e_3$};

\node (acq3) at (0.75, 0.75) { \blue{$\boldfrown$}};
\node (acq3name) at (1, 0.75) {\tiny $\lk_3$};

\node (rel3) at (0.75, -0.75) { \blue{$\boldsmile$}};
\node (rel3name) at (1, -0.75) {\tiny $\lk_3$};

\node (rel3p) at (0.75, -1.5) { \yellow{$\boldsmile$}};
\node (rel3pname) at (1, -1.5) {\tiny $\lk_3$};

\draw[densely dotted] (0,-2.25) -- (0,-1.25);
\draw (0,-1.25) -- (0,-0.25);
\draw (0,-0.25) -- (1.5,-0.25);
\draw (1.5,-0.25) -- (1.5,-1.25);
\draw[densely dotted] (1.5,-1.25) -- (1.5,-2.25);

\node (three) at (0.75, -3.5) {(3)};
\end{scope}

\begin{scope}[shift={(6.25,0)}]
\draw[densely dotted] (0,2.25) -- (0,1.25);
\draw (0,1.25) -- (0,0.25);
\draw (0,0.25) -- (1.5,0.25);
\draw (1.5,0.25) -- (1.5,1.25);
\draw[densely dotted] (1.5,1.25) -- (1.5,2.25);

\draw[dashed] (0.75, 2) -- (0.75, -2);
\node (t) at (0.6, 2) {\tiny $t$};

\node (acq3p) at (0.75, 1.5) { \yellow{$\boldfrown$}};
\node (acq3pname) at (1, 1.5) {\tiny $\lk_3$};

\node (acq3) at (0.75, 0.75) { \blue{$\boldfrown$}};
\node (acq3name) at (1, 0.75) {\tiny $\lk_3$};

\node (rel3) at (0.75, -0.75) { \blue{$\boldsmile$}};
\node (rel3name) at (1, -0.75) {\tiny $\lk_3$};

\node (e3) at (0.75, -1.5) {\tiny  $\bullet$};
\node (e3name) at (1, -1.5) {\tiny $e_3$};

\draw[densely dotted] (0,-2.25) -- (0,-1.25);
\draw (0,-1.25) -- (0,-0.25);
\draw (0,-0.25) -- (1.5,-0.25);
\draw (1.5,-0.25) -- (1.5,-1.25);
\draw[densely dotted] (1.5,-1.25) -- (1.5,-2.25);

\node (four) at (0.75, -3.5) {(4)};
\end{scope}

\node (one_and_two) at (6.25, -2.875) {\tiny $\lk_3 \in \locksheld{A}(e_3)$};

\begin{scope}[shift={(-0.5,0)}]
\node at (0, 2.2) {-};
\draw [-{stealth[scale=3.0]}] (0, 1.25) -- (0, 2.25);
\draw [-{stealth[scale=3.0]}] (0, 1.25) -- (0, 0.25);
\node[fill=white, inner sep=0.05cm] at (0, 1.25) {\tiny $B$};
\node at (0, 0.2) {-};
\end{scope}

\begin{scope}[shift={(-0.5,-2.5)}]
\node at (0, 2.2) {-};
\draw [-{stealth[scale=3.0]}] (0, 1.25) -- (0, 2.25);
\draw [-{stealth[scale=3.0]}] (0, 1.25) -- (0, 0.25);
\node[fill=white, inner sep=0.05cm] at (0, 1.25) {\tiny $C$};
\node at (0, 0.2) {-};
\end{scope}

\begin{scope}[shift={(8,0)}]
\node at (0, 2.2) {-};
\draw [-{stealth[scale=3.0]}] (0, 0) -- (0, -2.25);
\draw [-{stealth[scale=3.0]}] (0, 0) -- (0, 2.25);
\node[fill=white, inner sep=0.05cm] at (0, 0) {\tiny $A$};
\node at (0, -2.3) {-};
\end{scope}

\end{tikzpicture}
\bigskip
\caption{
Unmatched acquire and release events protect all 
the events of the same thread in the neighboring chunk 
when not matched in the entire chunk ((1) and (2)).
Re-entrant locks protect the neighboring chunk when the outermost
unmatched acquire/release is unmatched ((3) and (4)).
`$\boldfrown \lk_i$' represents $\acq{\lk_i}$, `$\boldsmile \lk_i$' is $\rel{\lk_i}$.
}  
\figlabel{lockset-openlocks}
\end{figure}

%% file: experiments.tex
%!TEX root = main.tex

\section{Evaluation}
\seclabel{experiments}

% \input{results_table}
\input{bigtable}

In order to gauge the effect of compression on the size of
traces, and the subsequent effect on time taken
to analyze these compressed traces for races,
we conducted experiments on
a large variety of benchmarks and evaluated our algorithms empirically.
In this section, we describe the details
of our implementation and experimental setup,
and analyze the results of these experiments.

\subsection{Implementation and Setup}

\begin{myparagraph}{Implementation}
Our algorithms for detecting races on compressed
traces, discussed in \secref{hb} and \secref{lockset}
have been implemented in our tool~\tool, which is
publicly available at~\cite{ziptrack}.
\tool~is written primarily in Java and analyzes traces generated by Java programs.
% Our techniques and algorithms are language independent 
% and can be implemented to analyze executions of programs written in
% any language that uses threads, locks and shared memory
% for concurrent computation, such as C, C++ and Java.
% In our prototype implementation, we analyze traces generated by Java programs.
% \tool~uses trace files generated by~\rvpredict~(version 1.8.2)~\cite{rvpredict}
% and ignores all events apart from those corresponding to 
% read/write to memory locations, acquire/release of locks,
% and join/fork of threads.
\tool~firsts collects trace logs as sequence of events, which include
read/write to memory locations, acquire/release of locks,
and join/fork of threads.
For this, we use the logging library provided by the
commercial tool \textsf{RVPredict}~\cite{rvpredict}.
After having generated the trace logs, 
% we filter out events corresponding to
% read/write to local memory accesses.
\tool~calls the Sequitur algorithm (available at~\cite{sequitur}) to compress 
these traces as straight line programs (see \secref{prelim}). 
\tool~then analyzes these SLPs to detect the presence of HB 
races and lockset discipline violations.
\end{myparagraph}
% \subsubsection{Optimizations}

% Apart from the algorithms mentioned in \secref{hb} and \secref{lockset},
% our tool~\tool~also implements the following optimization for the
% Happens Before analysis.
% When a production rule in the compressed trace is comprised
% only of terminal symbols, we use the vector clock algorithm
% to compute $\before{}$ and $\after{}$ sets for HB computation.
% \todo{Describe in detail.}
% We further transform the grammar produced by Sequitur
% to replace contiguous sequences of terminals of length $>2$
% by additional production rules.
% Both these optimizations help accelerate the detection of
% HB races on compressed traces.
\begin{myparagraph}{Optimizations}
The SLPs generated using the Sequitur algorithm are not strictly 
CNF grammars; production rules in the grammar can have length $> 2$ as well.
This is similar to the grammar shown in \figref{example-trace-grammar},
where both the non-terminals $F$ and $D$ have production rules of length $4$.
For detecting an HB-race on SLPs, \tool~employs the following optimizations
that rely on existence of such long production rules.
% \begin{itemize}
% \item \textbf{Vector Clocks for terminal-only rules.}
For a rule of the form $A \rightarrow a_1a_2\cdots a_k$, where each of
$a_1, \ldots a_k$ are terminals, our tool {\tool} uses a slight
modification of the basic HB vector clock algorithm and uses the
vector clock values to (i) determine if $\race(A)$ holds, and (ii)
compute the various sets associated with $A$ (such as $\ar{A}$,
$\br{A}$, etc.).
% Checking if $\race(A)$ is similar to the usual vector clock algorithm.
% To compute the other data-structures, the basic idea is to 
% infer the sets $\after{A}(e)$ and $\before{A}(e)$ 
% for an event $e \in \sem{A}$ using the vector time-stamp assigned to $e$, and
% the value of the vector clocks of the different threads, locks and variables,
% at the beginning and at the end of the algorithm.
% \ucomment{More details in appendix.}
% \item \textbf{Transforming grammar.}
Next, for production rules where the right hand side has both 
terminals and non-terminals and has
long contiguous sequences (or substrings) of terminals, 
we introduce new production
rules in the grammar, with fresh non-terminals
corresponding to these long sequences.
For example, for a rule of the form $A \rightarrow b_1\cdots b_k C d_1\cdots d_m$,
where $b_i$s and $d_i$s are terminals, we will
introduce two new non-terminals $B$ and $D$, with production rules
$B \rightarrow b_1\cdots b_k$ and $D \rightarrow d_1\cdots d_k$,
and replace the production rule of $A$ by $A \rightarrow BCD$.
This allows us to better exploit the vector-clock optimization.
% The idea here is to explicitly identify terminal-only rules
% to better exploit the vector-clock optimization.
% \end{itemize}
\end{myparagraph}

% \subsection{Experimental Setup}

\begin{myparagraph}{Setup and Benchmarks}
Our experiments were conducted on an 8-core 2.6GHz 64-bit Intel Xeon(R) Linux
machine, with 
% HotSpot 1.8.0 64-But Server as the JVM and 
30GB heap space.
To compare against Happens-Before and LockSet based analysis on
uncompressed traces, we use~\rapid~\cite{rapid}, which implements
the standard \textsc{Djit}+~\cite{ps03} vector clock algorithm,
epoch optimizations like in \textsc{FastTrack}~\cite{ff09},
the Goldilocks algorithm~\cite{eqt07}, and Eraser's lockset
algorithm~\cite{sbnsa97}, as described in~\cite{ps03}. 
%  The above
% implementations are fair in the sense that they avoid redundant checks
% and operations whenever possible.
% \ucomment{Say that the HB algorithm is not so naive. It is similar to DJIT+
% and that it does not do unnecessary clock increments.}
% 
% \ucomment{Maybe talk about grammar transformation, and native HB for all-terminal rules}
% 
% \input{results_table}
Our evaluation benchmarks (Column 1 in Table~\ref{tab:results}) are carefully chosen with the
goal of being comprehensive, and have been primarily derived from~\cite{hmr14}.
% the lines of code range from 60 to 0.5M,
% and the length of the executions generated vary from a few hundred
% events to 11.8M events.  
The first set of small-sized (LOC $\sim$ 50-300) benchmarks 
(\textsf{account} to \textsf{pingpong}) is  derived 
from the IBM Contest benchmark suite~\cite{fnu03}.  
The second set of medium sized (LOC $\sim$ 3K) benchmarks 
(\textsf{moldyn} to \textsf{raytracer}) is
derived from the Java Grande Forum benchmark suite~\cite{sbo01}.  
The third set (\textsf{derby} to \textsf{xalan}) of benchmarks 
(LOC $\sim$ 30K-500K) comes from the DaCaPo benchmark suite (version
9.12)~\cite{dacapo} and large real world software including Apache
FTPServer, W3C Jigsaw web server and Apache Derby.
Columns 3, 4 and 5 in Table~\ref{tab:results}
report the number of threads, locks and variables in the traces generated
from the corresponding programs in Column 1.
\end{myparagraph}
% \vspace{-0.1in}
\subsection{Results}
% Our results are summarized in Table~\ref{tab:comp} and Table~\ref{tab:time}.
% Additional details can be found in the supplementary material provided.
% \input{tab1}
\vspace{-0.1in}
\myparagraph{Compression Ratio}{
To analyze the effect of compression on the size of traces, consider
the compression ratios (ratio of the size of the original trace and
the size of the grammar representation) shown in Column 7 in Table~\ref{tab:results}. 
The compression ratios are not significant for the small and medium sized benchmarks,
barring \textsf{boundedbuffer} (compression ratio = 1.74), 
\textsf{moldyn} (compression ratio = 1.86) 
and the most notable \textsf{bufwriter} (compression ratio > 40,000).  
The compression ratios for the large benchmarks are impressive; as large as 3.25.  
This can be attributed to the fact that in large executions, 
the large amount of redundancies make them amenable to larger compression.
Despite smaller lines of code in the source code of \textsf{bufwriter},
the size of the execution observed is quite large, and thus the excellent compression ratio.
% Clearly, compression can give a significant space savings.
}
% 1. Large number of threads + locks => exponential dependence on 
% threads and locks not a bottleneck. 
%
% 2. Does give a lot of speedup. Especially when compression is high.

% \ucomment{Say our experiments reveal the power of compression and analysis on SLPs.}
% Let us first analyze the performance of \tool~for \textsf{HB} computation.
% \blue{Since the dependence of the asymptotic worst case
% running time of \tool~'s HB race detection algorithm
% on on the number of threads and locks is not polynomial,
% it would be interesting to see if \tool~indeed beats
% the performance of the widely adopted linear time 
% vector clock algorithm.}

\myparagraph{HB race detection}{
% Let us evaluate the effect of compression
% on the analysis time for HB race detection. 
Columns 8, 9 and 10 in Table~\ref{tab:results} represent the time taken
to detect the presence of an HB race by
respectively, \textsc{Djit}+,
\textsc{FastTrack} and Goldilocks.
Column 11 denotes the time taken by our HB race detection
algorithm for analyzing the traces compressed as SLPs
and Column 12 reports the speedup achieved over the best of the
three values in Columns 8, 9 and 10.
% The \textsc{FastTrack} algorithm
% almost always performs better than both \textsc{Djit}+
% and Goldilocks. We will henceforth compare the speedup/slowdown
% of our algorithm (Column 5) over \textsc{FastTrack} (Column 3).

First, in the smaller examples (\textsf{account} - \textsf{pingpong}),
the speed-up is not significant for most examples.
This can be attributed to the low compression ratios, and significant initial set-up times.
In particular, the \textsf{bubblesort} example has a significant slow-down.
% This example has $10$ threads and $53$ memory locations,
% and the slow-down can be attributed to the polynomial dependence 
% of our algorithm on the number of threads and memory locations, as
% compared to the linear dependence of vector clock algorithm. 
% However, it must be noted that there is only little slowdown 
% for the rest of the small examples.
One noteworthy small example that shows the power of compression
is \textsf{bufwriter} where the compression ratio and the resulting speedup
for race detection is very high ($>2500$x).

For the medium sized examples, the compression ratios range in $1.1-1.86$.
The speedup for \textsf{moldyn} and \textsf{montecarlo} is about $3-8$x, while for 
\textsf{raytracer}, we encounter a large slowdown.
A possible explanation for the degraded performance in both \textsf{bubblesort}
and \textsf{raytracer} is that, while the first race pair $(e_1, e_2)$
occurs very early in the uncompressed trace,
the SLP generated is such that, in order to discover any race, the entire
grammar needs to be processed.
% \todo{Try to explain why the slowdown}.

% \input{tab2}
The performance improvements for the large benchmarks
 are noteworthy and the speed ups
shoot to the order of $100$x. 
% One must note that the performance of the Goldilocks
% algorithm degrades as the benchmarks gets larger.
The \textsf{FastTrack} vector clock algorithm~\cite{ff09} is the gold standard
for detecting HB races, and our evaluation indicates
that analysis on compressed traces beats the advantages
offered by vector-clocks and further epoch-like optimizations. 
% In fact, our algorithm is closer, in spirit, to the Goldilocks algorithm (for which 
% the performance degradation intensifies on larger benchmarks),
% and all the speed-up over the vector-clock algorithm
% must be attributed to  the non-trivial compression ratios achieved.
% The performance degradation for the Goldilocks algorithm intensifies for larger traces,
% as compared to both the vector clock algorithms.
In fact our algorithm is, in spirit, closer to
the Goldilocks algorithm, for which 
the performance degradation deeply intensifies on larger benchmarks 
(also noted before in~\cite{ff09}).
The speedups (over \textsc{FastTrack}) achieved by our approach, 
despite this similarity, must be attributed to  the non-trivial compression ratios achieved.
Overall, the average speed-up is about $2.9$x over \textsc{FastTrack},
and around $200$x over the Goldilocks algorithm.
}

\myparagraph{Lockset violation detection}{
Columns 13 and 14 denote the time for detecting lockset violations
on uncompressed and compressed traces respectively.
% This is because, the theoretical complexity
% of race detection on compressed traces is linear
% in the size of the generated SLP, and there is no additional
% exponential overhead.
% Clearly, this is reflected in all our examples.
%
Since, the compression on smaller examples is not large,
we can observe that the speedup in such examples is not extraordinary.
However, there is little or almost no slowdown.
For the medium and large examples,
\tool~detects violations of lockset discipline on compressed traces
much faster than on uncompressed traces.
In fact, the speed-ups shoot upto more than $20,000$x,
and the time taken is almost always of the order of a few milliseconds.
The average speed-up achieved over the Eraser algorithm
is around $173$x.
% This speedup can be attributed to the simplicity of our algorithm for 
% analyzing compressed traces to detect locket violations.
}

Clearly, these large real-world examples illustrate the benefit of
compression; compression can be leveraged not only for
smaller storage spaces, but also for 
a more efficient race detection analysis.

%% file: bigtable.tex
%!TEX root = main.tex

% \usepackage{multirow}

\begin{table*}[t]
\centering
\scalebox{0.8}{
\begin{adjustbox}{center}

\begin{tabular}{!{\VRule[1pt]}c|c|c|c|c|c|c!{\VRule[1pt]}c|c|c|c|c!{\VRule[1pt]}c|c|c!{\VRule[1pt]}}
% \hline
\specialrule{1pt}{0pt}{0pt}
1 & 2 & 3 & 4 & 5 & 6 & 7 & 8 & 9 & 10 & 11 & 12 & 13 & 14 & 15\\ 
\specialrule{1pt}{0pt}{0pt}
  \cellcolor[HTML]{EFEFEF} 
& \cellcolor[HTML]{EFEFEF}
& \cellcolor[HTML]{EFEFEF}
& \cellcolor[HTML]{EFEFEF}
& \cellcolor[HTML]{EFEFEF} Memory
& \cellcolor[HTML]{EFEFEF} Grammar
& \cellcolor[HTML]{EFEFEF} Compr. 
& \multicolumn{5}{c!{\VRule[1pt]}}{\cellcolor[HTML]{EFEFEF}{\textsf{HB} (ms)}}
& \multicolumn{3}{c!{\VRule[1pt]}}{\cellcolor[HTML]{EFEFEF}{LockSet (ms)}} \\ 
\cmidrule{8-15}
% \hline 
  \cellcolor[HTML]{EFEFEF} Program
& \cellcolor[HTML]{EFEFEF} Events
& \cellcolor[HTML]{EFEFEF} Threads
& \cellcolor[HTML]{EFEFEF} Locks
& \cellcolor[HTML]{EFEFEF} Loc.
& \cellcolor[HTML]{EFEFEF} Size
& \cellcolor[HTML]{EFEFEF} Ratio 
& \cellcolor[HTML]{EFEFEF} \textsc{Djit}+
& \cellcolor[HTML]{EFEFEF} \textsc{F.Track}
& \cellcolor[HTML]{EFEFEF} Goldi.
& \cellcolor[HTML]{EFEFEF} Compr.
& \cellcolor[HTML]{EFEFEF} Speedup
& \cellcolor[HTML]{EFEFEF} Eraser 
& \cellcolor[HTML]{EFEFEF} Compr.
& \cellcolor[HTML]{EFEFEF} Speedup
\\
\specialrule{1pt}{0pt}{0pt}

account &  130 &  4 &  3 &  41 &  107 &  1.21 &  6 &  5 &  4 &  4 &  1 &  3 &  3 &  1 \\
airline &  137 &  4 &  0 &  44 &  132 &  1.04 &  8 &  4 &  5 &  6 &  0.67 &  2 &  1 &  2 \\
array &  47 &  3 &  2 &  30 &  47 &  1 &  5 &  4 &  4 &  3 &  1.33 &  3 &  1 &  3 \\
boundedbuffer &  337 &  2 &  2 &  63 &  194 &  1.74 &  8 &  8 &  2 &  18 &  0.11 &  3 &  2 &  1.5 \\
bubblesort &  4.2K &  10 &  2 &  167 &  3.3K &  1.29 &  14 &  13 &  12 &  92 &  0.13 &  2 &  4 &  0.5 \\
bufwriter &  11.8M  &  6 &  1 &  56 &  293 &  40238 &  20s &  15.5s &  36.3s &  6 &  2600 &  1 &  4 &  0.25 \\
critical &  55 &  4 &  0 &  30 &  55 &  1 &  3 &  4 &  3 &  5 &  0.6 &  2 &  1 &  2 \\
mergesort &  3028 &  5 &  3 &  621 &  2795 &  1.08 &  5 &  7 &  5 &  13 &  0.38 &  3 &  6 &  0.5 \\
pingpong &  146 &  4 &  0 &  51 &  135 &  1.08 &  10 &  9 &  9 &  3 &  3 &  2 &  1 &  2 \\
moldyn &  164K &  3 &  2 &  1197 &  88K &  1.86 &  53 &  60 &  57 &  6 &  8.83 &  2 &  2 &  1 \\
montecarlo &  7.2M &  3 &  3 &  876K &  6.1M &  1.18 &  317 &  271 &  302 &  87 &  3.11 &  300 &  1 &  300 \\
raytracer &  16.2K &  3 &  8 &  3879 &  14.6K &  1.11 &  58 &  32 &  32 &  315 &  0.1 &  25 &  133 &  0.19 \\
derby &  1.3M &  4 &  1112 &  186K &  735K &  1.83 &  1006 &  1011 &  26s &  592 &  1.7 &  848 &  <1 &   >1000 \\
eclipse &  90.6M &  19 &  8300 &  11.2M &  42.5M &  2.13 &  34.6s &  31.5s &  3776s &  17.4s &  1.8 &  21737 &  1 &  21737 \\
ftpserver &  49K &  11 &  301 &  5461 &  30K &  2.13 &  49 &  44 &  91 &  23 &  1.9 &  34 &  1 &  34 \\
jigsaw &  3M &  13 &  280 &  103K &  908K &  3.37 &  2432 &  2309 &  1888 &  195 &  9.7 &  12 &  4 &  3 \\
lusearch &  216M &  7 &  118 &  5.2M &  66.6K  &  3.25 &  1392 &  968 &  700 &  7 &  100 &  814 &  2 &  407 \\
xalan &  122M &  6 &  2491 &  4.4M &  71M &  1.7 &  5183 &  3008 &  3709 &  109 &  27.6 &  2779 &  1 &  2779 \\
% \hline
\specialrule{1pt}{0pt}{0pt}
\end{tabular}
\end{adjustbox}
}
\bigskip
\caption{
 Columns 1-5 describe the benchmarks and traces.
Columns 6 and 7 describe the size of the compressed traces and the compression ratios achieved.
Columns 8-10 describe the performance of various HB-race detection algorithms on uncompressed traces.
Column 11-12 reports the performance of~\tool's HB race detection on compressed traces and the resulting speedup achieved.
Column 13, 14 and 15 report the performance of respectively
Eraser's lockset algorithm on uncompressed traces, \tool's performance on compressed traces,
and the resulting speedup achieved.
}
\label{tab:results}
\end{table*}

%% file: conclusions.tex
\section{Conclusions}
\seclabel{conclusions}

We considered the problem of detecting races in traces compressed by
SLPs. We presented algorithms that detect {\hbrace}s and violations of
the lockset discipline in time that is linear in the size of the
compressed traces. Experimental evaluation of our implementation of
these algorithms in the tool {\tool}, demonstrated that analyzing
compressed traces can lead to significant speedups.

%% file: app-complexity-prelim.tex
%!TEX root = main.tex

\section{Complexity Analysis of Standard Algorithms}
\applabel{complexity}

\begin{myparagraph}{Complexity of Vector Clock Algorithm}
% \begin{reptheorem}{thm:vc_complexity}
\begin{theorem}\thmlabel{vc_complexity}
Consider a trace $\tr$ of length $n$ with $T$ threads, $L$ locks, and
$V$ variables. The vector clock algorithm for detecting {\hbrace}{s}
runs in time $O(nT\log n)$ and uses space $O((V+L+T)T\log n)$.
\end{theorem}
% \end{reptheorem}
\begin{proof}[Proof Sketch]
Each vector clock is a vector with $T$ components, where each entry is
a natural number $\leq n$. Thus, each vector clock uses space $O(T\log n)$ and
each vector clock operation takes $O(T\log n)$ time. Processing each
event in the trace involves constantly many vector clock
operations. Finally, the algorithm maintains a vector clock for each
thread, lock, and variable. All these observations together give us
the bounds in the theorem.
\end{proof}
\end{myparagraph}

\begin{myparagraph}{Complexity of Goldilocks Algorithm}
% \begin{reptheorem}{thm:gls-complexity}
\begin{theorem}\thmlabel{gls-complexity}
Consider a trace $\tr$ length $n$ with $T$
threads, $L$ locks, and $V$ variables. The Goldilocks algorithm for
detecting {\hbrace}{s} runs in time $O(n(L+TV))$ and uses space
$O(TV(T+L))$.
% \todo{Maybe use $\Theta(\cdot)$ instead of $O(\cdot)$}
\end{theorem}
% \end{reptheorem}
\begin{proof}[Proof Sketch]
Let us compute the space requirement first.
For every variable $x$ the algorithm maintains the set
$\glsw{}(x)$, which has at most $T+L$ elements.
Similarly, for every $(\thr,x)$ it maintains
the set $\glsr{}(\thr,x)$ of size at most $T+L$.
Thus, the total space usage is $O(V(T+L) + TV(T+L)) = O(TV(T+L))$.

Now let us compute the running time.
When a write event $\ev{\thr}{\wt{x}}$ is performed,
the algorithm checks if $\thr \in \glsw{}(x)$, 
or if $\bigvee_{\thr' \in \threads{}} \thr \in \glsr{}(\thr', x)$.
This check takes time $O(1+T)$.
After this check, the algorithm updates $\glsw{}(x)$
to the singleton set $\set{\thr}$.
This takes time $O(T+L)$ assuming we use a data structure
like \emph{bit-vectors} of length $T+L$ to maintain these sets.
The total time, thus, for processing a write event is $O(T+L)$.
When a read event $\ev{\thr}{\rd{x}}$ is performed,
the algorithm check if $\thr \in \glsw{}(x)$, and then updates
$\glsr{}(x)$ to the set $\set{\thr}$. Thus, a read event takes
time $O(T+L)$.

For an acquire event $\ev{\thr}{\acq{\lk}}$, the algorithm,
(i)~for every variable $x$ for which $\lk \in \glsw{}(x)$,
adds $\thr$ to $\glsw{}(x)$,
and (ii)~for every pair $(x, \thr')$ for which
$\lk \in \glsr{}(\thr',x)$,
adds $\thr$ to $\glsr{}(\thr', x)$.
This takes $O(V(1 + 1) + TV(1+1)) = O(TV)$ time.
For a release event, the converse happens; if
the thread $\thr$ is present in any of these sets, the lock $\lk$
is added to them.
Again, this is $O(TV)$ time.
A similar update happens on a fork or a join, and the
same bound applies for such events too.
So the overall time complexity is $O(n(T+L + TV)) = O(n(L+TV))$.
\end{proof}
\end{myparagraph}

\begin{myparagraph}{Complexity of Eraser's Lockset Algorithm}
% \begin{reptheorem}{thm:eraser-complexity}
\begin{theorem}\thmlabel{eraser-complexity}
Let $\tr$ be a trace $\tr$ of length $n$ with $T$
threads, $L$ locks, and $V$ variables. Let $r$ be the maximum number of
times a thread acquires a lock without releasing it.
The Eraser algorithm detects violations of the locket 
discipline on $\tr$ in time $O(n(L + \log r))$ and uses space 
%$O(TL(V + \log r))$.
$O(TL\log r + V(T+L))$.
\end{theorem}
% \end{reptheorem}
\begin{proof}[Proof Sketch]
Let us analyze the space usage first.
For every thread $\thr$ and every lock $\lk$, 
the algorithm maintains an integer value
representing the number of unmatched acquires
of lock $\lk$ in thread $\thr$.
Since each of these values do not exceed $r$,
this contributes $O(TL\log r)$ the space usage.
Additionally, for every variable $x$, the algorithm
maintains a set of locks, whose size can
be at most $L + |\cD| = L + T + 1$.
This amounts to a space usage of $O(V(T+L+1))$.
The total space usage is thus $O(TL\log r + V(T+L+1))$
= $O(TL\log r + V(T+L)).$

Let us now analyze the running time.
For every event corresponding to an acquire/release 
of lock $\lk$ by thread $\thr$, the algorithm
updates the number of unmatched acquires of $\lk$ in $\thr$.
This takes time $O(\log r)$.
For every read/write event $\ev{\thr}{a(x)}$ ($a \in \set{\rd{}, \wt{}}$), 
the algorithm checks if (i) $\thr$ is in the lockset
of $x$, (ii) if the dummy lock $\dmlock$ is contained in the lockset of $x$
(when $a = \rd{}$), 
(iii) if the disjunction of two sets, of size at most $L$, is empty.
This takes a time of $O(L)$.
The total time, therefore, is $O(n(L + \log r))$.
\end{proof}
\end{myparagraph}

%% file: app-proofs.tex
%!TEX root = main.tex

\section{Proofs from \secref{hb}}
\seclabel{proofs-hb-algo}

\begin{myparagraph}{Proof of \lemref{hb-before-after}}
\begin{replemma}{lem:hb-before-after}
Consider events $e \in \sem{B}$ and $e' \in \sem{C}$. $e \HB[BC] e'$
iff $\after{B}(e) \cap \before{C}(e') \neq \emptyset$.
\end{replemma}

\begin{proof}
Recall that if $e \HB[BC] e'$ then there is a sequence $e = e_1, e_2,
\ldots e_n = e'$ that satisfies the conditions in
\defref{happens-before}. Let $j$ be such that for all $i \leq j$, $e_i
\in \sem{B}$ and for all $i \geq j+1$, $e_i \in \sem{C}$. Also, we
have $e \HB[B] e_j$ and $e_{j+1} \HB[C] e'$.

If $e_j, e_{j+1}$ satisfy condition \itemref{thread-order}
 of \defref{happens-before}, then $\thr \in \after{B}(e)
\cap \before{C}(e')$, where $\thr$ is the thread performing both $e_j$
and $e_{j+1}$. If $e_j, e_{j+1}$ satisfy condition \itemref{acq-rel},
then $\lk \in \after{B}(e) \cap \before{C}(e')$, where $\lk$ is the
lock released/acquired by event $e_j$ and $e_{j+1}$,
respectively. Finally, if $e_j,e_{j+1}$ satisfy condition
\itemref{fork} or \itemref{join}, $\thr' \in \after{B}(e) \cap \before{C}(e')$, 
where $\thr'$ is the thread forked by $e_j$ and the one performing $e_{j+1}$
or the child thread joined in $e_{j+1}$.

On the other hand if $\after{B}(e) \cap \before{C}(e') \neq
\emptyset$, then we can demonstrate that $e \HB[BC] e'$. Let $u \in
\after{B}(e) \cap \before{C}(e')$. Let us first consider the case when
$u$ is a lock $\lk$. Since $u = \lk \in \after{B}(e)$, then there is
an event $e_1 \in \sem{B}$ such that $e_1$ is $\rel{\lk}$ event and $e
\HB[B] e_1$. Similarly, since $u = \lk \in \before{C}(e')$, there is
an event $e_2 \in \sem{C}$ such that $e_2$ is an $\acq{\lk}$ event and
$e_2 \HB[C] e'$. Putting these together, we get that $e \HB[BC]
e'$. Let us now consider the case when $u = \thr$. Again, $u = \thr
\in \after{B}(e)$ means that there is an event $e_1 \in \sem{B}$ such
that $e \HB[B] e_1$ and either $e_1$ is an event performed by $\thr$,
or $e_1$ is a $\fork{\thr}$ event.  Similary, $u = \thr \in
\before{C}(e')$, means that there is an event $e_2 \in \sem{C}$ such
that $e_2 \HB[C] e'$ and either $e_2$ is an event performed by $\thr$,
or $e_2$ is a $\join{\thr}$ event. In all these four cases, we can
conclude that $e \HB[BC] e'$.
\end{proof}

\end{myparagraph}

\begin{myparagraph}{Proof of \lemref{first_and_last_events}}

Let us first note the following observation:
\begin{lemma}
\lemlabel{HB_implies_after_before}
Let $\tr$ be a trace $\tr$, and let $e_1, e_2$ be events
such that $e_1 \HB[\tr] e_2$.
Then, $\after{\tr}(e_2) \subseteq \after{\tr} (e_1)$
and $\before{\tr}(e_1) \subseteq \before{\tr} (e_2)$.
\end{lemma}
\begin{proof}
The set $\after{\tr}(e_2) $ is
\begin{equation*}
\begin{array}{rll}
\after{\tr}(e_2) = & \setpred{\thr \in \threads(\tr)}{\exists e' =
  \ev{\thr}{o}.\ e_2 \HB[\tr] e'}  \\ 
\quad\quad\cup & \setpred{\thr \in \threads(\tr)}{\exists e' =
  \ev{\thr'}{\fork{\thr}}.\ e_2 \HB[\tr] e'}  \\
\cup & \setpred{\lk \in \locks(\tr)}{\exists e' = \ev{\thr}{\rel{\lk}}.\ e
  \HB[\tr] e'}
\end{array}
\end{equation*}
Since the HB relation $\HB[\tr]$ is transitive, we have that 
for all the events $e'$ for which $e_2 \HB[\tr] e'$,
we must have $e_1 \HB[\tr] e'$.
Clearly, $\after{\tr}(e_2) \subseteq \after{\tr} (e_1)$.

A similar reasoning proves $\before{\tr}(e_1) \subseteq \before{\tr} (e_2)$.
\end{proof}

Let us now prove \lemref{first_and_last_events}

\begin{replemma}{lem:first_and_last_events}
If there is no HB race in $\sem{B}$ or in $\sem{C}$,
and if there is an HB race between events $e \in \sem{B}$ and $e' \in \sem{C}$
then, there is an HB race between $\s{last}_B^e$ and $\s{first}_C^{e'}$,
where
\begin{equation*}
\s{last}_{B}^e = 
  \left\{\begin{array}{ll}
    \last{B}(\reads{B}(\thr, x)) & \text{if } e = \ev{\thr}{\rd{x}} \\
    \last{B}(\writes{B}(x)) & \text{if } e = \ev{\thr}{\wt{x}}
  \end{array} \right.
\end{equation*}
and
\begin{equation*}
\s{first}_{C}^{e'} = 
  \left\{\begin{array}{ll}
    \first{C}(\reads{C}(\thr, x)) & \text{if } e' = \ev{\thr'}{\rd{x}} \\
    \first{C}(\writes{C}(x)) & \text{if } e' = \ev{\thr'}{\wt{x}}
  \end{array} \right.
\end{equation*}
\end{replemma}

\begin{proof}
% Let us consider the case when $e = \ev{\thr}{\rd{x}}$ is a read event,
% and $e' = \ev{\thr'}{\wt{x}}$ is a write event.
% The other two cases ($e$ being a write, and $e'$ being a read, or, $e$ begin a write and
% $e'$ being a write) are similar.

% First note that 
First, notice that, for a given trace $\tr$, and for $(\thr, x) \in \rvars{\tr}$,
the elements of the set $\reads{\tr}(\thr, x)$ are totally ordered by
the relation $\HB[\tr]$. 
This follows from the rule \itemref{thread-order} of \defref{happens-before}.
Thus, for an event $e\in \reads{\tr}(\thr, x)$, we have
$\first{\tr}(\reads{\tr}(\thr, x)) \HB[\tr]  e \HB[\tr] \last{\tr}(\reads{\tr}(\thr, x))$.
Using \lemref{HB_implies_after_before}, we must also have
\begin{itemize}
\item $\after{\tr}(\last{\tr}(\reads{\tr}(\thr, x))) \subseteq \after{\tr}(e)$,
\item $\after{\tr}(e) \subseteq \after{\tr}(\first{\tr}(\reads{\tr}(\thr, x)))$,
\item $\before{\tr}(\first{\tr}(\reads{\tr}(\thr, x))) \subseteq  \before{\tr}(e)$, and 
\item $\before{\tr}(e) \subseteq  \before{\tr}(\last{\tr}(\reads{\tr}(\thr, x)))$
\end{itemize}

Next observe that for a trace $\tr$, if $\race(\tr)$ is false,
the elements of the set $\writes{\tr}(x)$ for some $x \in \wvars{\tr}$
must be totally ordered by $\HB[\tr]$, and thus for such a trace $\tr$,
\begin{itemize}
\item $\after{\tr}(\last{\tr}(\writes{\tr}(x))) \subseteq \after{\tr}(e)$,
\item $\after{\tr}(e) \subseteq \after{\tr}(\first{\tr}(\writes{\tr}(x)))$,
\item $\before{\tr}(\first{\tr}(\reads{\tr}(x))) \subseteq  \before{\tr}(e)$, and 
\item $\before{\tr}(e) \subseteq  \before{\tr}(\last{\tr}(\writes{\tr}(x)))$
\end{itemize}

Now since there is a race between $e \in \sem{B}$ and $e' \in \sem{C}$,
by \lemref{hb-before-after}, we have
$\after{B}(e) \cap \before{C}(e') = \emptyset$.
Clearly, this means that
$\after{B}(\s{last}_B^e) \cap \before{C}(\s{first}_C^{e'}) = \emptyset$.
Again, from \lemref{hb-before-after}, we have
that $\s{last}_B^e \hbconc \s{first}_C^{e'}$.
Also, $\s{last}_B^e \conflict \s{first}_C^{e'}$ because $e \conflict e'$.
Thus, there is an HB race between $\s{last}_B^e$ and $\s{first}_C^{e'}$.

% \ucomment{The other direction does not hold. 
% The last and first events can have a race, but $e$ and $e'$ might be ordered !}
\end{proof}
\end{myparagraph}

\begin{myparagraph}{Proof of \thmref{race-correctness}}

Before we prove \thmref{race-correctness},
let us first note the following simple observation.
\begin{proposition}
\proplabel{subchunk-race}
Let $\tr$ be a trace and $e_1, e_2 \in \tr$.
$e_1\HB[\tr] e_2$ iff for any trace
$\tr' = \tr_1\tr\tr_2$, we have $e_1\HB[\tr'] e_2$.
\end{proposition}
Informally, this means that the presence/absence of
an HB-race between two events in a trace $\tr$, does
not get affected by the ``context'' in which $\tr$
is placed.
The proof of the above proposition follows easily
from \defref{happens-before}, and is skipped.

We now move on to the proof of \thmref{race-correctness}.

\begin{reptheorem}{thm:race-correctness}
For any non-terminal $A$, $\race(A) =
\s{true}$ if and only if there are events $e_1,e_2 \in \sem{A}$ such
that $e_1 \conflict e_2$ and $e_1 \hbconc e_2$.
\end{reptheorem}
\begin{proof}
The proof is by induction on the ranking of the non-terminals. 
In the base case, the theorem clearly holds. 
In the inductive step, consider a non-terminal $A \rightarrow BC$. 
Suppose $\race(A) = \s{true}$. 
Based on \equref{race}, this means either $\race(B)$ or
$\race(C)$ are $\s{true}$ or some pair of after and before sets are disjoint. 
If $\race(B)$ (or $\race(C)$) is $\s{true}$, then by induction hypothesis, 
there is an {\hbrace} in $\sem{B}$ (or $\sem{C}$), 
and so there is a race in $\sem{A}$. 
This follows from \propref{subchunk-race}.
If not (that is, if both $\race(B)$ and $\race(C)$ are false), 
the condition in one of line 2, 3 or 4 in \equref{race} is true. 
The proof is the same no matter which of these conditions hold.
Therefore, let us assume without loss of generality, 
that $\ar{B}(\thr,x) \cap \bw{C}(x) = \emptyset$ 
for some thread $\thr$ and variable $x$. 
Since $(\thr, x) \in \reads{B}$ and $x \in \writes{C}$,
both $\ar{B}(\thr,x)$ and $\bw{C}(x)$ are non-empty sets.
Thus, both the events $e,e'$ given by $e = \last{B}(\reads{B}(\thr, x))$
and $e' = \first{C}(\writes{C}(x))$ are well defined.
Also, by definition, $\after{B}(e) = \ar{B}(\thr, x)$
and $\before{C}(e') = \bw{C}(x)$.
Thus, we have $\after{B}(e) \cap \before{C}(e') = \emptyset$.
Then, by \lemref{hb-before-after}, $e \hbconc e'$ 
in the chunk $\sem{BC} = \sem{A}$.
Also, $e, e'$ are clearly conflicting (read and write of same variable).
Thus, $(e, e')$ is an HB race in the chunk $A$.

Conversely, suppose there are a pair of events $e_1,e_2 \in \sem{A}$
such that $e_1 \conflict e_2$ and $e_1 \hbconc e_2$. Without loss of
generality, let us assume that $e_1$ is $\rd{x}$-event, and $e_2$ is a
$\wt{x}$-event. If $e_1, e_2 \in \sem{B}$ or $e_1, e_2 \in \sem{C}$
then by induction hypothesis, $\race(B)$ or $\race(C)$ is
$\s{true}$. By \equref{race} this would imply that $\race(A)$ is
$\s{true}$. 
Now consider the case when there are no races in $B$ or $C$, 
and there are $e_1 \in \sem{B}$ and $e_2 \in
\sem{C}$ which are in race. 
Without loss of generality, let us assume that
$e_1 = \ev{\thr}{\rd{x}}$ is a read event,
and $e_2 = \ev{\thr'}{\wt{x}}$ is a write event on the same variable $x$.
Then, by \lemref{first_and_last_events}, the events
$e = \s{last}^{e_1}_{B} = \last{B}(\reads{B}(\thr, x))$ and 
$e' = \s{first}^{e_2}_{C} \first{C}(\writes{C}(x))$
are in race too.
Then, again using  \lemref{hb-before-after},
we must have that $\after{B}(e) \cap \before{C}(e') = \emptyset$.
Now, by definition of the sets $\ar{}$ and $\bw{}$, 
we have $\ar{B}(\thr, x) = \after{B}(e)$
and $\bw{C}(x)= \before{C}(e')$.
Thus, clearly, $\race(A) = \s{true}$.
\end{proof}
\end{myparagraph}

\begin{myparagraph}{Proof of \lemref{after-cross-chunk} and \lemref{before-cross-chunk}}

The proof of \lemref{after-cross-chunk}
and \lemref{before-cross-chunk} depend on a simple observation that we
establish first.

\begin{lemma}
\lemlabel{hb-af-bl} Let $A$ be a non-terminal with rule $A \rightarrow
BC$. Let $e_1 \in \sem{B}$ and $e_2 \in \sem{C}$. The following three
statements are equivalent.
\begin{enumerate}
\item\itemlabel{one} $e_1 \HB[A] e_2$.
\item\itemlabel{two} There is $u \in \after{B}(e_1)$ such that $\s{first}_C^u$ is defined and $\s{first}_C^u \HB[C] e_2$.
\item\itemlabel{three} There is $u \in \before{C}(e_2)$ such that $\s{last}_B^u$ is defined and $e_1 \HB[B] \s{last}_B^u$.
\end{enumerate}
where
\begin{align*}
\s{first}_C^u =
\begin{cases}
\first{C}(\tjevents{C}(u)) & \text{if } u \in \threads(C) \\
\first{C}(\acquires{C}(u)) & \text{if } u \in \locks(C)
\end{cases}
\end{align*}
and
\begin{align*}
\s{last}_B^u =
\begin{cases}
\last{B}(\tfevents{B}(u)) & \text{if } u \in \threads(B) \\
\last{B}(\releases{B}(u)) & \text{if } u \in \locks(B)
\end{cases}
\end{align*}
\end{lemma}

\begin{proof}
First observe that set of events
$\tjevents{C}(\thr)$, for some $\thr \in \threads(C)$,
is totally ordered with respect to $\HB[C]$.
This is because all the events in this set are either of the form
$\ev{\thr}{o}$ or $\ev{\thr'}{\join{\thr}}$, and
(i) $\HB[C]$ orders all events in a single thread thread 
(\defref{happens-before}~\itemref{thread-order}),
and (ii) also orders all events of a thread before the 
join event of that thread (\defref{happens-before}~\itemref{join}).
Similarly, the events of the sets $\tfevents{B}(\thr)$,
$\acquires{C}(\lk)$ and $\releases{B}(\lk)$ are
also ordered totally with respect to $\HB[B]$, $\HB[C]$ and $\HB[B]$
respectively.

Let us start by showing \itemref{one} $\Rightarrow$ \itemref{two}. As in
the discussion on cross-races in \secref{hb-cross-race}, this means
there is a pair of events $e_3 \in \sem{B}$ and $e_4 \in \sem{C}$ such
that $e_1 \HB[B] e_3$, $e_4 \HB[C] e_2$, and one of the following four
conditions holds.
\begin{itemize}
\item $e_3$ and $e_4$ are events performed by the same thread (say)
  $\thr$. 
  Thus, $\tjevents{C}(\thr)$ is non-empty (contains $e_4$), and $\s{first}_{C}^\thr$
  is defined. 
  Since $e_1 \HB[B] e_3$, we have $\thr \in
  \after{B}(e_1)$. 
  Moreover,  $\s{first}_{C}^\thr \HB[C] e_4 \HB[C] e_2$,
  establishing the claim.
\item $e_3 = \ev{\thr'}{\fork{\thr}}$ and $e_4 = \ev{\thr}{o}$ for
  some $\thr', \thr, o$. 
Thus, $\tjevents{C}(\thr)$ is non-empty (contains $e_4$), and $\s{first}_{C}^\thr$
  is defined.
  Then by definition, we have $\thr \in
  \after{B}(e_1)$, and $\s{first}_{C}^\thr \HB[C] e_4 \HB[C] e_2$, which
  proves the claim.
\item $e_3 = \ev{\thr}{o}$ and $e_4 = \ev{\thr'}{\join{\thr}}$, for
  some $\thr,\thr'$, and $o$.
 Thus, $\tjevents{C}(\thr)$ is non-empty (contains $e_4$), and $\s{first}_{C}^\thr$
  is defined. 
   We have $\thr \in \after{B}(e_1)$. And,
  $\s{first}_{C}^\thr \HB[C] e_4 \HB[C] e_2$.
\item $e_3 = \ev{\thr}{\rel{\lk}}$ and $e_4 = \ev{\thr}{\acq{\lk}}$
  for some $\thr,\lk$. 
 Thus, $\acquires{C}(\lk)$ is non-empty (contains $e_4$), and $\s{first}_{C}^\lk$
  is defined. 
  In this case we have $\lk \in \after{B}(e_1)$,
  and $\s{first}_{C}^\lk \HB[C] e_4 \HB[C] e_2$, which establishes the
    claim as well.
\end{itemize}

Now we show that \itemref{two} $\Rightarrow$ \itemref{three}.
Let us assume that there is $u \in \after{B}(e_1)$ such that $\s{first}_{C}^u$
is defined and $\s{first}_{C}^u \HB[C] e_2$. 
Since $u \in
\after{B}(e_1)$, let $e_3$ be the event that ``witnesses'' the
membership of $u \in \after{B}(e_1)$. By definition, we have $e_1
\HB[B] e_3$. First observe that no matter what $u$ (i.e., lock/thread)
and $\s{first}_C^u$ (i.e., $\acq{\cdot}$, event, or $\join{\cdot}$)
are, we have $u \in \before{C}(e_2)$ because of $\s{first}_C^u$. Next,
no matter what $u$ and $e_3$ are, $e_3 \HB[B] \s{last}_{B}^u$. Thus, $e_1
\HB[B] \s{last}_B^u$ which establishes the claim.

We complete the proof by showing that \itemref{three} $\Rightarrow$
\itemref{one}. Suppose there is $u \in \before{C}(e_2)$ such that
$\s{last}_B^u$ is defined and $e_1 \HB[B] \s{last}_B^u$. 
Let $e_3 = \s{last}_{B}^u$. We have $e_1 \HB[B] e_3$. Let
$e_4 \in \sem{C}$ be the event that witnesses $u \in \before{C}(e_2)$;
thus, $e_4 \HB[C] e_2$. Since $\HB[B] \subseteq \HB[A]$ and $\HB[C]
\subseteq \HB[A]$, we can conclude that $e_1 \HB[A] e_3$ and $e_4
\HB[A] e_2$. Finally, since $\HB[A]$ is a transitive relation, 
it would be enough to establish that $e_3 \HB[A] e_4$ in order to complete the
proof. Considering the various possibilities for $u,e_3$, and $e_4$,
we will have four sub-cases like in the proof of \itemref{one}
$\Rightarrow$ \itemref{two}, and in all cases we can conclude that
$e_3 \HB[A] e_4$.
\end{proof}

Now let us prove \lemref{after-cross-chunk}:
\begin{replemma}{lem:after-cross-chunk}
Let $A$ be a non-terminal with rule $A \rightarrow BC$ and let $e \in
\sem{B}$. Then
\[
\after{A}(e) = \after{B}(e) \cup \smashoperator[r]{\bigcup\limits_{u \in \after{B}(e)}} \af{C}(u)
\]
\end{replemma}
\begin{proof}
We need to prove that for a non-terminal $A$ with rule $A \rightarrow
BC$ and $e \in \sem{B}$, 
\[
\after{A}(e) = \after{B}(e) \cup \bigcup\limits_{u \in \after{B}(e)} \af{C}(u)
\]
We prove containment in each direction. Consider $u' \in
\after{A}(e)$. Let us assume that event $e'$ ``witnesses'' 
the membership of $u'$ in the set $\after{A}(e)$. 
If $e' \in \sem{B}$ then $u' \in
\after{B}(e)$ because of $e'$. 
On the other hand, if $e' \in \sem{C}$, then $e \HB[A] e'$ and
then by \lemref{hb-af-bl}, there is a $u \in \after{B}(e)$ such that $e'
\in \setpred{e \in \sem{C}}{\s{first}_{C}^{u} \HB[C] e}$, 
where $\s{first}_{C}^{u}$ is defined as in \lemref{hb-af-bl}. 
Thus, $u' \in \after{C}(\s{first}_{C}^{u}) = \af{C}(u)$, 
This establishes the containment from left to right.

Let us now consider the other direction. Observe that $\after{B}(e)
\subseteq \after{A}(e)$. Consider $u' \in \cup_{u \in \after{B}(e)}
\af{C}(u)$. Let $u_1$ be such that $u_1 \in \after{B}(e)$ and $u' \in
\af{C}(u_1)$. 
Further, let $e'$ be the event witnessing the membership of $u'$ in
$\af{C}(u_1)$. By \lemref{hb-af-bl}, $e \HB[A] e'$ and therefore, $u'
\in \after{A}(e)$ which establishes the containment from right to
left.
\end{proof}

Now consider the dual \lemref{before-cross-chunk}.

\begin{replemma}{lem:before-cross-chunk}
Let $A$ be a non-terminal with rule $A \rightarrow BC$ and let $e \in
\sem{C}$. Then
\[
\before{A}(e) = \before{C}(e) \cup 
      \smashoperator[r]{\bigcup\limits_{u \in \before{C}(e)}} \bl{B}(u)
\]
\end{replemma}
\begin{proof}
The proof is similar to the proof of \lemref{after-cross-chunk} and is
therefore skipped.
\end{proof}
\end{myparagraph}

\begin{myparagraph}{Complexity Analysis}%{Proof of \thmref{hbalgo-complexity}}
% \begin{reptheorem}{thm:hbalgo-complexity}
\begin{theorem}\thmlabel{hbalgo-complexity}
Let $G$ be an SLP of size $g$ representing a string $\tr$, having $T$
threads, $L$ locks, and $V$ variables. 
The inductive algorithm described in this section to 
detect {\hbrace}s runs in time
$ O(g((T+L)^2(L+TV)))$ and uses space
$O(g((T+L)(L + TV)))$.
\end{theorem}
% \end{reptheorem}
%
\begin{proof}
Let us begin with the space requirements of the algorithm. 
For each non-terminal $A$, 
the algorithm maintains a bit $\race(A)$, the sets
$\af{A}(u)$ and $\bl{A}(u)$ for each lock and thread $u$, 
and the collections $\ar{A}(\thr, x)$, $\aw{A}(x)$, $\br{A}(\thr, x)$, 
and $\bw{A}(x)$ for each variable $x$ and thread $\thr$. 
The sets $\af{A}(u)$ and $\bl{A}(u)$ have at most $T+L$ elements. 
Each of the sets $\ar{A}(\thr, x)$, $\aw{A}(x)$, $\br{A}(\thr,x)$
and $\bw{A}(x)$ have size at most $T+L$. 
Putting all of this together, the total space used is $O(g(1 +
(T+L)(T+L) + V(T+L) + (TV)(T+L))) = O(g(T+L)(T+L+V+TV)) = O(g((T+L)(L + TV)))$.

Let us now analyze the time requirements. The algorithm computes the
various sets it maintains for each non-terminal. The time to compute
$\af{A}(u)$ (or $\bl{A}(u)$) is $O((L+T)(L+T))$, because it involves
taking the union of at most $(L+T+1)$ $\af{}$ (or $\bl{}$) sets, each
of which has size at most $L+T$. Next, let us analyze the time to
compute $\ar{A}(\thr, x)$ set; the time for $\aw{A}(x)$, $\br{A}(\thr, x)$, and
$\bw{A}(x)$ is the same. Let us recall the inductive definition of
$\ar{A}(\thr,x)$ for the rule $A \rightarrow BC$.
\begin{equation}
\ar{A}(\thr, x)  = \left\{ 
      \begin{array}{ll}
        \ar{C}(\thr, x) & \hspace{-0.5in} \mbox{if } \ar{C}(\thr, x) \\
                        &  \hspace{-0.3in}\neq \emptyset \\
        \ar{B}(\thr, x)  \cup & \hspace{-0.3in} \mbox{otherwise} \\
        \bigcup\limits_{u \in \ar{B}(\thr, x)}  \af{C}(u)
      \end{array} \right.
\end{equation}
In the worst case, when  $\ar{C}(\thr, x)$ is $\emptyset$,
one has to take the union of a maximum of $T+L+1$ sets,
each of whose size is bounded by $T+L$. So the time to
compute $\ar{A}(\thr, x)$ is $O((T+L)(T+L))$. 

Finally, let us analyze the time to compute $\race(A)$. 
From \equref{race}, this requires checking the
disjointness of sets of size at most $L+T$.
This takes time $O(L+T)$.

All these observations together give a total
running time of the entire algorithm to be 
\begin{equation*}
\begin{array}{ll}
O(g[(T+L)(T+L)^2 + V(T+L)^2 + TV(T+L)^2 + V(T+L) \\ + TV(T+L)])
 = O(g((T+L)^2(L+TV)))
 \end{array}
\end{equation*}
\end{proof}
\end{myparagraph}

% \clearpage

\section{Proofs from \secref{lockset}}
\seclabel{proofs-lockset-algo}

\begin{myparagraph}{Proof of \lemref{locksheld}}

Before we present the proof of \lemref{locksheld},
it will be useful to formally state the definition of
the function $\locksheld{}$.

\begin{definition}[$\locksheld{}$]
\deflabel{locksheld_def}
For a trace $\tr$, and an event $e = \ev{\thr}{o} \in \tr$,
the set of locks held by thread $\thr$ in trace $\tr$
when $e$ is being performed, is defined as
\begin{align*}
\locksheld{\tr}(e) =& \setpred{\lk \in \locks(\tr)}{\exists e' = \ev{\thr}{\acq{\lk}} \in \tr \\
& \text{ such that } 
 e' \TO[\tr] e \text{ and } \neg(\match(e') \TO[\tr] e ) \,} \\
 \cup & \setpred{\lk \in \locks(\tr)}{\exists e' = \ev{\thr}{\rel{\lk}} \in \tr \\
& \text{ such that } 
 e \TO[\tr] e' \text{ and } \neg(e \TO[\tr] \match(e') ) \,}
\end{align*}
\end{definition}

Let us now present the proof of \lemref{locksheld}

\begin{replemma}{lem:locksheld}
Let $A$ be a non-terminal with rule $A \rightarrow BC$.
Let $e \in \proj{B}{\thr}$ and $e' \in \proj{C}{\thr'}$
be read/write events performed by threads $\thr,\thr'$.
Then,
\vspace{-0.05in}
\begin{equation*}
\begin{array}{ll}
\locksheld{A}(e)  = & \locksheld{B}(e) \\
          & \cup \setpred{\lk}{\orel{C}(\thr,\lk) > \oacq{B}(\thr,\lk)} 
\end{array}
\end{equation*}
% \vspace{-0.05in}
% and,
\vspace{-0.05in}
\begin{equation*}
\begin{array}{ll}
\locksheld{A}(e') = & \locksheld{C}(e') \\
          & \cup \setpred{\lk}{\oacq{B}(\thr',\lk) > \orel{C}(\thr',\lk)} 
\end{array}
\end{equation*}
\end{replemma}

\begin{proof}
We will prove the first equation. 
The proof is similar for the second equation. 

We have event $e \in \proj{B}{\thr}$.
Let us first prove the containment
\begin{equation*}
\begin{array}{ll}
\locksheld{A}(e)  \subseteq & \locksheld{B}(e) \\
          & \cup \setpred{\lk}{\orel{C}(\thr,\lk) > \oacq{B}(\thr,\lk)} 
\end{array}
\end{equation*}
Let $\lk \in \locksheld{A}(e)$.
Then, one of the following cases is possible.
\begin{enumerate}
  \item there is an $e' = \ev{\thr}{\acq{\lk}} \in \sem{A}$ such that
  $e' \TO[A] e$ and $e'$ is either not released at all in $A$, 
  or $\match{e'} \in \proj{C}{\thr}$.
  In either case, $\lk \in \locksheld{B}(e)$.
  \item there is an $e' = \ev{\thr}{\rel{\lk}} \in \sem{A}$ such that
  $e \TO[A] e'$, and either $\match{e'}$ is not in $A$ or is acquired in $\proj{B}{\thr}$.
  In the first case when $e'$ is unmatched in $A$, we have 
  $\orel{C}(\thr,\lk) > \oacq{B}(\thr,\lk)$. 
  Otherwise, $\lk \in \locksheld{B}(e)$.
\end{enumerate}
In either case, $\lk$ is contained in the set on the right.

Now, let us prove the following containment:
\begin{equation*}
\begin{array}{ll}
\locksheld{A}(e)  \supseteq & \locksheld{B}(e) \\
          & \cup \setpred{\lk}{\orel{C}(\thr,\lk) > \oacq{B}(\thr,\lk)} 
\end{array}
\end{equation*}
First, consider $\lk \in \locksheld{B}(e)$. Clearly, $\lk \in \locksheld{A}(e)$.
Otherwise, consider $\lk$ such that
$\orel{C}(\thr,\lk) > \oacq{B}(\thr,\lk)$.
Thus, there is atleast one release event $e' = \ev{\thr}{\rel{\lk}} \in \sem{C}$
such that $\match{e'} \not\in \sem{A}$.
In this case again, $\lk \in \locksheld{A}(e)$.
\end{proof}
\end{myparagraph}

\begin{myparagraph}{Complexity Analysis}%{Proof of \thmref{lockset-complexity}}
% \begin{reptheorem}{thm:lockset-complexity}
\begin{theorem}\thmlabel{lockset-complexity}
Let $G$ be an SLP of size $g$ representing a string $\tr$, having $T$
threads, $L$ locks, and $V$ variables. 
Let $r$ be the maximum number of
times a thread acquires a lock without releasing it, in the trace generated by $G$.
The inductive algorithm described in this section to detect violations of lockset discipline
runs in time $O(gTL(\log r + V))$ and uses space $O(gTL(\log r + V))$.
\end{theorem}
% \end{reptheorem}

\begin{proof}
Let us first analyze the space complexity.
For every non-terminal, we need to store $\lockset{}(\thr,x)$ 
having size atmost $L$, for every pair of thread $\thr$ and variable $x$.
For every non-terminal, we also need to store integers 
$\oacq{}(\thr, \lk)$ and $\oacq{}(\thr, \lk)$, for every pair of 
lock $\lk$ and thread $\thr$.
The size of every such integer is bounded by $O(\log r)$.
% \footnote{A 
% crude upper bound on the value of, say, $\oacq{}(\thr, \lk)$ is the size of the string generated by the SLP,
% which is upper bounded by $2^g$. 
% A simple induction on the rank of non-terminals can be used to establish this.
% The size of thse numbers therefore is $O(\log 2^g) = O(g)$}.
The total space usage per non-terminal therefore is $O(TLV + TL\log r)$, and
the overall space requirement is $O(gTL(V+\log r))$.

Let us now evaluate the time complexity.
For every non-terminal, the algorithm detects lockset violation
by checking if for some variable $x$, 
the intersection of atmost $T$ sets, each of size $L$ is empty.
This takes time $O(TLV)$ per non-terminal.
To compute $\lockset{}$, the algorithm, for every thread $\thr$
and variable $x$ performs set operations on
constantly many sets of size $O(L)$. This takes time $O(TLV)$ 
for every non-terminal.
Lastly, to compute $\oacq{}$ and $\orel{}$, the algorithm
performs arithmetic operations on constantly many integers
(of size $O(\log r)$) for every thread and lock, taking time
$O(TL\log r)$ per non-terminal.
The overall time complexity therefore is $O(gTL(V+\log r))$.
\end{proof}
\end{myparagraph}

%% file: app-hb-algo.tex
%!TEX root = main.tex

\section{Complete Algorithm for Happens Before with Proof of Correctness}
\applabel{hb-app}

Here, we will first present the complete algorithm for detecting HB races,
and then argue about correctness. 

\subsection{Algorithm}
\applabel{hb-algo}

We present the complete inductive algorithm for detecting {\hbrace}s
on traces represented by SLPs. The algorithm is inductive, and
computes various sets for each non-terminal in the SLP. The
definitions of the various sets being computed is given in
\secref{hb}. Here we only present the algorithms that compute these
sets. Each of these sets is computed for a non-terminal $A$ and the
computation depends on the rule associated with $A$. Recall that there
are two types of rules --- either $A \rightarrow a$, where $a$ is an
event, or $A \rightarrow BC$, where $B$ and $C$ are non-terminals. In
what follows, $\thr$ and $\thr'$ will be used to denote threads, $\lk$
to denote a lock, and $u$ and $u'$ to denote either a lock or thread.

\begin{myparagraph}{Race predicate}
For rule $A \rightarrow a$, $\race(A) = \s{false}$. For $A \rightarrow BC$,
\begin{equation*}
\begin{array}{ll}
\race(A) = & \race(B) \vee \race(C) \vee\\
 \bigvee_{x \in \wvars{B} \cap \wvars{C}} 
   &  \aw{B}(x) \cap \bw{C}(x) = \emptyset \\ %\s{RaceWW}_{B,C}(x)\\
\bigvee_{x \in \wvars{B}, (\thr, x) \in \rvars{C}}
  & \aw{B}(x) \cap \br{C}(\thr, x) = \emptyset \\
\bigvee_{(\thr,x) \in \rvars{B}, x \in \wvars{C}}
  & \ar{B}(\thr, x) \cap \bw{C}(x) = \emptyset\\
\end{array}
\end{equation*}
\end{myparagraph}

\begin{myparagraph}{After First Sets} 
For rule $A \rightarrow a$, we have
\begin{align*}
\af{A}(\thr) &= 
\begin{cases}
  \set{\thr, \lk} &\text{if } a = \ev{\thr}{\rel{\lk}}\\
  \set{\thr, \thr'} &\text{if } a = \ev{\thr}{\fork{\thr'}}\\
  \set{\thr}  &\text{if } a = \ev{\thr}{o} \text{ and } 
                o\not\in \set{\rel{\cdot}, \fork{\cdot}} \\
  \set{\thr'} &\text{if } a = \ev{\thr'}{\join{\thr}}\\
  \emptyset &\text{otherwise}
\end{cases} \\
\af{A}(\lk) &= 
\begin{cases}
  \set{\thr} &\text{if } a = \ev{\thr}{\acq{\lk}}\\
  \emptyset &\text{otherwise}
\end{cases}
\end{align*}
For rule $A \rightarrow BC$ (and $u$ either a thread $\thr$ or lock $\lk$)
\begin{align*}
\af{A}(u) &= \af{B}(u) \\ & \cup 
	\bigcup\limits_{
	\scriptsize
	\begin{array}{rl}
	u' \in \set{u} \cup \af{B}(u) 
	\end{array}
	}
	\af{C}(u')
\end{align*}
\end{myparagraph}

\begin{myparagraph}{Before Last Sets}
For rule $A \rightarrow a$, 
\begin{align*}
\bl{A}(\thr) &=
\begin{cases}
  \set{\thr, \lk} &\text{if } a = \ev{\thr}{\acq{\lk}}\\
  \set{\thr, \thr'} &\text{if } a = \ev{\thr}{\join{\thr'}}\\
  \set{\thr}  &\text{if } a = \ev{\thr}{o} \text{ and }
                o\not\in \set{\acq{\cdot}, \join{\cdot}} \\
  \set{\thr'} &\text{if } a = \ev{\thr'}{\fork{\thr}}\\
  \emptyset &\text{otherwise}
\end{cases} \\
\bl{A}(\lk) &=
\begin{cases}
  \set{\thr} &\text{if } a = \ev{\thr}{\rel{\lk}}\\
  \emptyset &\text{otherwise}
\end{cases}
\end{align*}
For rule $A \rightarrow BC$
\begin{align*}
\bl{A}(u) &= \bl{C}(u) \\ &\cup
        \bigcup\limits_{
        \scriptsize
        \begin{array}{rl}
        u' \in \set{u} \cup \bl{C}(u)
        \end{array}
        }
        \bl{B}(u')
\end{align*}
\end{myparagraph}

\begin{myparagraph}{After Read Sets}
For rule $A \rightarrow a$,
\begin{align*}
\ar{A}(\thr, x) = 
  \begin{cases}
    \{t\}, &\text{ if } a = \ev{t}{\rd{x}}\\
    \emptyset, &\text{ otherwise }
    \end{cases}
\end{align*}
For rule $A \rightarrow BC$,
\begin{equation*}
\ar{A}(\thr, x)  = \left\{ 
      \begin{array}{ll}
        \ar{C}(\thr, x) & \hspace{-0.5in} \mbox{if } \ar{C}(\thr, x) \\
                        & \hspace{-0.3in}  \neq \emptyset \\
        \ar{B}(\thr, x)  \cup & \hspace{-0.3in} \mbox{otherwise} \\
        \bigcup\limits_{u \in \ar{B}(\thr, x)}  \af{C}(u)
      \end{array} \right.
\end{equation*}

\end{myparagraph}

\begin{myparagraph}{After Write Sets}
For rule $A \rightarrow a$,
\begin{align*}
\aw{A}(x) =
  \begin{cases}
    \set{\thr} &
    \text{if } = \ev{\thr}{\wt{x}}\\
    \emptyset &\text{ otherwise }
  \end{cases}
\end{align*}
For rule $A \rightarrow BC$,
\begin{equation*}
\aw{A}(x)  = \left\{ 
      \begin{array}{ll}
        \aw{C}(x) & \hspace{-0.5in} \mbox{if } \aw{C}(x) \\
                        & \hspace{-0.3in}  \neq \emptyset \\
        \aw{B}(x)  \cup & \hspace{-0.3in} \mbox{otherwise} \\
        \bigcup\limits_{u \in \aw{B}(x)}  \af{C}(u)
      \end{array} \right.
\end{equation*}
\end{myparagraph}

\begin{myparagraph}{Before Read Sets}
For rule $A \rightarrow a$,
\begin{align*}
\br{A}(\thr, x) = 
  \begin{cases}
    \{t\}, &\text{ if } a = \ev{t}{\rd{x}}\\
    \emptyset, &\text{ otherwise }
    \end{cases}
\end{align*}
For rule $A \rightarrow BC$,
\begin{equation*}
\br{A}(\thr, x)  = \left\{ 
      \begin{array}{ll}
        \br{B}(\thr, x) & \hspace{-0.5in} \mbox{if } \br{B}(\thr, x) \\
                        & \hspace{-0.3in}  \neq \emptyset \\
        \br{C}(\thr, x)  \cup & \hspace{-0.3in} \mbox{otherwise} \\
        \bigcup\limits_{u \in \br{C}(\thr, x)}  \bl{B}(u)
      \end{array} \right.
\end{equation*}

\end{myparagraph}

\begin{myparagraph}{Before Write Sets}
For rule $A \rightarrow a$,
\begin{align*}
\bw{A}(x) =
  \begin{cases}
    \set{\thr} &
    \text{if } = \ev{\thr}{\wt{x}}\\
    \emptyset &\text{ otherwise }
  \end{cases}
\end{align*}
For rule $A \rightarrow BC$,
\begin{equation*}
\bw{A}(x)  = \left\{ 
      \begin{array}{ll}
        \bw{B}(x) & \hspace{-0.5in} \mbox{if } \bw{B}(x) \\
                        & \hspace{-0.3in}  \neq \emptyset \\
        \bw{C}(x)  \cup & \hspace{-0.3in} \mbox{otherwise} \\
        \bigcup\limits_{u \in \bw{C}(x)}  \bl{B}(u)
      \end{array} \right.
\end{equation*}
\end{myparagraph}

% \begin{myparagraph}{Before Read Sets}
% For rule $A \rightarrow a$,
% \begin{align*}
% \br{A}(x) =
%   \begin{cases}
%     \set{\set{\thr}} &
%     \text{if } = \ev{\thr}{\rd{x}}\\
%     \emptyset &\text{ otherwise }
%   \end{cases}
% \end{align*}
% For rule $A \rightarrow BC$,
% \begin{align*}
% \br{A}(x) &= \br{B}(x) \\ &\cup
%         \setpred{v \cup \bigcup\limits_{u \in v} \bl{B}(u)}{v \in \br{C}(x)}
% \end{align*}
% \end{myparagraph}

% \begin{myparagraph}{Before Write Sets}
% For rule $A \rightarrow a$,
% \begin{align*}
% \bw{A}(x) =
%   \begin{cases}
%     \set{\set{\thr}} &
%     \text{if } = \ev{\thr}{\wt{x}}\\
%     \emptyset &\text{ otherwise }
%   \end{cases}
% \end{align*}
% For rule $A \rightarrow BC$,
% \begin{align*}
% \bw{A}(x) &= \bw{B}(x) \\ &\cup
%         \setpred{v \cup \bigcup\limits_{u \in v} \bl{B}(u)}{v \in \bw{C}(x)}
% \end{align*}
% \end{myparagraph}

\subsection{Correctness of $\s{HB}$ algorithm}
\applabel{hb-correct}

In this section, we prove that the inductive algorithm for $\s{HB}$
described in the previous section is correct. The proof will rely on
Lemmas~\ref{lem:after-cross-chunk} and~\ref{lem:before-cross-chunk},
and \thmref{race-correctness}. 

\begin{theorem}
\thmlabel{hbalgo-correctness}
The algorithm presented in \appref{hb-algo} correctly detects
{\hbrace}s.
\end{theorem}
\begin{proof}
\thmref{race-correctness} already establishes the correctness of the
definition of the predicate $\race(\cdot)$. Thus, to complete the
proof of the correctness of our inductive algorithm, all we need to
prove is that the sets $\af{}$, $\bl{}$, $\ar{}$, $\aw{}$,
$\br{}$, and $\bw{}$ are
all being computed correctly by the inductive definitions in
\appref{hb-algo}. We will argue the correctness of the sets $\af{}$
and $\ar{}$; the proof of correctness for the remaining cases is
similar and skipped.

Let us begin the computation of $\af{A}(u)$ for a non-terminal $A$ and
lock/thread $u$. When $A$ has a rule $A \rightarrow a$, where $a$ is
an event, the computation of $\af{A}$ is consistent with the
definition in \equref{after-first-thread} and \equref{after-first-lock} . In the inductive step, consider
$A$ with rule $A \rightarrow BC$. We need to prove that
\begin{align*}
\af{A}(u) &= \af{B}(u) \cup
        \bigcup\limits_{
        \scriptsize         
        \begin{array}{rl}
        u' \in \set{u} \cup \af{B}(u)
        \end{array}
        }
        \af{C}(u')
\end{align*}
Let us begin by considering the simple case when 
$\s{first}_{A}^u = \s{undefined}$. 
In this case, by definition, $\af{A}(u) =
\emptyset$. Also, since $\s{first}_{A}^u = \s{undefined}$, 
$\s{first}_{B}^u = \s{undefined}$ and $\s{first}_{C}^u = \s{undefined}$. 
Therefore, inductively, $\af{B}(u) = \emptyset$ and $\af{C}(u) = \emptyset$,
giving us that the right hand side will also evaluate to
$\emptyset$. 
So let us now consider the case where 
$\s{first}_{A}^u = e \in \sem{A}$ is defined. 
There are two possibilities to consider. 
If $e \in \sem{C}$ (and therefore $e \not\in \sem{B}$) 
then $\af{B}(u) = \emptyset$, and
\begin{align*}
& \af{B}(u) \cup
        \bigcup\limits_{
        \scriptsize
        \begin{array}{rl}
        u' \in \set{u} \cup \af{B}(u)
        \end{array}
        }
        \af{C}(u') \\
= \:
& \emptyset \cup \af{C}(u) = \af{A}(u)
\end{align*}
which is indeed correct. The second sub-case to consider is when $e
\in \sem{B}$. Then, by \lemref{after-cross-chunk}, we have
\begin{align*}
\af{A}(u) & = \after{A}(e)  = \after{B}(e) \cup \bigcup\limits_{u' \in
  \after{B}(e)} \af{C}(u')\\
 & = \af{B}(u) \cup
        \bigcup\limits_{
        \scriptsize
        \begin{array}{rl}
        u' \in \cup \af{B}(u)
        \end{array}
        }
        \af{C}(u')
\end{align*}
Since $u \in \af{B}(u)$, we have $\af{B}(u) = \set{u} \cup \af{B}(u)$, and thus,
\begin{align*}
\af{A}(u) & = \af{B}(u) \cup
        \bigcup\limits_{
        \scriptsize
        \begin{array}{rl}
        u' \in \set{u} \cup \af{B}(u)
        \end{array}
        }
        \af{C}(u')
\end{align*}

Let us now establish the correctness of the inductive definition of
$\ar{A}(\thr, x)$ for a non-terminal $A$ and variable $x$. 
Again the base case of the non-terminal having rule 
$A \rightarrow a$, where $a$ is
an event, is clearly correct as per the definition in
\equref{ar-aw-br-bw}. 
In the inductive step of a rule $A \rightarrow
BC$, we need to prove the correctness of the following equation:
\begin{equation*}
\ar{A}(\thr, x)  = \left\{ 
      \begin{array}{ll}
        \ar{C}(\thr, x) & \hspace{-0.5in} \mbox{if } \ar{C}(\thr, x) \\
                        & \hspace{-0.3in} \neq \emptyset \\
        \ar{B}(\thr, x)  \cup & \hspace{-0.3in} \mbox{otherwise} \\
        \bigcup\limits_{u \in \ar{B}(\thr, x)}  \af{C}(u)
      \end{array} \right.
\end{equation*}

Let $e = \last{A}(\reads{A}(\thr, x))$.
If $e$ is $\s{undefined}$, then both $\ar{C}(\thr, x)$
and $\ar{B}(\thr, x)$ are $\emptyset$, and so is $\ar{A}(\thr, x)$.

Next consider the case when $e$ is defined and belongs to the chunk $C$
(i.e., $e \in \sem{C}$).
Then, clearly, $\after{A}(e) = \after{C}(e) = \ar{C}(\thr, x)$.
Inductively, $\ar{C}(\thr, x)$ will indeed by non-empty and thus
$\ar{A}(\thr, x) = \ar{C}(\thr, x)$.

The last case is when $e \in \sem{B}$.
The correctness then follows from \lemref{after-cross-chunk}.
% $\ar{A}(x)$ is the collection of all $\after{A}(e)$, where $e$ is a
% $\rd{x}$-event in $\sem{A}$. The $\rd{x}$ events in $\sem{A}$ can
% divided into two categories --- those that belong to $\sem{B}$ and
% those that belong to $\sem{C}$. For $e \in \reads{C}(x)$, observe that
% since $\setpred{e'}{e \HB[C] e'} = \setpred{e'}{e \HB[A] e'}$, we have
% $\after{A}(e) = \after{C}(e)$. On the other hand, for $e \in
% \reads{B}(x)$, $\after{A}(e)$ is given by
% \lemref{after-cross-chunk}. These two observations together establish
% the correctness of the $\ar{A}(x)$ in the inductive case.
\end{proof}

%% file: app-lockset-algo.tex
%!TEX root = main.tex

\section{Complete Algorithm for Lockset Computation on Compressed Traces with proof of Correctness}
\applabel{app-lockset-algo}

Here, we will first present the complete algorithm for detecting violations of the lockset discipline,
and then argue about correctness. 

\subsection{Algorithm}
\seclabel{lockset-algo}

We present the complete inductive algorithm for detecting 
violations of lockset discipline.
on traces represented by SLPs. The algorithm is inductive, and
computes various data-structures for each non-terminal in the SLP. 
The definitions of $\oacq{}$ and $\orel{}$ is given in \secref{lockset}.
Here we will first precisely state the definition for $\lockset{}$:

\begin{definition}
\deflabel{lockset_def}
For a trace $\tr$, thread $\thr$ and variable $x$,
we define 
\begin{align*}
\lockset{\tr}(\thr, x) = 
\begin{cases}
\top & \text{if } \access{\tr}(\thr, x) = \emptyset\\
\s{RealLocks}_\tr^{(\thr, x)} \cup \set{\dmlock,\dmlockthr{\thr}}  & \text{if } x \not\in \wvars{\proj{\tr}{\thr}} \\
\s{RealLocks}_\tr^{(\thr, x)} \cup \set{\dmlockthr{\thr}} & \text{otherwise}
\end{cases}
% 
% & 
% \setpred{\lk \in \locks(\tr)}{\forall e \in \access{\tr}(\thr, x), \\
% & \lk \in \locksheld{\tr}(e)} \\
% \cup \bigcap\limits_{\thr \in }
\end{align*}
where
\begin{align*}
\s{RealLocks}_\tr^{(\thr, x)} &= \bigcap\limits_{e \in \access{\tr}(\thr, x)} \locksheld{\tr}(\thr, x)
\end{align*}
\end{definition}
The set $\top$ in the above definition, is a universal set such that for any set $S$, $S \cap \top = S$,
and $S \cup \top = \top$. Also, $\top \cap \top = \top$ and $\top \cup \top = \top$.
This is introduced only for notational convenience.
When the set of all locks (including dummy locks)
is known, it can be used instead of using $\top$.

We will now present the algorithms that compute these data. 
Each of these data-structures is computed for a non-terminal $A$ and the
computation depends on the rule associated with $A$. Recall that there
are two types of rules --- either $A \rightarrow a$, where $a$ is an
event, or $A \rightarrow BC$, where $B$ and $C$ are non-terminals. 

\begin{myparagraph}{Computing  $\mathbf{\oacq{A}}$}

For rule $A \rightarrow a$, 
\begin{align*}
		\oacq{A}(\thr, \lk) &= 
		\begin{cases}
		1, &\text{ if } a = \ev{\thr}{\acq{\lk}}\\
		0, &\text{otherwise}
		\end{cases}
\end{align*}
For $A \rightarrow BC$,
\begin{equation*}
\begin{array}{rcl}
\oacq{A}(\thr, \lk) \!\!\!\!\!\!&=& \!\!\!\!  \oacq{C}(\thr, \lk) \\ 
& + & \!\!\!\! \max\{0, \oacq{B}(\thr, \lk) - \orel{C}(\thr, \lk)\}
\end{array}
\end{equation*}
\end{myparagraph}

\begin{myparagraph}{Computing $\orel{A}$}

For rule $A \rightarrow a$, 
\begin{align*}
		\orel{A}(\thr, \lk) &= 
		\begin{cases}
		1, &\text{ if } a = \ev{\thr}{\rel{\lk}}\\
		0, &\text{otherwise}
		\end{cases}
\end{align*}
For $A \rightarrow BC$,
\begin{equation*}
\begin{array}{rcl}
\orel{A}(\thr, \lk) \!\!\!\!\!\!&=& \!\!\!\!  \orel{B}(\thr, \lk) \\ 
& + & \!\!\!\! \max\{0, \orel{C}(\thr, \lk) - \oacq{B}(\thr,\lk)\}
\end{array}
\end{equation*}
\end{myparagraph}

\begin{myparagraph}{Computing $\lockset{A}$}

For rule $A \rightarrow a$, 
\begin{align*}
		\lockset{A}(\thr, \lk) &= 
		\begin{cases}
		\set{\dmlock,\dmlockthr{\thr}}, &\text{ if } a = \ev{\thr}{\rd{x}}\\
		\set{\dmlockthr{\thr}}, &\text{ if } a = \ev{\thr}{\wt{x}}\\
		\top, &\text{otherwise}
		\end{cases}
\end{align*}

For $A \rightarrow BC$,
\begin{equation*}
\begin{array}{ll}
	\lockset{A}(\thr, x) \\ 
	= \big( \lockset{B}(\thr, x) \cup \setpred{\lk}{\orel{C}(\thr,\lk) > \oacq{B}(\thr,\lk)}\big) \\
	\cap \big( \lockset{C}(\thr, x) \cup \setpred{\lk}{\oacq{B}(\thr',\lk) > \orel{C}(\thr',\lk)}\big)
\end{array} 
\end{equation*}
\end{myparagraph}

\begin{myparagraph}{Checking lockset violation}

For a non-terminal $A$, violation is reported if 
there is a variable $x$ accessed in $A$ for which
the following condition holds
\begin{equation*}
\bigcap\limits_{\thr \in \threads(A)} \lockset{A}(\thr, x) = \emptyset
\end{equation*}

\end{myparagraph}

\subsection{Correctness of $\lockset{}$ algorithm}
\applabel{lockset-correct}

Here, we will prove that the inductive algorithm
presented in \secref{lockset-algo} is correct.

We first note the following simple but important observation,
which is similar to \propref{subchunk-race},
and informally states that, the presence/absence of lockset violations in a trace $\tr$, does
not get affected by the ``context'' in which $\tr$
is placed.

\begin{proposition}
\proplabel{subchunk-lockset}
Let $\tr$ be a trace and $x$ be a variable
accessed in $\tr$.
The following two statements are equivalent
\begin{enumerate}
	\item 
	\begin{equation*}
\bigcap\limits_{\thr \in \threads(\tr)} \lockset{\tr}(\thr, x) = \emptyset
\end{equation*}
	\item For any trace
$\tr' = \tr_1\tr\tr_2$,
\begin{equation*}
\bigcap\limits_{\thr \in \threads(\tr')} \lockset{\tr'}(\thr, x) = \emptyset
\end{equation*}
\end{enumerate}
\end{proposition}

Thus if a non-terminal different from the
start symbol $S$ of the grammar $G$
reports a violation, there will be a violation in $S$ too.
Finally, the correctness of the inductive algorithm will
be complete if we prove that each of the inductive formulations
in the previous section are correct:

\begin{theorem}
The algorithm presented in \secref{lockset-algo} correctly detects
violations of lockset discipline.
\end{theorem}

\begin{proof}
We will prove that all the inductive formulations are correct.

First, let us prove the correctness of the inductive
formulation for $\oacq{A}(\thr, \lk)$.
We will skip the proof for $\orel{A}(\thr, \lk)$ because it is similar.

Let us fix some notation before we begin.
For non-terminal $D$, thread $\thr$ and lock $\lk$,
let 
\[\s{unmatchedAcq}^{(\thr, \lk)}_D = 
\setpred{e = \ev{\thr}{\acq{\lk}} \in D}{\match(e) \not\in \sem{D}}.
\]
and
\[\s{unmatchedRel}^{(\thr, \lk)}_D = 
\setpred{e = \ev{\thr}{\rel{\lk}} \in D}{\match(e) \not\in \sem{D}}.
\]
Then, $\oacq{D}(\thr, \lk) = |\s{unmatchedAcq}^{(\thr, \lk)}_D| $
and $\orel{D}(\thr, \lk) = |\s{unmatchedRel}^{(\thr, \lk)}_D| $.

In the base case, we have the rule $A \rightarrow a$.
The set $\s{unmatchedAcq}^{(\thr, \lk)}_A$
is non-empty only if $a = \ev{\thr}{\acq{\lk}}$ is an acquire event
in which case $|\s{unmatchedAcq}^{(\thr, \lk)}_A| = 1$. 
In all other cases $|\s{unmatchedAcq}^{(\thr, \lk)}_A| = 0$.
This justifies the base case.

In the inductive case, we have the rule
$A \rightarrow BC$.
Then, the set $\s{unmatchedAcq}^{(\thr, \lk)}_A$ can be expressed as
a disjoint union $S_B \cup S_C$ ($S_B \cap S_C = \emptyset$), where
\[
S_B = \setpred{e \in \s{unmatchedAcq}^{(\thr, \lk)}_A}{ e\in \proj{B}{\thr}}
\]
and
\[
S_C = \setpred{e \in \s{unmatchedAcq}^{(\thr, \lk)}_A}{ e\in \proj{C}{\thr}}
\]
First, note that for every event $e \in S_C$, $\match{e} \not\in A$
and thus $\match{e} \not\in C$.
Thus we have $S_C \subseteq \s{unmatchedAcq}^{(\thr, \lk)}_C$.
Next, see that every event $e \in \s{unmatchedAcq}^{(\thr, \lk)}_C$
is clearly unmatched even in $A$, and thus belongs to $\s{unmatchedAcq}^{(\thr, \lk)}_A$,
and thus to $S_C$. This gives $\s{unmatchedAcq}^{(\thr, \lk)}_C \subseteq S_C$.
Hence, $S_C = \s{unmatchedAcq}^{(\thr, \lk)}_C$.
The cardinality of $S_C = \s{unmatchedAcq}^{(\thr, \lk)}_C = \oacq{C}(\thr, \lk)$.
% Now, notice that $S_B \subseteq \s{unmatchedAcq}^{(\thr, \lk)}_B$,
% because for an event $e \in B$ which is not matched in $A$ will also be unmatched
% in $A$.
Next, consider the following set
\[
S^1_B = \setpred{e \in \s{unmatchedAcq}^{(\thr, \lk)}_B}{
\match(e) \in \s{unmatchedRel}^{(\thr, \lk)}_C}
\]
Also, let $S^2_B = \s{unmatchedAcq}^{(\thr, \lk)}_B \setminus S^1_B$.
Then, every event $e \in S^2_B$ is unmatched in $A$, and thus belongs
to $\s{unmatchedAcq}^{(\thr, \lk)}_A$ and hence to $\s{unmatchedAcq}^{(\thr, \lk)}_B$.
This means $S^2_B \subseteq S_B$.
Also, consider an event $e \in S_B$. Since it is unmatched in $A$,
it must also be unmatched in $B$, and thus $e \in \s{unmatchedAcq}^{(\thr, \lk)}_B$.
Also, it does not have a matching release in $C$
and thus, $e \not\in S^1_B$.
Thus, $e \in S^2_B$.
Clearly, $S_B = S^2_B$.
Also, the cardinality of $S^2_B$
is just $\max\{0, \oacq{B}(\thr, \lk) - \orel{C}(\thr, \lk)\}$.

Since $S_B \cap S_C = \emptyset$, we have the desired result. \\

Let us now prove the correctness of the
inductive formulation of $\lockset{}$.
The base case $A \rightarrow a$ indeed matches \defref{lockset_def}.

We consider the case $A \rightarrow BC$.
If $\access{A}(\thr, x) = \emptyset$, both
$\access{B}(\thr, x) = \emptyset$ and $\access{C}(\thr, x) = \emptyset$.
In this case, using the inductive hypothesis,
$\lockset{B}(\thr, x) = \top$ and $\lockset{C}(\thr, x) = \top$.
Thus, $\lockset{A}(\thr, x)$ evaluates to $\top$, which is correct.

Now consider the case when $\access{A}(\thr, x)$ is non-empty.
If $\access{B}(\thr, x)$ is empty,
then inductively, $\lockset{B}(\thr, x) = \top$ 
and $\lockset{A}(\thr, x)$ evaluates to $\lockset{C}(\thr, x)$.
This is correct since $\access{A}(\thr, x) = \access{C}(\thr, x)$
and $\wvars{\proj{A}{\thr}} = \wvars{\proj{C}{\thr}}$
 in this case.
The case when $\access{C}(\thr, x)$ is similar.
Next, consider the case when 
$\access{B}(\thr, x) \neq \emptyset$
and $\access{C}(\thr, x) \neq \emptyset$.
By the inductive hypothesis,
$\dmlockthr{\thr} \in \lockset{B}(\thr, x)$
and $\dmlockthr{\thr} \in \lockset{C}(\thr, x)$.
Clearly, $\dmlockthr{\thr} \in \lockset{A}(\thr, x)$, as expected.
Now we have two cases.
First, the case when
$\access{A}(\thr, x) = \reads{A}(\thr, x)$,
that is, there is no write event to $x$ in $\proj{A}{\thr}$.
Then, $\access{B}(\thr, x) = \reads{B}(\thr, x)$
and $\access{C}(\thr, x) = \reads{C}(\thr, x)$.
By induction, $\dmlock \in  \lockset{B}(\thr, x)$
and $\dmlock \in \lockset{C}(\thr, x)$.
Thus, $\dmlock \in \lockset{A}(\thr, x)$, which is correct.
Otherwise, atlast one of $\lockset{B}(\thr, x)$
and $\lockset{C}(\thr, x)$ does not have $\dmlock$.
As a result, $\dmlock \not\in \lockset{A}(\thr, x)$
Finally, 
\begin{equation*}
\begin{array}{ll}
\s{RealLocks}_A^{(\thr, x)}    
& =  \smashoperator[r]{\bigcap_{e \in \access{A}(\thr,x)}} \locksheld{A}(e) \\
& =  \smashoperator[r]{\bigcap_{e \in \access{B}(\thr,x)}} \locksheld{A}(e) \cap \smashoperator[r]{\bigcap_{e \in \access{C}(\thr,x)}} \locksheld{A}(e) \\
% & = 

% \locksheld{B}(e) \\
%           & \cup \setpred{\lk}{\orel{C}(\thr,\lk) > \oacq{B}(\thr,\lk)}

% & = \big( \lockset{B}(t, x) \cup \setpred{\lk}{\orel{C}(t,\lk) > \oacq{B}(t,\lk)}\big) \\
% 	& \cap \big( \lockset{C}(t, x) \cup \setpred{\lk}{\oacq{B}(t',\lk) > \orel{C}(t',\lk)}\big)
\end{array}
\end{equation*}

By \lemref{locksheld}, for an event $e \in B$, we have
\begin{align*}
\locksheld{A}(e) = & \locksheld{B}(e) \\
          & \cup \setpred{\lk}{\orel{C}(\thr,\lk) > \oacq{B}(\thr,\lk)}
\end{align*}
Thus, 
\begin{align*}
\smashoperator[r]{\bigcap_{e \in \access{B}(\thr,x)}} \locksheld{A}(e)
= &
\smashoperator[r]{\bigcap_{e \in \access{B}(\thr,x)}} \Big( \locksheld{B}(e) \cup \\
& \setpred{\lk}{  \orel{C}(\thr,\lk) > \oacq{B}(\thr,\lk)} \Big) \\
= & \Big( \smashoperator[r]{\bigcap_{e \in \access{B}(\thr,x)}}  \locksheld{B}(e) \Big) \\
\cup & \setpred{\lk}{  \orel{C}(\thr,\lk) > \oacq{B}(\thr,\lk)} \\
= & \s{RealLocks}_B ^{(\thr, x)} \cup \\
& \setpred{\lk}{  \orel{C}(\thr,\lk) > \oacq{B}(\thr,\lk)}
\end{align*}

Similarly, 
\begin{align*}
\smashoperator[r]{\bigcap_{e \in \access{C}(\thr,x)}} \locksheld{A}(e)
= & \s{RealLocks}_C ^{(\thr, x)} \cup \\
& \setpred{\lk}{  \oacq{B}(\thr,\lk) > \orel{B}(\thr,\lk)}
\end{align*}

Thus, we have
\begin{equation*}
\begin{array}{ll}
	\lockset{A}(\thr, x) \\ 
	= \big( \lockset{B}(\thr, x) \cup \setpred{\lk}{\orel{C}(\thr,\lk) > \oacq{B}(\thr,\lk)}\big) \\
	\cap \big( \lockset{C}(\thr, x) \cup \setpred{\lk}{\oacq{B}(\thr',\lk) > \orel{C}(\thr',\lk)}\big)
\end{array} 
\end{equation*}
\end{proof}